\documentclass[12pt]{ociamthesis}  

\usepackage{latexsym}
\usepackage{epsfig,amssymb,euscript}
\usepackage{amsmath}
\usepackage{mathrsfs}
\usepackage{setspace}

\def\be{\begin{equation}}
\def\ee{\end{equation}}
\def\bea{\begin{eqnarray}}
\def\eea{\end{eqnarray}}

\newcommand{\nn}{\nonumber}

\def\Dslash{\,\,{\raise.15ex\hbox{/}\mkern-12mu D}}
\def\Dbarslash{\,\,{\raise.15ex\hbox{/}\mkern-12mu {\bar D}}}
\def\delslash{\,\,{\raise.15ex\hbox{/}\mkern-9mu \partial}}
\def\delbarslash{\,\,{\raise.15ex\hbox{/}\mkern-9mu {\bar\partial}}}
\def\pslash{\,\,{\raise.15ex\hbox{/}\mkern-9mu p}}
\def\calDslash{\,\,{\raise.15ex\hbox{/}\mkern-12mu {\cal D}}}

\newcommand\R{\mathbb{R}}
\newcommand\Z{\mathbb{Z}}

\newcommand\C{\mathbb{C}}

\newcommand\diff{\mathrm{d}}
\newcommand\ex{\mathrm{e}}

\newcommand{\vol}{\mathrm{vol}}
\newcommand{\dd}{\mathrm{d}}
\newcommand{\ii}{\mathrm{i}}

\newcommand{\me}{\mathrm{e}}
\newcommand{\comment}[1]{}
\newcommand{\secref}[1]{\S\ref{#1}}

\usepackage{cite} 

\usepackage{setspace}

\title{The AdS/CFT Correspondence\\ \vspace{.4cm} 
        and Symmetry Breaking}   

\author{Nessi Benishti}             
\college{St John's College}  

\degree{Doctor of Philosophy}     
\degreedate{Trinity 2011}         

\begin{document}

\baselineskip=20pt plus1pt

\setcounter{secnumdepth}{3}
\setcounter{tocdepth}{3}

\maketitle                  
\begin{dedication}

\emph{I dedicate this thesis to my family in gratitude for their constant support.}

\end{dedication}        
\begin{acknowledgements}

It is a pleasure to thank the people who made this thesis possible. It is difficult to overstate my gratitude to my supervisor James Sparks whose patience and kindness, as well as his immense knowledge, have been invaluable to me. I appreciate very much his enthusiasm, his inspiration, and his great efforts to explain everything clearly. 

I am truly indebted and thankful to my supervisor Yang-Hui He for his patience, motivation, enthusiasm, and academic experience. Throughout my research course period, he provided encouragement, sound advice, and excellent teaching.

I would like to show my gratitude to Andre Lukas for his willingness to act as my supervisor in the absence of Yang. I owe sincere and earnest thankfulness to my collaborator Diego Rodriguez-Gomez from whom I learned a great deal. I also wish to thank Dario Martelli for insightful discussions and email correspondence on issues concerning the work presented here.

I would like to thank the members of the Oxford string theory group for providing a rich and exciting environment.

\end{acknowledgements}   
\pagestyle{empty} 
\begin{titlepage}
\begin{originality}

In this thesis we present selected results taken from the author's published work \cite{Benishti:2009ky, Benishti:2010jn, Benishti:2011ab}. Due to space constraints not all the relevant material is presented and the reader is referred where necessary to extra results in the original papers.

\secref{sec:Gflux} and \secref{sec:unhiggsing} are based on the article \cite{Benishti:2009ky} written in collaboration with Yang-Hui He and James Sparks. \secref{sec:2}, \secref{sec:3}, \secref{sec:4-1}, \secref{sec:generalwarped}, \secref{sec:5}, \secref{sec:6} and \secref{sec:D} are based on the article \cite{Benishti:2010jn} with Diego Rodriguez-Gomez and James Sparks. \secref{sec:B3} and \secref{sec:B4} are based on the article \cite{Benishti:2011ab}. \secref{sec:exo} is based on unpublished work.

Original contributions reported in this thesis include:

\begin{itemize}
\item
The examination of baryonic-type symmetries in the AdS$_4$/CFT$_3$ correspondence and the study of different choices of quantization of gauge fields in AdS$_4$ that lead to different field theory duals to the same gravitational background (\secref{sec:3}). 
\item
The classification of isolated Calabi-Yau four-fold singularities with no vanishing six-cycles and the extension of AdS$_4$/CFT$_3$ with proposed candidate dual gauge theories for such singularities (\secref{sec:gen}).
\item
The study of the physics of vacua in which the above-mentioned baryonic symmetries are spontaneously broken, showing that a dual gravity analysis involving resolutions of Calabi-Yau singularities, baryonic condensates, Goldstone bosons and global strings matches with field theory expectations (\secref{sec:4-1} and \secref{sec:generalwarped}).
\item
The derivation of a general formula for the action of a Euclidean M5 brane which is wrapped on a minimal six-submanifold (\secref{sec:M5general}).
\item
The exploration of the role of supergravity fluxes in the dual description of the Higgs effect (\secref{sec:Gflux}).
\item
The identification of the importance of the M-theory circle in the supergravity dual to Higgsing of the field theories, and the study of the related exotic baryonic branches and renormalization group flow scenarios (\secref{sec:exo}).
\item
The study of non-perturbative corrections due to M5 instantons wrapped on six-cycles in Calabi-Yau four-folds and the implications for the AdS$_4$/CFT$_3$ correspondence (\secref{sec:5}). 
\item
The derivation of a general formula relating the M5 instanton action to L$^2$ normalizable harmonic two-forms in the resolved backgrounds (\secref{s:general-V}).
\item
The development of a general un-Higgsing algorithm that allows one to construct quiver-Chern-Simons theories by blowing up (\secref{sec:unhiggsing}).
\item
The study of resolutions of backgrounds that contain four-cycles in the AdS$_5$/CFT$_4$ correspondence and the demonstration of the periodicity of the directions in the moduli space of the field theory that are dual to the Neveu-Schwarz-Neveu-Schwarz (NS-NS) B-field moduli (\secref{sec:B3} and \secref{sec:B4}).
\item
The discovery and study of the phenomena in which critical NS-NS B-field vacuum expectation values in such backgrounds correspond to Higgsings in the field theory that lead to confinement and emergence of baryonic symmetries during the renormalization group flow (\secref{sec:B3} and \secref{sec:B4}).
\end{itemize}

\end{originality}
\begin{abstract}

In the first part of this thesis we study baryonic $U(1)$ symmetries dual to Betti multiplets in the AdS$_4$/CFT$_3$ correspondence for M2 branes at Calabi-Yau four-fold singularities. Such short multiplets originate from the Kaluza-Klein compactification of eleven-dimensional supergravity on the corresponding Sasaki-Einstein seven-manifolds. Analysis of the boundary conditions for vector fields in AdS$_4$ allows for a choice where wrapped M5 brane states carrying non-zero charge under such symmetries can be considered. We begin by focusing on isolated toric singularities without vanishing six-cycles, which we classify, and propose for them field theory duals. We then study in detail the cone over the well-known Sasaki-Einstein space $Q^{111}$, which is a $U(1)$ fibration over $\mathbb{CP}^1 \times \mathbb{CP}^1 \times \mathbb{CP}^1$. The boundary conditions considered are dual to a CFT where the gauge group is $U(1)^2 \times SU(N)^4$. We find agreement between the spectrum of gauge-invariant baryonic-type operators in this theory and M5 branes wrapping five-cycles in the $Q^{111}$ space. Moreover, the physics of vacua in
which these symmetries are spontaneously broken precisely matches a dual gravity analysis involving resolutions of the singularity, where we are able to match condensates of the baryonic operators, Goldstone bosons and global strings. We then study the implications of turning on a closed three-form with non-zero periods through torsion three cycles in the Sasaki-Einstein manifold. This three-form, otherwise known as torsion $G$-flux, non-trivially affects the supergravity dual of Higgsing, and we show that the supergravity and field theory analyses precisely match in an example based on the Sasaki-Einstein manifold $Y^{1,2}(\mathbb{CP}^2)$, which is a $S^3$ bundle over $\mathbb{CP}^2$. We then explain how the choice of M-theory circle in the background can result in exotic renormalization group flows in the dual field theory, and study this in detail for the Sasaki-Einstein manifold $Y^{1,2}(\mathbb{CP}^2)$. We also argue more generally that theories where the resolutions have six-cycles are expected to receive non-perturbative corrections from M5 brane instantons. We give a general formula relating the instanton action to normalizable harmonic two-forms, and compute it explicitly for the Sasaki-Einstein $Q^{222}$ example, which is a $\Z_2$ orbifold of $Q^{111}$ in which the free $\Z_2$ quotient is along the R-symmetry $U(1)$ fibre. The holographic interpretation of such instantons is currently unclear.

In the second part of this thesis we study the breaking of baryonic symmetries in the AdS$_5$/CFT$_4$ correspondence for D3 branes at Calabi-Yau three-fold singularities. This leads, for particular vacuum expectation values, to the emergence of non-anomalous baryonic symmetries during the renormalization group flow. We identify these vacuum expectation values with critical values of the NS-NS B-field moduli in the dual supergravity backgrounds. We study in detail the $\C^3/\Z_3$ orbifold theory and the dual supergravity backgrounds that correspond to the breaking of the emerging baryonic symmetries, and identify the expected Goldstone bosons and global strings in the
infra-red. In doing so we confirm the claim that the emerging symmetries are indeed non-anomalous baryonic symmetries.
\end{abstract}          
\end{titlepage}

\pagestyle{plain}
\begin{romanpages}          
\tableofcontents            
\end{romanpages}            



\chapter*{Introduction}

The AdS/CFT correspondence is a conjectured duality between string theory (and hence gravity) in AdS space and a conformal field theory on the boundary of this space \cite{Maldacena:1997re}. The duality has produced major progress in our understanding of the intimate relationship between the dynamics of gauge theories and strings. A remarkable fact about the duality is that whenever one description is strongly coupled the dual description is weakly coupled. Thus, besides being of great theoretical interest, the duality is becoming a useful tool for studying strongly coupled gauge theories. Our main interest will be to study various symmetry considerations, that dominate modern fundamental physics, in this framework. 

The AdS/CFT correspondence is one realization of the holographic principle, in which a theory that includes gravity is dual to a non-gravitational field theory on the boundary. This is believed to be a general property of quantum gravity, thus more general realizations of string/gauge dualities are expected. One immediate extension of the correspondence is to non-AdS spaces that are dual to field theories in which the conformal invariance is broken. This can be achieved by deforming the action with gauge invariant operators or by giving vacuum expectation values (VEVs) to operators in the field theory. Such operations induce renormalization group (RG) flows that could end at an infra-red (IR) fixed point, or develop non-trivial IR non-conformal dynamics, like confinement. The energy scale in the field theory is geometrized and encoded as the radial coordinate in the gravity dual background. Thus, the correspondence translates the study of RG flow in the field theory to the analysis of gravity equations of motion in this background. This makes the study of such RG flows in strongly coupled field theories highly feasible. No less important is the fact that such flows are translated to interesting physics in the string theory dual.

The breaking of conformal invariance of interest to us will be induced by non-vanishing VEVs for operators that carry charge under gauge and global symmetries in supersymmetric field theories. As we will show, information about the condensates of these operators can be gained by studying certain instantonic brane configurations in the dual string theory. The breaking of symmetries by giving non-vanishing VEVs to scalar operators is the main component in the idea of spontaneous symmetry-breaking (SSB). In realistic systems, such broken symmetries are not manifest to us because the vacuum state is not invariant. This idea plays a crucial role in understanding various phenomena such as ferromagnetism, superconductivity, low energy interactions of pions, and electroweak unification of the Standard Model. In the cases in which the field theories are strongly coupled, there are generally no efficient tools with which symmetry-breaking can be studied, and therefore the AdS/CFT correspondence might be an important tool for filling this gap.
 
The main interest in this thesis will be Type IIB string theory in AdS$_5 \times Y_5$ and M-theory in AdS$_4 \times Y_7$, where $Y_5$ and $Y_7$ are Sasaki-Einstein five-manifolds and seven-manifolds, respectively. One important point to note is that the field theory in the AdS$_5$/CFT$_4$ correspondence admits a Lagrangian description since the Yang-Mills coupling can be tuned to a small value. This is because this coupling, which is related to the dilation in the string background, is exactly marginal. In this limit one expects to be able to describe the theory classically. Considering the AdS$_4 \times S^7$ background, it seems that the dual field theory is always strongly coupled. This originates from the fact that in M-theory one does not have the dilaton that can be set small. In order to overcome this apparent problem, Aharony, Bergman, Jafferis and Maldacena (ABJM) \cite{Aharony:2008ug} considered instead the AdS$_4 \times S^7/\Z_k$ geometries. The large $k$ limit reduces the geometry to AdS$_4 \times \mathbb{CP}^3$ in Type IIA string theory, which is weakly coupled. The existence of this limit guarantees the existence of a Lagrangian description for the dual field theory. This theory was shown to be a Chern-Simons (CS) theory coupled to matter in \cite{Aharony:2008ug}, where the orbifold rank $k$ is encoded by the CS levels. 

The Type IIB and M-theory backgrounds just discussed originate from the near-horizon limit of branes probing the cone over a Sasaki-Eintein manifold, which is a Calabi-Yau (CY) cone by definition. Thus the moduli space of the field theory contains a branch that corresponds to the position of the stack of branes on the CY space. One can study breaking of symmetries by moving the stack of branes away from the tip of the cone. In addition, one may break gauge symmetry, global symmetry and conformal invariance by moving in the K\"ahler moduli space or by giving VEVs to form-fields in supergravity. These will be the scenarios of interest to us in this thesis. This was first studied in the AdS$_5$/CFT$_4$ correspondence by Klebanov-Witten \cite{Klebanov:1999tb}. There it was shown that the RG flow in the field theory can be described by a string theory background with two asymptotically AdS boundaries that correspond to the IR and UV conformal fixed points. With such solutions at hand, many properties of the strongly coupled RG flows can be studied. 

We begin in \secref{part1} with the study of symmetry-breaking in AdS$_4$/CFT$_3$ correspondence, which has been little researched owing to the fact that the dual theories have been discovered only recently. We discover that the analysis required is much more involved than that found in the AdS$_5$/CFT$_4$ correspondence (see \secref{sec:intro} for an overview). In \secref{part2} we study symmetry-breaking in the AdS$_5$/CFT$_4$ correspondence for toric CY three-folds with vanishing four-cycles. In these examples, we identify an interesting and important phenomenon, namely the emergence of global symmetries during the RG flows (see \secref{overview-p2} for an overview).
\part{Symmetry Breaking in AdS$_4$/CFT$_3$} \label{part1}
\chapter{Overview}\label{sec:intro}

Over the last two years there have been major advances towards understanding the AdS$_4$/CFT$_3$ duality. Elaborating on \cite{Gustavsson:2007vu, Bagger:2007vi}, ABJM\cite{Aharony:2008ug} proposed a theory conjectured to be dual to M2 branes probing a $\mathbb{C}^4/\mathbb{Z}_k$ singularity, where $\Z_k$ acts with weights $(1,1,-1,-1)$ on the coordinates of $\C^4$. This low energy theory on the world-volume of $N$ coincident M2 branes is a $U(N)_k\times U(N)_{-k}$ quiver Chern-Simons (QCS) theory.
Motivated by this progress in understanding the maximally SUSY case, it is natural to consider M2 branes moving in less symmetric spaces, leading to versions of the duality with reduced SUSY. Inspired by ABJM \cite{Aharony:2008ug}, the theories considered are $\prod_{a=1}^G U(N_a)_{k_a}$ QCS theories with bifundamental and fundamental matter. This study was initiated in \cite{Martelli:2008si, Hanany:2008cd}, followed by a large number of works \cite{Ueda:2008hx, Martelli:2008rt,Jafferis:2009th,Benini:2011cm,Cremonesi:2010ae,Benini:2009qs, Imamura:2008qs, Hanany:2008fj, Franco:2008um, Hanany:2008gx, Amariti:2009rb, Franco:2009sp, Davey:2009sr, Davey:2009qx, Martelli:2009ga, Davey:2009bp, Hewlett:2009bx, Taki:2009wf, Davey:2011mz}. It has been argued in \cite{Gaiotto:2009mv, Gaiotto:2009yz} that the sum $\sum_{a=1}^G k_a$ corresponds to the Type IIA supergravity Romans mass parameter, which is just the Ramond-Ramond (R-R) zero-form $F_0$. In this thesis we will focus entirely on the case in which the Romans mass vanishes and the system admits an M-theory lift. 

Systematic ways to obtain field theories dual to M2 branes on toric conical CY four-folds were developed recently. These constructions exploit the fact that these CYs can be written as a $U(1)$ fibration over a seven-manifold. Considering this $U(1)$ as the M-theory circle and reducing on this direction one obtains D2 branes probing $\R \times \text{CY}_3$ in Type IIA string theory with R-R two-form flux. After compactifying the $\R$ direction to a circle and T-dualizing over this circle one obtains D3 branes probing a CY three-fold in Type IIB. The dual circle should shrink to zero size, thus the field theory on the D3 branes is effectively $(2+1)$d. Fortunately, the field theories living on such D3 branes probing toric CY three-folds are known. To obtain the theories on the M2 branes one needs to add CS terms to such quivers \cite{Martelli:2008si,Hanany:2008cd}. These CS terms originate from the coupling of the R-R flux to the D-branes \cite{Aganagic:2009zk, HananyTalk}. The above description is valid when the $U(1)$ does not degenerate over the base. Degenerations, however, should result in extra objects in the Type IIA picture. In particular, degenerations on non-compact co-dimension two sub-manifolds of the $U(1)$ correspond to D6 branes in the Type IIA reduction. In \cite{Benini:2009qs,Jafferis:2009th} it was suggested that flavours should be added to the quivers in order to obtain candidate duals. These proposals were added to the already known theories that do not contain flavours; however, it is still not clear if these different types of theories are indeed connected by duality. In this thesis we will always discuss the latter type of candidates whenever a comparison with field theory will be made. Applying our discussion to the theories with flavours is left for future work. If the M-theory circle degenerates over compact divisor, the D6 branes in the Type IIA background wrap compact four-cycles in the CY three-fold. In \cite{Benini:2011cm} the authors suggested dual field theories for some geometries with such degenerations. In these candidates the ranks of the quivers, that are inferred from the Type IIA reduction described above, are not equal in general as a result of the presence of D6 branes. The M-theory circle in these backgrounds affects the Higgsing of the field theories in an interesting way, as we will discuss in \secref{sec:exo}. 

In general, the presence of global symmetries is of great help in classifying the spectrum of a gauge theory. One particularly important example of a global symmetry is the R-symmetry. In three dimensions a theory preserving $\mathcal{N}$ supersymmetries admits the action of an $SO\left(\mathcal{N}\right)$ R-symmetry. Thus the existence of a non-trivial R-symmetry, which can then provide important constraints on the dynamics, requires that we focus on $\mathcal{N}\ge2$, implying there is at least a $U(1)_R$. In particular, assuming that the theory flows to an IR superconformal fixed point, it follows that the scaling dimensions of chiral primary operators coincide with their R-charges. We note that, generically, the $\mathcal{N}=2$ theories considered have classically irrelevant superpotentials. Strong gauge dynamics is required to give large anomalous dimensions, thus making it possible to reach a non-trivial IR fixed point. 

Recently an analogous version of $a$-maximization \cite{Intriligator:2003jj}, which for four-dimensional $\mathcal{N}=1$ theories allows one to determine the R-charge in the superconformal algebra at the IR fixed point, was suggested \cite{Kapustin:2009kz, Drukker:2010nc, Jafferis:2010un, Hama:2010av}. According to this suggestion the correct R-charges locally maximize the free energy $F$ on a three-dimensional sphere. This quantity, that reduces to a certain matrix integral by using localization, seems to be a good measure of the number of degrees of freedom in the field theory. Due to the fact that it can be calculated at strong coupling, both in the gauge theory and in AdS supergravity, it can be used to test the correspondence. 
In \cite{Martelli:2011qj, Jafferis:2011zi, Cheon:2011vi} this suggestion was confirmed for several examples of $\mathcal{N}=2$ CS theories. However, it is fair to say that this is still poorly understood, and more work should be done to generalize those computations to other theories.

The $\mathcal{N}=2$ QCS theories that we consider are expected to be dual to M2 branes moving in a CY four-fold cone over a seven-dimensional Sasaki-Einstein base $Y_7$, thus giving rise to an AdS$_4\times Y_7$ near horizon dual geometry. Such Sasaki-Einstein manifolds $Y_7$ will typically have  non-trivial topology, implying the existence of Kaluza-Klein (KK) modes obtained by reduction of supergravity fields along the corresponding homology
cycles. Of particular interest are five-cycles, on which one can reduce the M-theory six-form potential to obtain $b_2(Y_7)=\dim H_2(Y_7,\R)$ vector fields in AdS$_4$. These vector fields are part of short multiplets of the KK reduction on $Y_7$, known as \textit{Betti multiplets} \cite{D'Auria:1984vv, D'Auria:1984vy} (for a discussion relevant to the cases we will consider, see also \cite{Fabbri:1999hw, Merlatti:2000ed}). In analogy with the Type IIB case, where the constant gauge transformations in the bulk are well known to correspond to global baryonic symmetries on the boundary\cite{Klebanov:1999tb}, we will sometimes employ the same terminology here and refer to these as baryonic $U(1)$s. 

In this part of the thesis we set out to study the above symmetries in the AdS$_4$/CFT$_3$ correspondence. In the rather better-understood AdS$_5$/CFT$_4$ correspondence in Type IIB string theory, from the field theory point of view these baryonic $U(1)$ symmetries appear as non-anomalous combinations of the diagonal $U(1)$ factors inside the $U(N)$ gauge groups\footnote{This will be discussed in more detail in \secref{part2}.}. The key point is that, in four dimensions, Abelian gauge fields are IR free and thus become global symmetries in the IR. However, this is no longer true in three dimensions, thus raising the question of the fate of these Abelian symmetries. From the gravity perspective, in the dual AdS$_4$ the vector fields admit two admissible fall-offs at the boundary of AdS$_4$ \cite{Witten:2003ya, Marolf:2006nd}. This is in contrast to the AdS$_5$ case where only one of them is allowed, for which the interpretation as dual to a global current is required. That the two behaviours are permitted implies that the corresponding boundary symmetries remain either gauged or ungauged, respectively, defining in each case a different boundary CFT. This issue is closely related to the gauge groups being either $U(N)$ or $SU(N)$ in the case at hand. From the point of view of the QCS theory with $U(N)$ gauge groups, at lowest CS level $k=1$ there is no real distinction between $U(N)$ and $U(1)\times SU(N)$ gauge groups  \cite{Aharony:2008ug, Imamura:2008nn,  Lambert:2010ji} . Therefore the discussion in \cite{Witten:2003ya} can be applied to the Abelian part of the symmetry. In this way it is possible to connect the $SU(N)$ and the $U(N)$ theories while keeping track of the corresponding action on the gravity side, which amounts to selecting one particular fall-off for the vector fields in AdS$_4$. This provides motivation to look at the $SU(N)$ version of the theory as dual to a particular choice of boundary conditions in the dual gravity picture.

We start by focusing on the simplest class of examples, namely isolated toric CY four-fold singularities with no vanishing six-cycles. These are discussed in more detail in \secref{sec:gen}. In particular, we study in detail the example of the cone over the Sasaki-Einstein manifold $Q^{111}$, which from now on will be denoted as $\mathcal{C}(Q^{111})$. $Q^{111}$ is a regular Sasaki-Einstein seven-manifold that can be described as a principal $U(1)$ bundle over a $\mathbb{CP}^1 \times \mathbb{CP}^1 \times \mathbb{CP}^1$ base. It is called regular since the fibres all close and have the same length. Motivated by the analysis of the behaviour of gauge fields in AdS$_4$, we will choose boundary conditions where the diagonal $U(1)$ factors inside $b_2(Q^{111})=2$ of the $U(N)$ gauge factors in the field theory are ungauged. This amounts to focusing on a certain version of the theory with gauge group $U(1)^2\times SU(N)^4$. On the other hand, gauge fields in AdS$_4$ can have {\it a priori} both electric and magnetic sources. These are the M2 branes and M5 branes wrapping two-cycles and five-cycles in the Sasaki-Einstein manifold, respectively. These wrapped branes form particles in the AdS$_4$ space. It turns out that the boundary conditions necessary to define the AdS/CFT correspondence allow for just one of the two types at a time \cite{Witten:2003ya}. In particular, the chosen $U(1)^2\times SU(N)^4$ quantization allows only for electric sources; that is, wrapped supersymmetric M5 branes. In turn, these correspond to baryonic operators \cite{Imamura:2008ji} in the field theory that are charged under the global symmetries. We will analyse this correspondence in detail, finding the expected agreement.

On the other hand, magnetic sources correspond to M2 branes \cite{Imamura:2008ji}. While in the AdS geometry these wrap non-supersymmetric cycles, we can also consider resolutions of the corresponding cone where there are supersymmetric wrapped M2 branes. Along the lines of \cite{Klebanov:2007us, Klebanov:2007cx}, we will identify the relevant operator, responsible for the resolution, which is acquiring a VEV. It is possible to find an interpretation of these solutions as spontaneous symmetry-breaking backgrounds through the explicit appearance of a Goldstone boson in the supergravity dual. 

A natural next step is to enlarge the class of singularities under consideration by allowing dual geometries with exceptional six-cycles. One such example is a $\mathbb{Z}_2$ orbifold of $\mathcal{C}(Q^{111})$ known as $\mathcal{C}(Q^{222})$. The interpretation of such six-cycles is somewhat obscure holographically. Indeed, such six-cycles, when resolved, can support M5 brane instantons leading to non-perturbative corrections \cite{Witten:1996bn}. In \secref{sec:5} we set up the study of such corrections by finding a general expression for the Euclidean action of such branes in terms of normalizable harmonic two-forms, and compute this explicitly for $Q^{222}$. We leave a full understanding of such non-perturbative effects from the gauge theory point of view for future work.

An important difference between the M2 brane and D3 brane cases is that, typically for the background AdS$_4\times Y_7$, one is allowed to turn on \emph{torsion} $G$-flux in $H^{4}(Y_7,\Z)$; whereas for AdS$_5\times Y_5$ backgrounds, with $Y_5$ a toric Sasaki-Einstein five-manifold, there is never torsion in $H^{3}(Y_5,\Z)$. Indeed, typically $H^{4}_{\mathrm{tor}}(Y_7,\Z)$ is non-trivial, and each different choice of flux should give a physically distinct theory. Turning on a torsion $G$-flux is equivalent to turning on a closed three-form $C$ with non-zero periods through torsion three-cycles in the Sasaki-Einstein manifold.
This was first discussed in this context by \cite{Aharony:2008gk}, who considered the ABJM model with $Y_7=S^{7}/\Z_k$. In this case $H^{4}(Y_7,\Z)\cong\Z_k$, so there are $k$ distinct M-theory backgrounds corresponding to the $k$ choices of torsion $G$-flux. The authors of \cite{Aharony:2008gk} argued this corresponds to changing the \emph{ranks} of the ABJM theory from $U(N)_k\times U(N)_{-k}$ to $U(N+l)_k\times U(N)_{-k}$, where $0\leq l <k$. As we explain quite generally, theories with non-zero torsion $G$-flux have a richer behaviour under Higgsing than those without any flux. As for the D3 brane case, when there is no flux one can argue from the supergravity dual that one expects to obtain field theories for \emph{all} partial resolutions of a given singularity by Higgsing the original theory. However, once one turns on torsion flux the story is more complicated. 
This can lead to interesting predictions for the expected patterns of Higgsings observed in the dual field theory. 
We examine this in detail in the example where $Y_7=Y^{1,2}(\mathbb{CP}^2)$ is a certain non-trivial Sasaki-Einstein seven-manifold, finding precise agreement between the supergravity analysis and field theory analysis.

The organization of this part of the thesis is as follows. In \secref{sec:2} we review the Freund-Rubin-type solutions which are eleven-dimensional AdS$_4\times Y_7$  backgrounds. We then turn to KK reduction of the supergravity six-form potential on five-cycles in $Y_7$, leading to the Betti multiplets of interest. General analysis of gauge fields in AdS$_4$ shows that two possible fall-offs are admissible. We then review the construction in \cite{Witten:2003ya} relating these different boundary conditions for a single Abelian gauge field in AdS$_4$ to the action of $SL(2,\mathbb{Z})$. In \secref{sec:3} we turn in more detail to the field theory description. We start by reviewing general aspects of $U(N)$ QCS theories that have appeared in the literature, before turning in \secref{sec:Q111} to the example of interest. We then propose a set of boundary conditions dual to the $U(1)^2\times SU(N)^4$ theory. We identify the ungauged $U(1)$s via the electric M5 branes wrapping holomorphic divisors in the geometry. In  \secref{sec:4} we turn to the spontaneous breaking of these baryonic symmetries. We compute on the gravity side the baryonic condensate and identify the Goldstone boson of the SSB. In \secref{sec:Gflux} we explore the role of supergravity fluxes in the dual description of the Higgs effect. In \secref{sec:exo} we discuss the importance of the M-theory circle in the dual Higgsing of the field theories and the study of the related exotic baryonic branches and RG flow scenarios. In \secref{sec:5} we initiate the study of exceptional six-cycles. We compute the warped volume of a Euclidean brane in the resolved $\mathcal{C}(Q^{222})$ geometry. By extending our results on warped volumes to arbitrary geometries, both for the baryonic condensate and the Euclidean brane, we find general formulae for such warped volumes. We end with some concluding comments in \secref{sec:6}. In two appendices we present the un-Higgsing algorithm (\secref{sec:unhiggsing}), and a number of relevant calculations and formulae (\secref{sec:D}).
\chapter{AdS$_4$ backgrounds and Abelian symmetries} \label{sec:2}

We begin by reviewing general properties of Freund-Rubin AdS$_4$ backgrounds, and also introduce the $Q^{111}$ and $Q^{222}=Q^{111}/\Z_2$ examples of main interest. KK reduction of the M-theory potentials on topologically non-trivial cycles leads to gauge symmetries in AdS$_4$. We review their dynamics in the AdS/CFT context and the sources allowed, depending on the chosen quantization. Of central relevance for our purposes will be wrapped supersymmetric M5 branes.

\section{Freund-Rubin solutions}\label{sec:FR}

The AdS$_4$ backgrounds of interest are of Freund-Rubin type, with eleven-dimensional metric and 
four-form given by
\bea\label{AdSbackground} 
\diff s^2_{11} &= &R^2\left(\frac{1}{4}\diff s^2(\text{AdS}_4) + \diff s^2(Y_7)\right)~,\\ \nn 
G &=& \frac{3}{8}R^3 \diff\vol(\text{AdS}_4)~. 
\eea
Here $\mathrm{\diff\vol}(\text{AdS}_4)$ stands for the volume form of the AdS$_4$ space and the AdS$_4$ metric is normalized so that 
$R_{\mu\nu} = -3g_{\mu\nu}$. The Einstein equations 
imply that $Y_7$ is an Einstein 
manifold of positive Ricci curvature, with metric normalized so that 
$R_{ij} = 6g_{ij}$. With complete analogy to quantum electromagnetism, the generalized Dirac quantization condition requires
\bea
\frac{1}{(2\pi \ell_p)^6}\int_{Y_7} \star_{11}\, G = N\in \mathbb{Z}~.
\eea
This then leads to the relation
\bea 
R = 2\pi \ell_p\left(\frac{N}{6\vol(Y_7)}\right)^{1/6}~,
 \eea
where $\ell_p$ denotes the eleven-dimensional Planck length and $\vol(Y_7)$ is the volume of the Sasaki-Einstein manifold.

As is well-known, such solutions arise as the near-horizon limit of $N$ M2 branes placed at the tip $r=0$ of the Ricci-flat cone
\bea\label{cone}
\diff s^2({\cal C}(Y_7)) = \diff r^2 + r^2 \diff s^2(Y_7)~. 
\eea
More precisely, the eleven-dimensional solution is
\bea\label{background} 
\dd s^2_{11} &=& h^{-2/3} \dd s^2(\R^{1,2}) + h^{1/3} \dd s^2(X)~, \\ \nn 
G &=& \dd^3 x \wedge \dd h^{-1}~, 
\eea
where in the case at hand we take the eight-manifold to be the cone over the Sasaki-Einstein manifold $X={\cal C}(Y_7)$ with conical metric (\ref{cone}).
Placing $N$ Minkowski space-filling M2 branes at $r=0$ leads, after including their gravitational back-reaction, to the warp factor
\bea \label{warping}
h = 1 + \frac{R^6}{r^{6}}~. 
\eea
In the near-horizon limit, near to $r=0$, the background (\ref{background}) approaches the AdS$_4$ background (\ref{AdSbackground}).
In fact the warp factor $h=R^6/r^6$ is precisely the AdS$_4$ background in a Poincar\'e slicing. More precisely, writing 
\bea
\label{z_coordinate}
z= \frac{R^2}{r^2} \ , \qquad \diff s^2(\text{AdS}_4) = z^{-2}\left(\diff z^2 + \diff s^2(\R^{1,2})\right)~,
\eea
leads to the metric (\ref{AdSbackground}).

We restrict attention to the $\mathcal{N}=2$ Sasaki-Einstein case, which includes the $\mathcal{N}=3$ three-Sasakian geometry as a special case. 
It is then equivalent to say that the cone metric on ${\cal C}(Y_7)$ is K\"ahler as well as Ricci-flat, {\it i.e.} CY. 
Geometries with $\mathcal{N}\geq 4$ supersymmetries are necessarily quotients of $S^7$.

Until recently the only known examples of such Sasaki-Einstein 
seven-manifolds were homogeneous spaces. Since then there has been dramatic progress. 
Three-Sasakian manifolds, with $\mathcal{N}=3$, may be constructed via an analogue of the hyperK\"ahler quotient, leading 
to rich infinite classes of examples \cite{BG}. For $\mathcal{N}=2$ supersymmetry
one could take $Y_7$ to be one of the explicit $Y^{p,q}$  
manifolds constructed in \cite{Gauntlett:2004hh}, and further 
studied in \cite{metrics, Martelli:2008rt}, or any of their subsequent generalizations. These $\mathcal{N}=2$ examples 
are all toric, meaning that the isometry group contains
$U(1)^4$ as a subgroup. In fact, toric Sasaki-Einstein manifolds are now
completely classified thanks to the general existence and uniqueness result in 
\cite{FOW}. At the other extreme, there are also non-explicit metrics in which $U(1)_R$ is the only 
isometry \cite{BG}. 

However, for our purposes it will be sufficient to focus on two specific homogeneous examples, namely 
$Q^{111}$ and $Q^{222}=Q^{111}/\Z_2$, with 
$\Z_2\subset U(1)_R$ being along the R-symmetry of $Q^{111}$. These will turn out to be simple enough so that everything can be computed explicitly, and yet at the same time 
we shall argue that many of the features seen in these cases hold also for the more general geometries mentioned above.
In both cases the isometry group is
$SU(2)^3\times U(1)_R$, and in local coordinates the explicit metrics are
\bea\label{Qiiimetric}
\diff s^2 = \frac{1}{16}\left(\diff\psi+\sum_{i=1}^3 \cos\theta_i\diff\phi_i\right)^2 + \frac{1}{8}\sum_{i=1}^3 
\left(\diff \theta_i^2 + \sin^2\theta_i\diff\phi_i^2\right)~.
\eea
Here $(\theta_i,\phi_i)$ are standard coordinates on three copies of $S^2=\mathbb{CP}^1$, $i=1,2,3$, and $\psi$ has period $4\pi$ for $Q^{111}$ and period 
$2\pi$ for $Q^{222}$. The two Killing spinors are charged under $\partial_\psi$, which is dual to the $U(1)_R$ symmetry. The metric 
(\ref{Qiiimetric}) shows very explicitly the regular structure of a $U(1)$ bundle over the standard K\"ahler-Einstein metric on
$\mathbb{CP}^1\times\mathbb{CP}^1\times\mathbb{CP}^1$, where $\psi$ is the fibre coordinate and the Chern numbers are $(1,1,1)$ and $(2,2,2)$ respectively. These are hence 
natural generalizations\footnote{The other natural such generalization is the homogeneous space $V_{5,2}=SO(5)/SO(3)$, which has been studied in detail in \cite{Martelli:2009ga}.} to seven dimensions of the $T^{11}$ and $T^{22}$ manifolds. 

\section{$C$-field modes}\label{sec:Cfield}

One might wonder whether it is possible to turn on an internal $G$-flux $G_Y$ on $Y_7$, in addition to the $G$-field in (\ref{AdSbackground}), and still preserve supersymmetry, {\it i.e.}
\bea
G = \frac{3}{8}R^3 \diff\vol(\text{AdS}_4) + G_Y~.
\eea
In fact necessarily $G_Y=0$. This follows from the results of \cite{Becker:1996gj}: 
for any warped CY four-fold background with metric of the form (\ref{background}), one can turn on a 
$G$-field $G_X$ on $X$ without changing the CY metric on $X$ only if 
$G_X$ is self-dual. But for a cone, with $G_X=G_Y$ a pull-back from the base $Y_7$, this obviously implies that $G_X=0$. 

However, more precisely the $G$-field in M-theory determines a 
class\footnote{This is true since the membrane global anomaly described in \cite{Witten:1996md} is always zero on a seven-manifold $Y_7$ that is spin.} in $H^4(Y_7,\Z)$. The differential form part of $G$ captures 
only the image of this in $H^4(Y_7,\R)$, and so $G_Y=0$ still allows for a topologically non-trivial $G$-field classified by the torsion part $H^4_{\mathrm{tor}}(Y_7,\Z)$. This is also captured, up to gauge equivalence, by the holonomy of the corresponding flat $C$-field through dual torsion three-cycles in $Y_7$.
There are hence $|H^4_{\mathrm{tor}}(Y_7,\Z)|$ physically distinct AdS$_4$ Freund-Rubin backgrounds associated to the same geometry, which should 
thus correspond to physically inequivalent dual SCFTs. Different choices of this torsion $G$-flux have been 
argued to be dual to changing the \emph{ranks} in the quiver \cite{Martelli:2009ga, Aharony:2008gk,Benishti:2009ky,Cremonesi:2010ae,Benini:2011cm}. In particular, for example, one can compute $H^4(Q^{111},\Z)\cong\Z_2$, implying there are two distinct M-theory backgrounds with the same $Q^{111}$ geometry but different $C$-fields. 

More straightforwardly, if one has $b_3(Y_7)=\dim H_3(Y_7,\R)$ three-cycles in $Y_7$ then one can also turn on a closed three-form $C$ with non-zero periods through these 
cycles. Including large gauge transformations, this gives a space $U(1)^{b_3(Y_7)}$ of such flat $C$-fields. 
Since these are continuously connected to each other they would be dual to marginal deformations in the dual field theory. Indeed, the \emph{harmonic} three-forms on 
a Sasaki-Einstein seven-manifold are in fact paired by an almost complex structure \cite{Boyer:1998sf} and thus $b_3(Y_7)$ is always even, allowing these 
to pair naturally into complex parameters as required by $\mathcal{N}=2$ supersymmetry. However, for the class of toric singularities 
studied in this thesis, including $Q^{111}$ and $Q^{222}$, it is 
straightforward\footnote{There are, however, examples: the CY four-fold hypersurfaces $\sum_{i=1}^5 z_i^d=0$, where 
$d=3, 4$, are known to have CY cone metrics, and these have $b_3(Y_7)=10$, $60$, respectively \cite{Boyer:1998sf}.} to show that $b_3(Y_7)=0$ and there are hence no such marginal deformations associated to the $C$-field. 

Finally, since $H_6(Y_7,\R)=0$ for any positively curved Einstein seven-manifold \cite{Myers}, there are never
periods of the dual potential $C_6$ through six-cycles in $Y_7$.

\section{Baryonic symmetries and wrapped branes}\label{sec:baryons}

Of central interest in this thesis will be symmetries associated to the topology of $Y_7$, and the corresponding 
charged BPS states associated to wrapped M branes. By analogy with the corresponding situation in AdS$_5\times Y_5$ in Type IIB string theory, 
we shall refer to these symmetries as baryonic symmetries; the name will turn out to be justified.

Denote by $b_2(Y_7)=\dim H_2(Y_7,\R)$ the second Betti number of $Y_7$. By Poincar\'e duality we have $\dim H_5(Y_7,\R)=\dim H_2(Y_7,\R)=b_2(Y_7)$. Let $\alpha_1,\ldots,\alpha_{b_2(Y_7)}$ be a set of dual harmonic five-forms with integer periods. Then for the $AdS_4 \times Y_7$ Freund-Rubin background we may  write the KK ansatz
\bea 
\delta C_6 = \frac{2\pi}{T_5} \sum_{I=1}^{b_2(Y_7)} \mathcal{A}_I\wedge\alpha_I~,
\label{3-form-to-global}
\eea
where $T_5={2\pi}/{(2\pi \ell_p)^6}$ is the M5 brane tension. 
This gives rise to $b_2(Y_7)$ massless $U(1)$ gauge fields $\mathcal{A}_I$ in AdS$_4$. 
For a supersymmetric theory these gauge fields of course sit in certain multiplets, known as 
\emph{Betti multiplets}. See, for example, \cite{D'Auria:1984vv, D'Auria:1984vy, Fabbri:1999hw, Merlatti:2000ed}. 

\subsection{Vector fields in AdS$_4$, boundary conditions and dual CFTs} \label{sec:Vec-boun}

The AdS/CFT duality requires specifying the boundary conditions for the fluctuating fields in AdS. In particular, vector fields in AdS$_4$ admit different sets of boundary conditions \cite{Witten:2003ya, Marolf:2006nd} leading to different boundary CFTs. In order to see this, let us consider a vector field in AdS$_{4}$ using the coordinates in (\ref{z_coordinate}). In the gauge $A_z=0$ the bulk equations of motion set
\begin{equation}
\label{gauge_field_In_AdS4}
A_{i}=a_{i}+j_{i}\, z \ , \quad i=\{t,x,y\} \ ,
\end{equation}
where $a_{i}$ and $j_{i}$ satisfy the free Maxwell equation in Lorentz gauge in the Minkowski space. It is not hard to see that in $d=3$ both behaviours have finite action, and thus can be used to define a consistent AdS/CFT duality \cite{Witten:2003ya, Marolf:2006nd}. In order to have a well-defined variational problem for the gauge field in AdS$_4$ we should be careful with the boundary terms when varying the action. As discussed in \cite{Marolf:2006nd}, we need to impose boundary conditions where $a_{i}$ or $j_{i}$ is fixed on the boundary. 

Fixing $a_{i}$ on the boundary while leaving $j_{i}$ unfixed is interpreted as providing a generating functional for the global current correlators in the field theory 
\bea
\label{cur-ins}
\langle \exp\left(\int_{\R^{1,2}}\diff^3\,x\,\vec{A} \cdot \vec{J}\right) \rangle~,
\eea
where $\vec{J}$ is the conserved current of the boundary theory. If one does not wish to insert sources in the field theory then $a_{i}$ should in fact be set to zero. In addition, the one-point function of this current is related to the sub-leading behaviour of $A_i$ near the boundary. On the other hand, when $j_{i}$ is fixed on the boundary and $a_{i}$ is left unrestricted, the latter should be integrated over in the path integral. Thus, $a_{i}$ is interpreted as a dynamical gauge field in the field theory with the action perturbed as before \eqref{cur-ins}. Notice that the equation of motion of $a_i$ gives $\langle J_i \rangle=0$, which is exactly the same as fixing $j_i=0$ on the boundary.

Defining $\vec{B}=\frac{1}{2}\epsilon^{ijk}\, F_{jk}$ and $\vec{E}=F_{i z}$, we have
\begin{equation}
B^{i}=\epsilon^{ijk}\partial_{j}a_{k}+\epsilon^{ijk}\partial_{j}j_{k}\, z\,, \qquad E^{i}=j^{i} \ .
\end{equation}
The two sets of boundary conditions then correspond to either setting $E_{i}=0$ while leaving $a_{i}$ unrestricted, or setting $B_{i}=0$ while leaving $j_{i}$ unrestricted. 

As noted, $a_{i}$ and $j_{i}$ are naturally identified with a dynamical gauge field and a global current in the boundary, respectively. In accordance with this identification, eq. (\ref{gauge_field_In_AdS4}) and the usual AdS/CFT prescription show each field to have the correct scaling dimension for this interpretation: for a gauge field $\Delta(a_{i})=1$, while for a global current $\Delta(j_{i})=2$. Therefore, the quantization $E_{i}=0$ is dual to a boundary CFT where the $U(1)$ gauge field is dynamical; while the quantization $B_{i}=0$ is dual to a boundary CFT where the $U(1)$ is ungauged and is instead a global symmetry. Furthermore, as discussed in \cite{Klebanov:1999tb} for the scalar counterpart, once the improved action is taken into account the two quantizations are Legendre transformations of one another \cite{Klebanov:2010tj}, as can be seen by, for example, computing the free energy in each case. 

Electric-magnetic duality in the bulk theory, which exchanges $E_{i}\leftrightarrow B_{i}$, translates in the boundary theory into the so-called $\mathcal{S}$ \textit{operation} \cite{Witten:2003ya}. This is an operation on three-dimensional CFTs with a global $U(1)$ symmetry, taking one such CFT to another. In addition, in four-dimensional Abelian gauge theory, the following term can be added to the action 
\bea
I_{\theta}=\frac{\theta}{32\pi^2}\int\,\diff^4\,x\,\epsilon^{\mu\nu\lambda\rho}F_{\mu\nu}F_{\lambda\rho}~.
\eea
It is possible to construct a $\mathcal{T}$ \textit{operation}, which amounts to a shift of the bulk $\theta$-angle by $2\pi$ and thus leaves $\exp(\ii\,I_{\theta})$ invariant. Following \cite{Witten:2003ya}, we can be more precise in defining these actions in the boundary CFT. Starting with a three-dimensional CFT with a global $U(1)$ current $J^{i}$, one can couple this global current to a background gauge field $C$ resulting in the action $S[C]$. The $\mathcal{S}$ operation then promotes $C$ to a dynamical gauge field and adds a BF coupling of $C$ to a new background field $D$, while the $\mathcal{T}$ operation instead adds a CS term for the background gauge field $C$:
\begin{equation}
\mathcal{S}:\, S[C]\,\rightarrow\, S[C]+\frac{1}{2\pi} \int D\wedge \dd C~,\qquad \mathcal{T}:\,S[C]\,\rightarrow\, S[C]+\frac{1}{4\pi}\int C\wedge \dd C \ .
\end{equation}
As shown in \cite{Witten:2003ya}, these two operations generate the group $SL(2,\mathbb{Z})$\footnote{Even though we are explicitly discussing the effect of $SL(2,\mathbb{Z})$ on the vector fields, since these are part of a whole Betti multiplet we expect a similar action on the other fields of the multiplet. We leave this investigation for future work.}. In turn, as discussed above, the $\mathcal{S}$ and $\mathcal{T}$ operations have the bulk interpretation of exchanging $E_{\mu}\leftrightarrow B_{\mu}$ and shifting the bulk $\theta$-angle by $2\pi$, respectively. It is important to stress that these actions on the bulk theory change the boundary conditions. Because of this, the dual CFTs living on the boundary are different.

\subsection{Boundary conditions and sources for gauge fields: M5 branes in toric manifolds} \label{s:bc}

We are interested in gauge symmetries in AdS$_4$ associated to the topology of $Y_7$; that is, arising from KK reductions as in (\ref{3-form-to-global}). All KK modes, and hence their dual operators, carry zero charge under these $b_2(Y_7)$ $U(1)$ symmetries. However, there are operators associated to wrapped M branes that do carry charge under this group. In particular, an M5 brane wrapped on a five-manifold $\Sigma_5\subset Y_7$, such that the cone ${\cal C}(\Sigma_5)$ is a complex divisor in the K\"ahler cone ${\cal C}(Y_7)$, is supersymmetric and leads to a BPS particle propagating in AdS$_4$. Since the M5 brane is a source for $G$, this particle is electrically charged under the $b_2(Y_7)$ massless $U(1)$ gauge fields $\mathcal{A}_I$. One might also consider M2 branes wrapped on two-cycles in $Y_7$. However, such wrapped M2 branes are supersymmetric only if the cone ${\cal C}(\Sigma_2)$ over the two-submanifold $\Sigma_2\subset Y_7$ is calibrated in the CY cone, 
and there are no such calibrating three-forms. Thus these particles, although topologically stable, are not BPS. They are magnetically charged under the $U(1)^{b_2(Y_7)}$ gauge fields in AdS$_4$ \cite{Imamura:2008ji}.

As discussed above, the AdS/CFT duality instructs us to choose, for each $U(1)$ gauge field, a set of boundary conditions where either $E_{\mu}$ or $B_{\mu}$ vanishes. Clearly, only the latter possibility allows for the existence of the SUSY electric M5 branes, otherwise forbidden by the boundary conditions. In turn, this quantization leaves, in the boundary theory, the $U(1)$ symmetry as a global symmetry. Therefore, in this case we should expect to find operators in the field theory that are charged under the global baryonic symmetries and dual to the M5 brane states. We turn to this point in the next chapter. 

For toric manifolds there is a canonical set of such wrapped M5 brane states, where ${\cal C}(\Sigma_5)$ are taken to be the toric divisors. Each such state leads to a corresponding dual chiral primary operator that is charged under the $U(1)^{b_2(Y_7)}$ global symmetries, and will also have definite charge under the $U(1)^4$ flavour group dual to the isometries of $Y_7$. We refer the reader to the standard literature for a thorough introduction to toric geometry. 
However, the basic idea is simple to state. The cone ${\cal C}(Y_7)$ fibres over a polyhedral cone $\mathcal{C}_4$ in $\R^4$ with generic fibre $U(1)^4$. This polyhedral cone is by definition a convex set of the form $\bigcap\{\mathbf{x}\cdot \mathbf{v}_\alpha\geq 0\}\subset\R^4$, where $\mathbf{v}_\alpha\in\Z^4$ are integer vectors. 
This set of vectors is precisely the set of charge vectors specifying the $U(1)$ subgroups of $U(1)^4$ that have complex codimension one fixed point sets. 
These fixed point sets are, by definition, the toric divisors referred to above. 
The CY condition implies that, with a suitable choice of basis, we can write $\mathbf{v}_\alpha=(1,\mathbf{w}_\alpha)$, with $\mathbf{w}_\alpha\in\Z^3$. If we plot these latter points in $\R^3$ and take their convex hull, we obtain the \emph{toric diagram}. 

For the $Q^{111}$ example the toric divisors are given by $\Sigma_5=\{\theta_i=0\}$ or 
$\Sigma_5=\{\theta_i=\pi\}$, for any $i=1,2,3$, which are six five-manifolds in $Y_7$. The toric diagram for $Q^{111}$ is shown in 
Figure \ref{fig:GKZ1} on page 37, where one sees clearly these six toric divisors as the six external vertices. 
Notice that for $Q^{111}$ the full isometry group may be used to rotate $\{\theta_i=0\}$ into $\{\theta_i=\pi\}$, specifically using the $i$th
copy of $SU(2)$ in the $SU(2)^3\times U(1)_R$ isometry group. In fact these two five-manifolds are two points in an $S^2$ family of such five-manifolds related via the isometry group. 
Similar comments apply also to $Q^{222}$. 

\chapter{Baryonic symmetries in QCS theories}\label{sec:3}

In this chapter we turn to a more precise field theoretic description of the global $U(1)^{b_2(Y_7)}$. We begin with a brief review of the $U(N)$ theories considered in the literature, before turning to our $\mathcal{C}(Q^{111})$ example and considering the role of the Abelian symmetries in this case.

\section{$U(N)$ QCS theories}

Let us start by considering the $\prod_{a=1}^G U(N_a)$ theories. The Lagrangian, in $\mathcal{N}=2$ superspace notation, for a theory containing an arbitrary number of bifundamentals $X_{ab}$ in the representation $(\square_a,\, \bar{\square}_b)$ under the $(a,\, b)$-th gauge groups and a choice of superpotential $W$, reads 
\bea\label{action}
\mathcal{L}&=&\int \dd^4 \theta\, \mathrm{Tr}\left[ \sum\limits_{X_{ab}} X_{ab}^\dagger \, \ex^{-V_{a}} X_{ab} \, \ex^{V_{b}}
+ \sum\limits_{a=1}^G \frac{k_a}{2\pi} \int\limits_0^1 \dd t\, V_a \bar{D}^{\alpha}(\ex^{t V_a} D_{\alpha} \ex^{-tV_a})
\right] \nonumber \\
&& + 
\int \dd^2 \theta \, W(X_{ab}) \, + \, \mathrm{c.c.}~.
\eea
Here $k_a\in\Z$ are the CS levels for the vector multiplet $V_a$. For future convenience we define $k={\rm gcd}\{k_a\}$.

The classical vacuum moduli space (VMS) is determined in general by the following equations 
\cite{Martelli:2008si, Hanany:2008cd}
\begin{eqnarray}\label{VMSeqns}
\nn \partial_{X_{ab}} W &=& 0~,\\
\nn \mu_a := -\sum\limits_{b=1}^G {X_{ba}}^{\dagger} {X_{ba}} + 
\sum\limits_{c=1}^G  {X_{ac}} {X_{ac}}^{\dagger} &=& 
\frac{k_a\sigma_a}{2\pi}~, \\
\label{DF} \sigma_a X_{ab} - X_{ab} \sigma_b &=& 0~,
\end{eqnarray}
where $\sigma_a$ is the scalar component of $V_a$. Following \cite{Martelli:2008si}, upon diagonalization of the fields using $SU(N)$ rotations, one can focus on the branch where $\sigma_a=\sigma$, $\forall a$, so that the last equation is immediately satisfied\footnote{We stress that there might be, and indeed even in the $Q^{111}$ example there are, other branches of the moduli space where the condition $\sigma_a=\sigma$ for all $a$ is not met, and yet still the bosonic potential is minimized.}. Under the assumption that $\sum_{a=1}^G k_a=0$, the equations for the moment maps $\mu_a$ boil down to a system of $G-2$ independent equations for the bifundamental fields, analogous to D-term equations. Since for toric superpotentials the set of F-flat configurations, determining the so-called \textit{master space} $\mathscr{F}_{G+2}$ \cite{Forcella:2008bb}, is of dimension $G+2$, upon imposing the $G-2$ D-terms and dividing by the associated gauge symmetries we have a ${\rm dim}_\mathbb{C}\mathscr{M}=4$ moduli space $\mathscr{M}$ where the M2 branes move. 

However, due to the peculiarities of the CS kinetic terms, extra care has to be taken with the diagonal part of the gauge symmetry. At a generic point of the moduli space the gauge group is broken to $N$ copies of $U(1)^{G}$.  The diagonal gauge field $\mathcal{B}_G=\sum_{a=1}^G\mathcal{A}_a$ is completely decoupled from the matter fields, and only appears coupled to $\mathcal{B}_{G-1}=k^{-1}\,\sum_{a=1}^Gk_a\,\mathcal{A}_a$ through 
\begin{equation}\label{SBG}
S(\mathcal{B}_G)=\frac{k}{4\pi\, G}\int (\mathcal{B}_{G-1})_{\mu}\,\epsilon^{\mu\nu\rho}\, (\mathcal{G}_G)_{\nu\rho}~.
\end{equation}
Since $\mathcal{B}_G$ appears only through its field strength, it can be dualized into a scalar $\tau$. Following the standard procedure, it is easy to see that integrating out $\mathcal{G}_G=\dd\mathcal{B}_G$ sets
\begin{equation}
\label{identification}
\mathcal{B}_{G-1}=\frac{G}{k}\, \dd\tau~,
\end{equation}
such that the relevant part of the action becomes a total derivative
\begin{equation}
S(\mathcal{B}_G)=\int \dd\Big(\frac{\tau}{2\pi}\,\mathcal{G}_G\Big)~.
\end{equation}
Around a charge $n\in\Z$ monopole in the diagonal $U(1)$ gauge field $\mathcal{B}_G$ we then have $\int \mathcal{G}_G=2\pi\,G\, n$, so that $\tau$ must have period $2\pi/G$ in order for the above phase to be unobservable \cite{Martelli:2008si}. Gauge transformations of  $\mathcal{B}_{G-1}$ then allow one to gauge-fix $\tau$ to a particular value via (\ref{identification}), but this still leaves a residual discrete set of $\Z_k$ gauge symmetries that leave this gauge choice invariant. The space of solutions to (\ref{DF}) is then quotiented by gauge transformations where the parameters $\theta_a$ satisfy $\sum_{a=1}^G k_a\, \theta_a =0$, together 
with the residual discrete $\Z_k$ gauge transformations generated by $\theta_a = 2\pi/k$ for all $a$. Altogether this leads to a $U(1)^{G-2}\times \Z_k$ quotient. We refer to \cite{Martelli:2008si} for further discussion, and to  \cite{Franco:2009sp} for a discussion in the context of the $Q^{111}$ theory in particular.

\section{The $\mathcal{C}(Q^{111})$ theory}\label{sec:Q111}

\subsection{The theory and its moduli space}

\begin{figure}[ht]
\begin{center}
\includegraphics[scale=1.2]{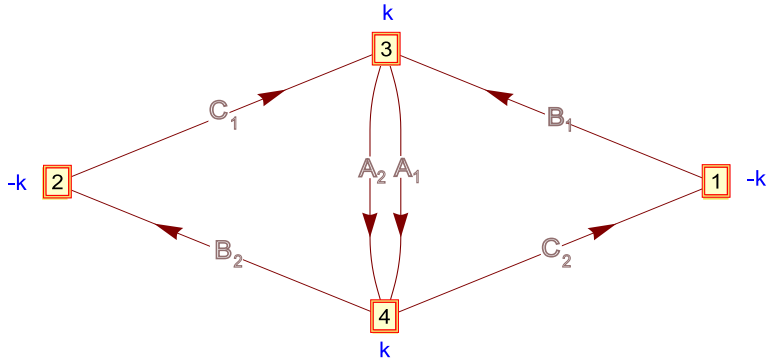}
\end{center}
\caption{The quiver diagram for a conjectured dual of $\mathcal{C}(Q^{111})$.}
\label{fig:quiverdiagramQ111}
\end{figure}

A field theory candidate dual to M2 branes probing $\mathcal{C}(Q^{111})/\Z_k$ was proposed in \cite{Franco:2008um} and further studied in \cite{Franco:2009sp}. The proposal in those references is a $U(N)^4$ Chern-Simons gauge theory with CS levels $(k,\, k,\, -k,\, -k)$, with matter content summarized by the quiver in Figure \ref{fig:quiverdiagramQ111}. 

In addition, there is a superpotential given by
\begin{equation}
\label{WQ111}
W={\rm Tr}\, \Big(\, C_2\, B_1\, A_1\, B_2\, C_1\, A_2\,-\,C_2\,B_1\, A_2\, B_2\, C_1\, A_1\, \Big)\ .
\end{equation}
As expected for a field theory dual to $N$ point-like branes moving in $\mathcal{C}(Q^{111})/\Z_k$, the moduli space contains a branch which is the symmetric product of $N$ copies of this conical singularity. To see this, let us begin with the Abelian theory in which all the gauge groups are 
$U(1)$. As shown in  \cite{Franco:2009sp}, after integrating out the auxiliary $\sigma$ scalar the geometric 
branch of the moduli space with $k=1$ is described by $G-2=2$ D-term equations. Recalling the special role played by $\mathcal{B}_{G-1}=\mathcal{B}_3,\, \mathcal{B}_G=\mathcal{B}_4$, it is useful to introduce the following basis for the $U(1)$ gauge fields:
\bea
\label{redefinedGF}
\nn
&&\mathcal{A}_{I}=\frac{1}{2}(\mathcal{A}_{1}-\mathcal{A}_{2}+\mathcal{A}_{3}-\mathcal{A}_{4})~, \quad
\mathcal{A}_{II}=\frac{1}{2}(\mathcal{A}_{1}-\mathcal{A}_{2}-\mathcal{A}_{3}+\mathcal{A}_{4})~,\\ \nn
&&\mathcal{B}_3=\mathcal{A}_{1}+\mathcal{A}_{2}-\mathcal{A}_{3}-\mathcal{A}_{4}~,\quad
\mathcal{B}_4=\mathcal{A}_{1}+\mathcal{A}_{2}+\mathcal{A}_{3}+\mathcal{A}_{4}~.
\eea
Then the two D-terms to impose are those for $\mathcal{A}_I,\,\mathcal{A}_{II}$. In turn, the charge matrix is
\bea
\label{gaugedU(1)charges}
\begin{array}{l | c c c c c c} 
 & A_1 & A_2 & B_1 & B_2 & C_1 & C_2 \\ \hline
 U(1)_{I} & 1 & 1 & 0 & 0 & -1 & -1 \\ 
 U(1)_{II} & -1 & -1 & 1 & 1 & 0 & 0 \\
 U(1)_{\mathcal{B}_3} & 0 & 0 & -2 & 2 & -2 & 2 \\ 
 U(1)_{\mathcal{B}_4} & 0 & 0 & 0 & 0 & 0 & 0 
\end{array}~.
\eea
Notice the appearance of the $SU(2)$ global symmetry, under which the pair $A_i$ transform as doublet. This is a subgroup of the expected $SU(2)^3$ global symmetry which is not manifest in the Lagrangian due to the choice of the M-theory circle\footnote{The M-theory circle can be deduced from the $U(1)_{\mathcal{B}_3}$ action on the moduli space.}. 

Since for the Abelian theory the superpotential is identically zero, one can determine the Abelian moduli space 
by constructing the gauge-invariants with respect to the gauge transformations for $\mathcal{A}_I,\, \mathcal{A}_{II}$. Borrowing the results from \cite{Franco:2009sp}, for CS level $k=1$ these are
\begin{equation}
\begin{array}{lclclcl}
w_1 = A_1\,B_2\,C_1~, &  \  & w_2 = A_2\,B_1\,C_2~, & \ &  w_3 = A_1\,B_1\,C_2~, &  \ & w_4 = A_2\,B_2\,C_1~, \\
w_5 = A_1\,B_1\,C_1~, &  \ & w_6 = A_2\,B_1\,C_1~, & \  & w_7 = A_1\,B_2\,C_2~, & \ & w_8 = A_2\,B_2\,C_2  \, .
\end{array}
\label{ws}
\end{equation}
One can then check explicitly that these satisfy the nine relations defining $\mathcal{C}(Q^{111})$ as an affine variety:
\begin{equation}
\begin{array}{cccccccc}
w_1\,w_2 - w_3\,w_4 & = & w_1\,w_2 - w_5\,w_8 & = & w_1\,w_2 - w_6\,w_7 & = & 0~, & \\
w_1\,w_3 - w_5\,w_7 & = & w_1\,w_6 - w_4\,w_5 & = & w_1\,w_8 - w_4\,w_7 & = & 0~, & \\
w_2\,w_4 - w_6\,w_8 & = & w_2\,w_5 - w_3\,w_6 & = & w_2\,w_7 - w_3\,w_8 & = & 0~.
\label{eq_Q11}
\end{array}
\end{equation}
This is an affine toric variety, with toric diagram given by Figure \ref{fig:GKZ1} on page 37. Indeed, we also notice 
that for the Abelian theory the description of the moduli space as a $U(1)^2$ K\"ahler quotient of $\C^6$ with 
coordinates $\{A_i,B_i,C_i\}$ is precisely the minimal gauged linear sigma model (GLSM) description. 
Thus the six toric divisors in Figure \ref{fig:GKZ1}, discussed in \secref{sec:baryons}, 
are defined by $\{A_i=0\}$, $\{B_i=0\}$, $\{C_i=0\}$, $i=1,2$.

For CS level $k>1$ one obtains an $\mathcal{N}=2$ supersymmetric $\Z_k\subset U(1)_{\mathcal{B}_{G-1}}$ orbifold of $\mathcal{C}(Q^{111})$. 
Notice that $\{w_i\mid i=1,\ldots,4\}$ are invariant under this action, while $\{w_5,w_6\}$ and $\{w_7,w_8\}$ are rotated with equal and opposite 
phase. On the other hand, for the non-Abelian theory with $N>1$ it was shown in \cite{Franco:2009sp} that for large $k$, where 
the use of still poorly-understood monopole operators is evaded, upon using the F-terms of the full non-Abelian superpotential 
(\ref{WQ111}) the chiral ring matches that expected for the corresponding orbifold. 
In this case the chiral primaries at the non-Abelian level are just the usual gauge-invariants given by
\begin{equation}
\label{nonabelianchirals}
{\rm Tr}\, \Big( \prod_{a=1}^r\, X^{\pm}_{i_a}\Big)\ ,\quad \mbox{where}\quad X^+_i=A_i\, C_2\, B_1\ , \quad X^-_i=A_i\, B_2\, C_1~.
\end{equation}
An important subtlety in this theory is that 
 $U(1)_{\mathcal{B}_{G-1}}$ does not act freely on $Q^{111}$: it fixes two disjoint copies of $S^3$ inside $Q^{111}$. Indeed, using (\ref{gaugedU(1)charges}) one sees that the corresponding two cones $\C^2=\mathcal{C}(S^3)$ are parametrized 
respectively by $\{w_1,w_4\}$ and $\{w_2,w_3\}$, with in each case all other $w_i=0$. Thus for $k>1$ the 
horizon $Y_7=Q^{111}/\Z_k$ has orbifold singularities in co-dimension four. This means that the supergravity approximation cannot be trusted for $k>1$. In fact these are $A_{k-1}$ singularities which can support ``fractional'' M2 branes wrapping the collapsed cycles, and one expects an $SU(k)$ gauge theory to be supported on these $S^3$s. A different perspective can be obtained by interpreting $U(1)_{\mathcal{B}_{G-1}}$ as the M-theory circle and reducing to Type IIA. This results in $k$ D6 branes wrapping these two $S^3$ sub-manifolds\footnote{In \cite{Benini:2009qs,Jafferis:2009th} this picture led the authors to propose a different $Q^{111}$ dual candidate that contains flavours and correspond to the same M-theory circle. It will be interesting to check if this theory is dual in the IR to the one without flavours that we just described.}. From now on we will therefore assume that $k=1$. 

\subsection{Gauged versus global Abelian subgroups and $SL(2,\mathbb{Z})$}

At $k=1$ the orbifold identification due to the CS terms is trivial. Indeed, in this case there is no real distinction between $U(N)$ and $SU(N)\times U(1)$ gauge groups, as discussed in \cite{Aharony:2008ug, Imamura:2008nn,  Lambert:2010ji} for the ABJM theory and orbifolds of it. We shall argue that ungauging some of the $U(1)$s is dual to a particular choice of boundary conditions on the gravity side. That is, we apply the general discussion in \secref{sec:baryons} to the $b_2(Q^{111})=2$ $U(1)$ gauge fields, and argue that the associated $U(1)$ symmetries are those in $SU(N)\times U(1)$, for appropriate gauge group factors. This raises the important problem of how to identify the relevant two $U(1)$ symmetries dual to the Betti multiplets in the QCS theory proposed above. The key is to recall that the boundary conditions which amount to ungauging these $U(1)$s in turn allow for the existence of supersymmetric M5 branes on the gravity side. As discussed in \secref{sec:baryons}, from an algebro-geometric point of view the corresponding divisors are easy to identify. In turn we notice that, for the Abelian theory, the fields $\{A_i, B_i, C_i\}$ are also the minimal GLSM coordinates. Setting each to zero therefore gives one of the six toric divisors that may be wrapped by an M5 brane. The charges of the resulting M5 brane states under $U(1)^{b_2(Y_7)}$ are then the \emph{same} as the charges of these fields under the $U(1)_I\times U(1)_{II}$ we quotient by in forming the Abelian moduli space -- this was shown for the D3 brane case in \cite{Franco:2005sm}, and the same argument applies here also. This strongly suggests that the gauge symmetries $U(1)_I$, $U(1)_{II}$ should in fact be dual to the Betti multiplets discussed in \secref{sec:baryons}. 

Once we have identified the relevant Abelian symmetries, we can consider acting with the $\mathcal{S}$ and $\mathcal{T}$ operations. We schematically write the action of the $U(N)^4$ $Q^{111}$ theory (which we will denote as $S_U$), separating the $U(1)$ sector from the rest, as
\begin{equation}
\label{S_gauged}
S_U\sim \int  \mathcal{B}_3\wedge \dd \mathcal{B}_4+ \mathcal{A}_I\wedge \dd \mathcal{A}_{II}+\int \mathcal{L}_R \ ,
\end{equation}
where $\int \mathcal{L}_R$ stands for the remaining terms. We can then consider a theory without the gauge fields $\mathcal{A}_I$, $\mathcal{A}_{II}$, constructed schematically as $S_{SU}=\int \mathcal{B}_3\wedge \dd \mathcal{B}_4+\int \mathcal{L}_R$. By construction, this theory has exactly two global symmetries satisfying all the properties expected as dual to Betti multiplets. Following \cite{Witten:2003ya}, we can introduce a background gauge field for one of them, which we can call $\mathcal{A}_I$. Then, as reviewed in \secref{sec:baryons}, the $\mathcal{S}$-operation amounts to regarding this field as dynamical, while at the same time introducing a coupling to another background field $\mathcal{C}_I$ as
\begin{equation}
S_{SU}\rightarrow S_{SU}[\mathcal{A}_I]+\int \mathcal{C}_I\wedge \dd\mathcal{A}_I \ .
\end{equation}
We can introduce yet another background gauge field $\mathcal{A}_{II}$ for the second global symmetry and perform  $\mathcal{S}$-operations on $\mathcal{A}_{II}$ and $\mathcal{C}_{I}$ such that the action becomes
\begin{equation} \label{tr2}
S_{SU}[\mathcal{A}_I,\, \mathcal{A}_{II}]+ \int \mathcal{C}_I\wedge \dd\mathcal{A}_I + \mathcal{D}_{I}\wedge \dd\mathcal{C}_I+\mathcal{D}_{II}\wedge \dd\mathcal{A}_{II} \ .
\end{equation}
Now we define $\mathcal{C}_{II}=\frac{1}{2}(\mathcal{D}_I-\mathcal{D}_{II})$ and $\mathcal{C}_{III}=\frac{1}{2}(\mathcal{D}_I+\mathcal{D}_{II})$ and rewrite \eqref{tr2} as
\begin{equation} 
S_{SU}[\mathcal{A}_I,\, \mathcal{A}_{II}]+\int \mathcal{C}_I\wedge \dd\mathcal{A}_I + (\mathcal{C}_{II}+\mathcal{C}_{III})\wedge \dd\mathcal{C}_I+(\mathcal{C}_{III}-\mathcal{C}_{II})\wedge\dd\mathcal{A}_{II} \ .
\end{equation}
We use the $\mathcal{S}$-operation to make $\mathcal{C}_{III}$ dynamical and couple it to $\mathcal{E}$. After integrating by parts the added term one obtains
\begin{equation} 
S_{SU}[\mathcal{A}_I,\, \mathcal{A}_{II}]+\int \mathcal{C}_I\wedge \dd\mathcal{A}_I + \mathcal{C}_{II}\wedge \dd(\mathcal{C}_I-\mathcal{A}_{II})+\mathcal{C}_{III}\wedge \dd(\mathcal{C}_I+\mathcal{A}_{II}+\mathcal{E}) \ .
\end{equation}
Notice that now the path integral is taken with respect to $\mathcal{A}_{I}$, $\mathcal{A}_{II}$, $\mathcal{C}_{I}$ and $\mathcal{C}_{III}$. Since $\mathcal{C}_{III}$ only appears linearly, its functional integral gives rise to a delta functional setting $\mathcal{E}=-\mathcal{C}_I-\mathcal{A}_{II}$ and the action reduces to
\begin{equation}
S_{SU}[\mathcal{A}_I,\, \mathcal{A}_{II}]+\int \mathcal{C}_I\wedge \dd\mathcal{A}_I + \int \mathcal{C}_{II}\wedge \dd(\mathcal{C}_I-\mathcal{A}_{II}) \ .
\end{equation}
Integrating by parts yields
\begin{equation}
S_{SU}[\mathcal{A}_I,\, \mathcal{A}_{II}]+\int \mathcal{C}_I\wedge \dd(\mathcal{C}_{II}+\mathcal{A}_I) - \int \mathcal{C}_{II}\wedge \dd\mathcal{A}_{II} \ .
\end{equation}
The functional integral with respect to $\mathcal{C}_I$ gives rise to a delta functional setting $\mathcal{C}_{II}=-\mathcal{A}_I$, which leads to an action of the precise form (\ref{S_gauged}). We have therefore been able to establish a connection between a theory where the gauge group is $U(1)^2\times SU(N)^4$, and whose action is $S_{SU}$, and the original $U(N)^4$ theory, whose action is given by $S_U$, via repeated action with the $\mathcal{S}$-operation. 

More generally, the whole of $SL(2,\Z)$ will act on the boundary conditions for the bulk gauge fields, leading in general to an infinite orbit of CFTs for each $U(1)$ gauge symmetry in AdS$_4$. This is a rich structure that deserves considerable further investigation. In the following, however, we will content ourselves to study the particular choice of boundary conditions described by the $S_{SU}$ theory. Since the dual to the $\mathcal{S}$ operation is the exchange of the $E_{\mu}\leftrightarrow B_{\mu}$ boundary conditions, we expect the gravity dual to the $S_{SU}$ theory to still be AdS$_4\times Q^{111}$, but with an appropriate choice of boundary conditions. In turn, these boundary conditions allow for the existence of the electrically charged M5 branes which we used to identify the symmetries. These M5 branes would not be allowed in the quantization $E_{\mu}=0$, which in turn would be dual to a CFT where the corresponding $U(1)$ factors would remain gauged. In agreement, the dual operators which we will propose below would not be gauge-invariant in that case.

Let us now consider the effect of the $U(1)^2\times SU(N)$ gauge group on the construction of the moduli space. The diagonalization of the $\sigma_a$ auxiliary fields in the equations defining the moduli space (\ref{DF}) relies on the non-Abelian part of the gauge symmetry, and therefore it applies even if we consider ungauging some of the diagonal $U(1)$ factors. More crucially, in order to obtain the correct four-fold moduli space we needed the $S(\mathcal{B}_4)$ piece (\ref{SBG}) of the CS action so that, upon dualizing the $\mathcal{B}_4$ field, the dual scalar $\tau$ is gauge-fixed via gauge transformations of $\mathcal{B}_{3}$. Thus provided we leave 
$\mathcal{B}_{4}$ and $\mathcal{B}_{3}$ gauged, with the same CS action, all of this discussion is unaffected if we ungauge the remaining $U(1)_I$, $U(1)_{II}$.
Correspondingly, we will still have the 8 gauge-invariants (\ref{ws}), which will give rise to the same 9 equations defining 
$\mathcal{C}(Q^{111})$ as a non-complete intersection ``mesonic'' moduli space. 
The remarks on the non-Abelian chiral ring elements spanned by (\ref{nonabelianchirals}) are also unchanged. 
However, with only a  $U(1)_{\mathcal{B}_3}\times U(1)_{\mathcal{B}_4}\times SU(N)^4$ gauge symmetry we also have additional 
chiral primary operators, charged under the now global $U(1)_I$, $U(1)_{II}$. Indeed, we have the following ``baryonic'' type operators:
\bea
\nn
\mathscr{B}_{A_{I_1...I_N}}&=&\frac{1}{N!}\, \epsilon^{i_1\cdots i_N}\, \epsilon_{j_1\cdots j_N}\, (A_{I_1})^{j_1}_{i_1}\cdots  (A_{I_N})^{j_N}_{i_N}~,\\ \nn
\mathscr{B}_{B_i}&=&\frac{1}{N!}\, \epsilon^{i_1\cdots i_N}\, \epsilon_{j_1\cdots j_N}\, (B_i)^{j_1}_{i_1}\cdots  (B_i)^{j_N}_{i_N}\,\me^{\ii\,(-1)^{i-1}\,N\,\tau}~,\\
\mathscr{B}_{C_i}&=&\frac{1}{N!}\, \epsilon^{i_1\cdots i_N}\, \epsilon_{j_1\cdots j_N}\, (C_i)^{j_1}_{i_1}\cdots  (C_i)^{j_N}_{i_N}\,\me^{\ii\,(-1)^{i-1}\,N\,\tau}~.
\label{b-operators}
\eea
In particular, for the six fields in the quiver there is a canonical set of six baryonic operators given by determinants of these fields, 
dressed by appropriate powers of the disorder operators $\me^{\ii \tau}$ to obtain gauge-invariants under $\mathcal{B}_3$. 
These operators are in 1-1 correspondence with the toric divisors in the geometry. This is precisely the desired mapping between 
baryonic operators in the field theory and M5 branes wrapping such toric sub-manifolds, with one M5 brane 
state for each divisor. Indeed, the charges of these operators under the two baryonic $U(1)$s are
\begin{displaymath}
\begin{array}{c | c c c} 
 & \mathscr{B}_{A_{I_1..I_N}} & \mathscr{B}_{B_i} & \mathscr{B}_{C_i} \\ \hline
 U(1)_{I} & N & 0 & -N \\ 
 U(1)_{II} & -N & N & 0 \\
 \end{array}~.
 \end{displaymath}
These are precisely the charges of M5 branes, wrapped on the five-manifolds corresponding to the divisors $\{A_i=0\}$, $\{B_i=0\}$, $\{C_i=0\}$, 
under the two $U(1)^{b_2(Y_7)}$ symmetries in $AdS_4$. Indeed, recall that the two two-cycles in $Q^{111}$ may be taken to be 
the anti-diagonal $S^2$s in two factors of $\mathbb{CP}^1\times \mathbb{CP}^1\times \mathbb{CP}^1$, at $\psi=0$. Let us choose these to be the anti-diagonal in the
first and third factor, and second and first factor, respectively. 
The charge of an M5 brane wrapped 
on a five-cycle $\Sigma_5\subset Y_7$ under each $U(1)$ is then the intersection number of $\Sigma_5$ with each corresponding two-cycle. 
Thus with this basis choice, the charges of the operator associated to an M5 brane wrapped on the base of one of the six toric divisors $\{A_i=0\}$, $\{B_i=0\}$, $\{C_i=0\}$ are precisely those listed in the above table. 

Being chiral primary, the conformal dimensions of these operators are given by $N\, \Delta[X]=N\, R[X]$, $R[X]$ being the R-charge of the field $X$. The conformal dimension 
of an M5 brane wrapping a supersymmetric five-cycle $\Sigma_5\subset  Y_7$ is given by the general formula 
\cite{Gubser:1998fp} $\Delta[\Sigma_5] = N\pi \vol(\Sigma_5)/(6\vol(Y_7))$. These volumes are easily computed for the $Q^{111}$ metric (\ref{Qiiimetric}): 
$\vol(Q^{111})=\frac{\pi^4}{8}$, $\vol(\Sigma_5)=\frac{\pi^3}{4}$, where 
$\Sigma_5$ is any of the 6 toric five-cycles. From this one obtains
$\Delta[\Sigma_5]=\frac{N}{3}$ in each case, giving conformal dimensions 
$\Delta[X]=\frac{1}{3}$ for each field. 
With this R-charge assignment we see that the superpotential 
(\ref{WQ111}) has R-charge 2, precisely as it must at a superconformal fixed point. Indeed, the 
converse argument was applied in \cite{Hanany:2008fj} to obtain this R-charge assignment. 
We thus regard this as further evidence in support of our claim in this section, as well as further support for 
these theories as candidate SCFT duals to AdS$_4\times Q^{111}$.

\section{QCS theories dual to isolated toric Calabi-Yau four-fold singularities}\label{sec:gen}

We would like to apply the preceding discussion to more general $\mathcal{N}=2$ CS-matter theories dual to M2 branes probing CY four-fold cones. With the exception of $\C^4$, the apex of the cone always corresponds to a \emph{singular point} $y$ in the toric variety. An important question is whether this is an \emph{isolated} singular point, or whether there are other singular loci that intersect it. In the former case, $X\setminus\{y\}\cong \R_+\times Y_7$ where $Y_7$ is a \emph{smooth} Sasakian seven-manifold. 

As was discussed in \secref{s:bc}, an affine toric four-fold variety is specified by a polyhedral cone $\mathcal{C}_4$. The condition for the singular point $y$ to be isolated is precisely the condition that $\mathcal{C}_4$ is \emph{good}, in the sense of \cite{Lerman}. This condition may be stated as follows. Let $F$ be a face of the cone, and let $\{v_{a_1},\ldots,v_{a_m}\}$ be the normals to the set of supporting hyperplanes meeting at the face $F$. Then the singularity is isolated if and only if for every face $F$ the $\{v_{a_1},\ldots,v_{a_m}\}$ may be extended to a $\Z$-basis for $\Z^4$. In particular, this means that necessarily $m=\mathrm{codim}\, F$. This translates into the following condition on the toric diagram $\Delta$:
\bea \label{conditions}
\nn
&&\text{1. Each face of }\Delta\text{ is a triangle.} \\
&&\text{2. There are no lattice points internal to any edge or face of }\Delta.
\eea
These are necessary and sufficient for the ``link'' $Y_7$ to be a smooth manifold. It was proven recently in 
\cite{FOW} that all such toric Sasakian manifolds admit a unique Sasaki-Einstein metric compatible with the complex structure of the cone.

An additional ingredient is the possible presence of vanishing six-cycles at the tip of the cone. In terms of the toric data, these six-cycles are signalled by internal lattice points in the toric diagram. These codimension two cycles, in very much the same spirit as their four-cycle Type IIB counterparts (these will be discussed in \secref{part2}), represent a further degree of complexity. We postpone the analysis of geometries with exceptional six-cycles to \secref{sec:5}.

We want to study such isolated CY singularities without vanishing six-cycles in more detail in this section, in particular classifying the singularities. In the cases where a Lagrangian description of the M2 brane theory exists, it turns out that for all these cases one can construct an appropriate toric superpotential, so that there is a toric gauge theory which realizes at the Abelian level the minimal GLSM.  This toric gauge theory has $b_2(Y_7)+2$ gauge group factors, and can be promoted to have $U(N)$ gauge groups.  Such quiver Chern-Simons theories have been considered in the past in \cite{Aharony:2008ug, Franco:2008um, Benishti:2009ky}. 

We would like to generalize our proposal to this simplest class of isolated singularities with no vanishing six-cycles. Indeed, we expect that a similar sequence of $\mathcal{T}$ and $\mathcal{S}$ operations amounts to ungauging of precisely $b_2(Y_7)$ $U(1)$ factors. In very much the same spirit as in the $\mathcal{C}(Q^{111})$ example, this should correspond to a particular choice of boundary conditions in the dual AdS$_4$. Furthermore, we conjecture the gauge group to be $U(1)^2\times SU(N)^{b_2(Y_7)+2}$, the two $U(1)$ factors being those corresponding to the $\mathcal{B}_{G}$ and $\mathcal{B}_{G-1}$ gauge fields. In this way we are naturally left with $b_2(Y_7)$ global $U(1)$ symmetries which exactly correspond to the $b_2(Y_7)$ expected $U(1)$ baryonic symmetries. Furthermore, the M5 branes would be naturally identified with the corresponding baryonic operators, constructed in a similar manner as in the $\mathcal{C}(Q^{111})$ example.
 
\subsection{The number of gauge nodes in the dual QCS theories}

Recall that, as explained in \secref{s:bc}, the $v_{\alpha}$ vectors define an affine toric CY four-fold. This definition is unique up to a unimodular transformation $\mathcal{R}$, where $\mathcal{R} \in GL(4,\Z)$ and $\det \mathcal{R}=\pm 1$. The vectors $v_{\alpha}$ may be written as $v_{\alpha} = (1,w_{\alpha})$ for an appropriate choice of basis, where $w_{\alpha} \in \Z^3$ are the vertices of the three-dimensional toric diagram. It is convenient to define the $G$ matrix that its columns are the $v_{\alpha}$ vectors.
 
We will be interested in singularities with no vanishing six-cycles. We therefore demand, in addition to the two conditions in \eqref{conditions}, that no lattice points appear inside the polytope. These toric diagrams are known as \textit{lattice-free} polytopes. Such lattice-free polytopes in three dimensions are characterized by the fact that they have \textit{width} one (see for example \cite{kantor-1997} and references therein). This is sometimes referred to as Howe's theorem, and is translated into the fact that the vertices of any lattice-free polytope are sitting in adjacent planes, {\it i.e.} two lattice planes with no lattice points inbetween. These planes can be chosen to be $\{z=0\}$ and $\{z=1\}$\footnote{If there are more vertices in one plane we choose it to be the $\{z=0\}$ plane without loss of generality.}. 

We want to show that for seven-dimensional simply-connected toric Sasakian manifolds the number of gauge fields of the dual theory can be read from the toric diagram. To show this, we refer to the results of \cite{Lerman2}, from which we learn that the first and second homotopy groups of a toric Sasakian manifold $Y_7$ can be read straightforwardly from the toric diagram. The results for CY four-folds are
\bea
\pi_1(Y_7)\cong \Z^4/L \quad , \quad \pi_2(Y_7)\cong\Z^{d-4} \ ,
\label{homotopy}
\eea
where $L=\mathrm{span}_\Z\{v_\alpha\}$ is the span over $\Z$ of the space of $d$ external vertices of the toric diagram. According to the Hurewicz Theorem $H_2(Y_7)\cong\pi_2(Y_7)$ whenever $H_1(Y_7)\cong\pi_1(Y_7)/[\pi_1(Y_7),\pi_1(Y_7)]$ is trivial. Therefore we see from \eqref{homotopy} that for simply-connected Sasakian manifolds $b_2(Y_7)=d-4$, which is also the number of gauge groups in the minimal GLSM describing this geometry. This immediately suggests that the number of gauge nodes of the corresponding field theory, in the case that the latter is identified with the minimal GLSM, is $b_2(Y_7)+2$. The two additional gauge nodes correspond to the $U(1)$s which are not quotiented by in forming the moduli space. The identification with the minimal GLSM is expected since the moduli are purely geometric in the type of singularities that we consider. This is as opposed, for example, to the types of cases that will be considered in \secref{part2}, in which additional NS-NS B-field and R-R two-form moduli should be added. 

Note from \eqref{homotopy} that $Y_7$ is simply-connected if and only if the external vertices span $\Z^4$. We want to show that any lattice-free three-dimensional toric diagram with more than four vertices corresponds to a simply-connected Sasakian manifold. Let us consider
\bea\label{AnB}
A=
\left(
\begin{array}{cccc}
 1 & 1 & 1 & 1  \\
 0 & x_1 & x_2 & x_3 \\
 0 & y_1 & y_2 & y_3 \\ 
 0 & 0 & 0 & 1
\end{array}\right) \ , \quad
B=
\left(
\begin{array}{cccc}
 1 & 1 & 1 & 1  \\
 0 & 1 & 0 & 0 \\
 0 & 0 & 1 & 0 \\ 
 0 & 0 & 0 & 1
\end{array}\right) \ .
\eea
For polytopes with more than four vertices, three of the vertices must be co-planar. Thus the matrix that describes four of the vertices is $A$ in \eqref{AnB}, where each column corresponds to a vertex (the $x$, $y$ and $z$ coordinates correspond to the second, third and fourth rows, respectively). It is easy to see that by an $\mathcal{R}$ transformation $A$ can be brought into $B$ in \eqref{AnB}.
To see that the matrices $A$ and $B$ are related by an $\mathcal{R}$ transformation one should show that $|\det (A^{-1}\,B)|=1/|(x_1\,y_2-x_2\,y_1)|$ is equal to one. The denominator is just the area of the parallelogram made up of two identical triangles defined by the first three columns in $A$. The area of this triangle is $1/2$ as this is the condition for a lattice-free triangle in two dimensions. Thus indeed $|\det (A^{-1}\,B)|=1$.

The four vertices described by $B$ span $\Z^4$. Therefore any three-dimensional lattice-free polytope with more than four vertices corresponds to a simply-connected Sasakian seven-manifold. This is not always true for diagrams with four vertices, since in this case each plane can contain two points. 

\subsection{Complete classification of the singularities}

To start our analysis we note that lattice-free polytopes with four vertices correspond to a type orbifold singularity that has been discussed intensively in the literature (see {\it e.g.} Section 3.1 in \cite{Morrison:1998cs}). This is a supersymmetric $\C^4/\Z_k$ orbifold corresponding to an isolated singularity that cannot be resolved, with the orbifold weights in this case being $(1,-1,q,-q)$ with $\gcd \{k,q\}=1$. As already noted in \cite{Jafferis:2009th}, if $q>1$, for any choice of $U(1)$ isometry to reduce on, one can show that $Y_7$ reduces in Type IIA to a space with orbifold singularities. It was suggested in \cite{Jafferis:2009th} that smooth AdS$_4 \times Y_7$ backgrounds that reduce to singular spaces are dual to field theories with no Lagrangian description. This suggestion is consistent with the results of \cite{Gaiotto:2008ak} that show that the matter sector, in the quiver theories that are dual to the orbifold spaces discussed above, appears to have no Lagrangian description. Thus we are left with the ABJM orbifolds, obtained by taking $q=1$, for which there are of course already field theory candidates.

We now continue with the classification of diagrams with five vertices. Recall that we have shown that the first four vertices are described by $B$. Since the fifth vertex should be in the $\{z=1\}$ plane, to prevent a face with four vertices, it can be written as $(1,x,y,1)$ with $x\geqslant y>0$ without loss of generality\footnote{As can be easily seen from (\ref{Qt-matrix}), toric diagrams with triangular faces obtained by picking other values of $x$ and $y$ are related by $\mathcal{R}$ transformations.}. The only way to break the lattice-free condition would be if there were points between $(1,x,y,1)$ and $(1,0,0,1)$. Thus we have to require $\gcd\{x,y\}=1$. This concludes the classification of polytopes with five vertices. There are no additional $\mathcal{R}$ transformations that connect between diagrams in this set; as we show later, the corresponding GLSM charge matrix that describes this toric diagram is unique for any choice of $x$ and $y$. 

For toric diagrams with six vertices one needs to add to $B$ two columns $\{1,x_1,y_1,1\}$ and $\{1,x_2,y_2,1\}$. The triangle in the $z=1$ plane will be lattice-free if its area is $1/2$, {\it i.e.} 
\bea \label{triang}
|x_1\,y_2-x_2\,y_1|=1~.
\eea
We can study first the cases in which there are four co-planar points. For these diagrams we can always set $\{x_1=0,y_1=1\}$ by an $\mathcal{R}$ transformation. From \eqref{triang} we get $|0\times y_2-x_2\times 1|=1$ and thus $\{x_2=-1,y_2 \neq 0\}$ to prevent a rectangular face. One can easily show that $y_2 \sim 2-y_2$ under an $\mathcal{R}$ transformation, thus we can consider just $y_2>0,\,y_2\neq 2$.

To complete our classification of toric diagrams with six nodes let us treat the toric diagrams with no four co-planar points. Thus $x_1,x_2,y_1,y_2 \neq 0$ and an $\mathcal{R}$ transformation can be used to set $y_2 > y_1 > 0$. In addition, the following cases $\{y_1=1,x_1 = -1\}$, $\{x_1=x_2\}$ and $\{y_1+1=y_2,x_1-1= x_2\}$ are subtracted. From \eqref{triang} we see that $0<x_1/x_2=(y_1\,\pm\,1/x_2)/y_2<1$ and thus $\text{sign}(x_1)=\text{sign}(x_2)$ and $|x_1|<|x_2|$. Using an $\mathcal{R}$ transformation one can show that
\bea \label{6nodes}
\left(
\begin{array}{cccccc}
 1 & 1 & 1 & 1 & 1 & 1 \\
 0 & 1 & 0 & 0 & x_1 & x_2 \\
 0 & 0 & 1 & 0 & y_1 & y_2 \\ 
 0 & 0 & 0 & 1 & 1 & 1
\end{array}\right) \sim
\left(
\begin{array}{cccccc}
 1 & 1 & 1 & 1 & 1 & 1 \\
 0 & 1 & 0 & 0 & -x_1-y_1 & -x_2-y_2 \\
 0 & 0 & 1 & 0 & y_1 & y_2 \\ 
 0 & 0 & 0 & 1 & 1 & 1
\end{array}\right)~;
\eea
thus we can always take $x_2<x_1<0$. So to summarize, toric diagrams with no four co-planar points are described by the left matrix in \eqref{6nodes}, for $\{x_i,y_i\}$ that satisfy \eqref{triang} with $x_2<x_1<0$, $y_2>y_1>0$ and excluding the cases in which $\{y_1=1,x_1 = -1\}$ or $\{y_1+1=y_2,x_1-1= x_2\}$. This concludes our classification of diagrams with six vertices.

Six vertices is also the maximal number since, otherwise, it is not possible to arrange the vertices in two adjacent planes with the constraint that all faces are triangular. 

\subsection{Candidate duals}

\begin{figure}[ht]
\begin{center}
\includegraphics[scale=0.8]{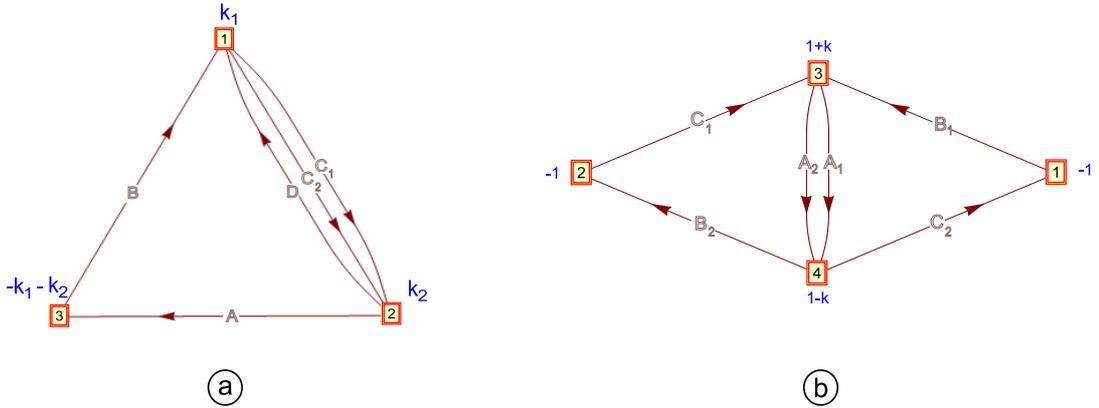}
\end{center}
\caption{Quiver diagrams with CS levels in blue.(a) The unique five-field quiver. (b) The six-field quiver.}
\label{General_quiver}
\end{figure}

We continue now with a discussion of the geometries that correspond to the polytopes obtained above. Recall that, given a toric diagram, one can recover the corresponding CY four-fold via Delzant's construction. In physics terms, this would be called a GLSM description of the four-fold. Let us discuss the toric diagrams with five vertices described above. The GLSM charge matrix can be computed by taking the null-space of the $G$ matrix, obtaining
\bea
\label{Qt-matrix}
Q_t=\left(x+y , -x , -y , -1 , 1 \right) \ .
\eea
Since the GLSM charge matrix contains one gauge group, we find that the corresponding quiver should have three nodes. However, it is not possible to find a QCS field theory for every value of $x$ and $y$. First, note that there are no zero entries in $Q_t$, therefore there should be no adjoint fields in the quiver. The most general way to construct a quiver with three nodes and five fields with no adjoints, such that there is an equal number of in-going and out-going arrows at each node, is given in Figure~\ref{General_quiver} (a).
This quiver was also discussed in \cite{Hanany:2008gx}. Since we are interested in field theories which reproduce the minimal GLSM, we must have a toric superpotential which vanishes in the Abelian case. The natural candidate is
\begin{equation}
 W = \mathrm{Tr}\, \epsilon^{ij} C_iDC_jAB~,
 \label{WW}
\end{equation}
where $\epsilon^{ij}$ is the usual alternating symbol.

In this theory the only contribution to the GLSM matrix comes from the D-term, which in this case reduces to 
\bea
Q_{\mathrm{quiver}}=\left(-k_1 - k_2, -k_1 - k_2, k_1 + k_2, k_2, k_1 \right) \ ,
\eea
Note that two of the entries are equal while in general there are no equal entries in $Q_t$. Therefore the only hope to reproduce $Q_t$ (up to an overall minus sign) is to choose $k_1+k_2=\pm\,1$. Substituting this back into $Q_{\mathrm{quiver}}$ we obtain
\bea
Q_{\mathrm{quiver}}=\left(\mp\,1, \mp\,1, \pm\,1, \pm\,1-k_1, k_1 \right) \ .
\eea
Obviously we can reproduce $Q_t$ only for $x=1$ or $y=1$. For other values, the geometries are not captured by the quiver that we have written. Indeed, it seems that the AdS$_4 \times Y_7$ spaces, where $Y_7$ is the base of the corresponding isolated CY singularity, reduce in Type IIA to singular spaces, for any choice of $U(1)$ isometry on $Y_7$. Thus the corresponding M2 brane theories apparently do not admit a Lagrangian description, according to 
\cite{Jafferis:2009th}. 

The toric diagram with six nodes that contains internal rectangular face describes the moduli space of the field theory described by the quiver in Figure~\ref{General_quiver} (b), which was also discussed in \cite{Hanany:2008gx}, with the superpotential given in \eqref{WQ111} and $k=y_2-1$. Indeed, the $Q^{111}$ field theory that we study is the special case in which $k=0$. The other toric diagrams with six nodes seem not to admit a Lagrangian description. 
\chapter{Gravity duals of baryonic symmetry-breaking}\label{sec:4}

In the rest of this part we study the 
quantization in AdS$_4$ which is more closely analogous to the case in Type IIB string theory, in which the 
$b_2(Y_7)$ $U(1)$s in the field theories are ungauged. As we have argued, 
in this theory M5 branes wrapped on supersymmetric cycles in $Y_7$ should appear as chiral primary baryonic-type operators 
in the dual SCFT. Indeed, at least for toric theories with appropriate smooth supergravity horizons $Y_7$ we expect the dual 
SCFT to be described by a QCS theory with gauge group $U(1)^2\times SU(N)^{b_2(Y_7)+2}$. The M5 brane states  
are then the usual gauge-invariant determinant-like operators in these theories, as we discussed in detail for the 
$Q^{111}$ theory in the previous chapter. 

We may then study the gravity duals to vacua in which the 
$b_2(Y_7)$ global $U(1)$s are (spontaneously) broken. On general grounds, these should correspond to 
supergravity solutions constructed from resolutions of the corresponding cone over $Y_7$. The baryonic operators are charged under the global 
baryonic symmetries, and vacua in which these operators obtain a VEV lead to 
spontaneous symmetry-breaking. By giving this VEV we pick a point in the moduli space of the theory, which at the same time introduces a scale and thus an RG flow, whose endpoint will be a different SCFT. The supergravity dual of this RG flow was first discussed in the Type IIB context by Klebanov-Witten \cite{Klebanov:1999tb}. 

More generally, different choices of 
boundary conditions will imply that some, or all, of the M5 brane states considered here are absent. It is then clearly very interesting 
to ask what is the dual field theory interpretation of these gravity backgrounds in such situations. Again, we leave this for future work.

\section{Baryonic symmetry-breaking in the $Q^{111}$ theory} \label{sec:4-1}

In this section we begin by discussing in detail the baryonic symmetry-breaking for the case of $Q^{111}$. In the next section we describe how to generalize this discussion for general CY four-fold singularities. In particular, we will obtain a general formula for M5 brane condensates, or indeed more generally still a formula for the on-shell action of a wrapped brane in a warped CY background. Essentially this formula appeared in \cite{Baumann:2006th}, where it was checked in some explicit examples. Here we provide a general proof of this formula.

\subsection{Resolutions of $\mathcal{C}(Q^{111})$}\label{sec:resQ111}

In this section we consider the warped resolved gravity backgrounds for $Q^{111}$. We begin by discussing this 
in the context of the GLSM, and then proceed to construct corresponding explicit supergravity solutions. 

\subsubsection{Algebraic analysis}\label{sec:GKZQ111}

\begin{figure}[ht]
\begin{center}
\includegraphics[scale=1.4]{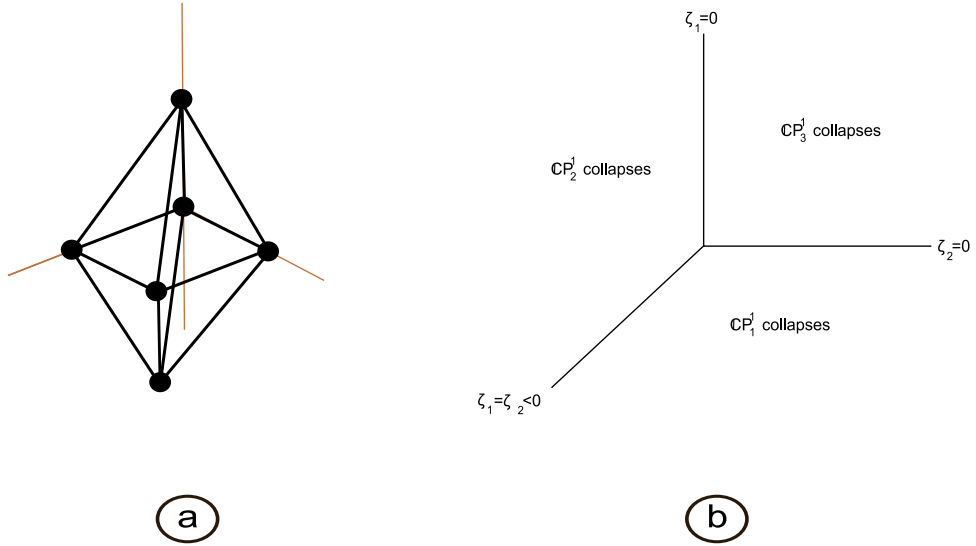}
\end{center}
\caption{(a) The toric diagram for $\mathcal{C}(Q^{111})$. (b) The GKZ fan for $Q^{111}$ is $\R^2$, divided into three cones.}
\label{fig:GKZ1}
\end{figure}

The toric singularity $\mathcal{C}(Q^{111})$ may be described by a GLSM with six fields, $a_i$, $b_i$, $c_i$, $i=1,2$, 
and gauge group $U(1)^2$. This is also the same as the Abelian QCS theory presented in \secref{sec:3}, but without the CS terms. 
The charge matrix is
\bea
\begin{array}{c|cccccc} & a_1 & a_2 & b_1 & b_2 & c_1 & c_2 \\ \hline U(1)_1 & -1 & -1 & 1 & 1 & 0 & 0 \\ U(1)_2 & -1 & -1 & 0 & 0 & 1 & 1 \end{array}~.
\eea
The singular cone $\mathcal{C}(Q^{111})$ is the moduli space of this GLSM where the FI parameters 
$\zeta_1=\zeta_2=0$ are both zero. However, more generally we may allow 
$\zeta_1, \zeta_2\in \R$, leading to different (partial) resolutions of the singularity. In fact 
since there are no internal points in the toric diagram in Figure
\ref{fig:GKZ1}, this 
GLSM in fact describes \emph{all} possible (partial) resolutions of the singular cone.

It is straightforward to analyse the various cases. 
Suppose first that $\zeta_1,\zeta_2>0$ are both positive. We may write the two D-terms of the GLSM as
\bea\label{Dpos}
|b_1|^2 + |b_2|^2 &=& \zeta_1 + |a_1|^2 + |a_2|^2>0~,\nonumber\\
|c_1|^2 + |c_2|^2 &=& \zeta_2 + |a_1|^2 + |a_2|^2>0~.
\eea
In particular, for $a_1=a_2=0$ we obtain $\mathbb{CP}^1\times\mathbb{CP}^1$ where the K\"ahler class of each factor is 
proportional to $\zeta_1$ and $\zeta_2$, respectively. Here $b_i$ and $c_i$ may be thought of as homogeneous coordinates on the $\mathbb{CP}^1$s. 
Altogether, this describes the total space of the bundle $\mathcal{O}(-1,-1)\oplus \mathcal{O}(-1,-1)\rightarrow\mathbb{CP}^1\times\mathbb{CP}^1$, with $a_1, a_2$ the two fibre coordinates on the $\C^2$ fibres. 

Suppose instead that $\zeta_1<0$. We may then rewrite the D-terms as
\bea
|a_1|^2 + |a_2|^2 &=& -\zeta_1 + |b_1|^2 + |b_2|^2>0~,\nonumber\\
|c_1|^2 + |c_2|^2 &=& \zeta_2-\zeta_1 + |b_1|^2 + |b_2|^2~.
\eea
Provided also $\zeta_2-\zeta_1>0$, we hence obtain precisely the same geometry as when $\zeta_1,\zeta_2>0$, but with the $\mathbb{CP}^1\times\mathbb{CP}^1$ zero section now parametrized by $a_i$ and $c_i$ and with K\"ahler classes proportional to $-\zeta_1$ and $\zeta_2-\zeta_1$, respectively. There is a similar situation with $\zeta_2<0$ and $\zeta_1-\zeta_2>0$.

Hence in total there are three different resolutions of $\mathcal{C}(Q^{111})$, corresponding to choosing which of the three $\mathbb{CP}^1$s in $Q^{111}$ collapses at the zero section in $\mathcal{O}(-1,-1)\oplus \mathcal{O}(-1,-1)\rightarrow\mathbb{CP}^1\times\mathbb{CP}^1$. We label these three $\mathbb{CP}^1$s as $\mathbb{CP}^1_a$, $a=1,2,3$, which in $Q^{111}$ are parametrized by $c_i$, $b_i$, $a_i$, respectively. This is shown in Figure \ref{fig:GKZ1}, which is known more generally as the \emph{Gel'fand-Kapranov-Zelevinsky (GKZ) fan}. Notice there is a $\Sigma_3$ permutation symmetry of the three $\mathbb{CP}^1$s in $Q^{111}$ and the three different resolutions are permuted by this symmetry.

The boundary edges between the regions correspond to collapsing another of the $\mathbb{CP}^1$s, leading only to a partial resolution 
of the singularity. Thus, for example, take $\zeta_1>0$ but $\zeta_2=0$. The D-terms are now
\bea
|b_1|^2 + |b_2|^2 &=& \zeta_1 + |a_1|^2 + |a_2|^2>0~,\nonumber\\
|c_1|^2 + |c_2|^2 &=& |a_1|^2 + |a_2|^2~.
\eea
The second line describes the conifold singularity, which is then fibred over a $\mathbb{CP}^1$, parametrized by the $b_i$, of K\"ahler class $\zeta_1$. 

\subsubsection{Supergravity analysis}\label{sec:sugraQ111}

For each of the resolutions of $\mathcal{C}(Q^{111})$ described above there is a corresponding Ricci-flat K\"ahler metric
that is asymptotic to the cone metric over $Q^{111}$. More precisely, there is a unique such metric for each choice of 
K\"ahler class, or equivalently FI parameter $\zeta_1,\zeta_2\in \R$. As we shall discuss later in this section, this is guaranteed
by a general theorem that has only just been proven in the mathematics literature. However, for $Q^{111}$ these 
metrics may in fact be written down explicitly. Denoting the (partially) resolved CY generically by $X$, the 
CY metrics are given by
\begin{eqnarray}\label{resolvedQ111}
\diff s^2(X) &=&\kappa(r)^{-1}\dd r^2+\kappa(r)\frac{r^2}{16}\Big(\dd\psi+\sum_{i=1}^3 \cos\theta_i \dd\phi_i\Big)^2+\frac{2a+r^2}{8}\Big(\dd\theta_2^2+\sin^2\theta_2 \dd\phi_2^2\Big)\nonumber \\ &&+\frac{2b+r^2}{8}\Big(\dd\theta_3^2+\sin^2\theta_3\dd\phi_3^2\Big)+\frac{r^2}{8}\Big(\dd\theta_1^2+\sin^2\theta_1\dd\phi_1^2\Big)~,
\end{eqnarray}
where 
\begin{equation}\label{kappa}
\kappa(r)=\frac{(2A_-+r^2)(2A_++r^2)}{(2a+r^2)(2b+r^2)}~,
\end{equation}
$a$ and $b$ are arbitrary constants, and we have also defined 
\begin{equation}
A_\pm=\frac{1}{3}\Big(2a+2b\pm\sqrt{4a^2-10ab+4b^2}\Big)~.
\end{equation}
One easily sees that at large $r$ the metric approaches the cone over the $Q^{111}$ metric (\ref{Qiiimetric}). This 
way of writing the resolved metric breaks the $\Sigma_3$ symmetry, since it singles out the $\mathbb{CP}^1$ parametrized by 
$(\theta_1,\phi_1)$ as that collapsing at $r=0$. Here we have an exceptional $\mathbb{CP}^1\times\mathbb{CP}^1$, parametrized 
by $(\theta_2,\phi_2)$, $(\theta_3,\phi_3)$, with K\"ahler 
classes proportional to $a>0$ and $b>0$, respectively. Thus setting $\{a=0, b>0\}$, or $\{b=0,a>0\}$, leads 
to a partial resolution with a residual $\mathbb{CP}^1$ family of conifold singularities at $r=0$.  We shall examine this 
in more detail below.

We are interested in studying supergravity backgrounds corresponding to M2 branes localized on one of these resolutions 
of $\mathcal{C}(Q^{111})$. We thus consider the following ansatz for the background sourced by such M2 branes
\begin{eqnarray}\label{metricagain}
\dd s^2_{11}&=&h^{-2/3}\, \dd s^2\left(\R^{1,2}\right) +h^{1/3}\dd s^2(X)~, \nonumber\\ G&=&\dd^3 x \wedge \dd h^{-1}~,
\end{eqnarray}
where $\dd s^2(X)$ is the CY metric (\ref{resolvedQ111}).
If we place $N$ spacetime-filling M2 branes at a point $y\in X$, we must then also solve the equation
\bea
\Delta_x h[y] = \frac{(2\pi \ell_p)^6N}{\sqrt{\det g_X}} \delta^8(x-y) \ ,
\eea
for the warp factor $h=h[y]$. Here $\Delta$ is the scalar Laplacian on $X$. 
Having the explicit form of the metric we can compute this Laplacian and solve for the warp factor to obtain the full supergravity solution. 
This is studied in detail in \secref{sec:D}. 

In the remainder of this subsection let us analyse the simplified case 
in which we partially resolve the cone, 
setting $a=0$ and $b>0$. This corresponds to one of the boundary lines in the GKZ fan in Figure \ref{fig:GKZ1}, with the point on the boundary 
$\R_{> 0}$ labelled by the metric parameter $b>0$.
Here one can solve explicitly for the warp factor in the case where we put the $N$ M2 branes 
at the north pole of the exceptional $\mathbb{CP}^1$ parametrized by $(\theta_3,\phi_3)$; this is the point with coordinates $y=\{r=0, \theta_3=0\}$. 
Notice the choice of north pole is here without loss of generality, due to the $SU(2)$ isometry acting 
on the third copy of $\mathbb{CP}^1$. We denote the corresponding warp factor in this case as simply $h[y=\{r=0,\theta_3=0\}]\equiv h$. 
As shown in \secref{sec:Q111-warp}, $h=h(r,\theta_3)$ is then given explicitly in terms of hypergeometric functions by
\begin{eqnarray}
h(r,\theta_3) &=&\sum_{l=0}^\infty\,H_{l}(r)\, P_{l}(\cos\theta_3)~,\nonumber \\
H_{l}(r)&=& \mathcal{C}_{l}\, \Big(\frac{8b}{3r^2}\Big)^{3(1+\beta)/2}\, _2F_1\left(-\tfrac{1}{2}+\tfrac{3}{2}\beta,\tfrac{3}{2}+\tfrac{3}{2}\beta,1+3\beta,-\tfrac{8b}{3r^2}\right)~,
\label{Q111-warp-factor}
\end{eqnarray}
where $P_{l}$ denotes the $l$th Legendre polynomial, 
\begin{equation}
 \beta=\beta(l)=\sqrt{1+\frac{8}{9}l(l+1)}~,
\end{equation}
and the normalization factor $\mathcal{C}_{l}$ is given by
\bea
\mathcal{C}_{l}&=&\frac{3\Gamma(\frac{3}{2}+\frac{3}{2}\beta)^2}{2\Gamma(1+3\beta)}\left(\frac{3}{8b}\right)^3\, (2l+1)\, R^6~,\\\label{anothereqn}
 R^6&=&\frac{(2\pi\ell_p)^6 N}{6\text{vol}(Q^{111})}=\frac{256}{3}\pi^2 N \ell_p^6~.
\eea
In the field theory this solution corresponds to breaking one combination of the two global $U(1)$ baryonic symmetries, rather than both of them. 
This will become clear in the next section.
The resolution of the cone can be interpreted in terms of giving an expectation value to a certain operator $\mathcal{U}$ in the field theory. This operator is contained in the same multiplet as the current that generates the broken baryonic symmetry, and couples to the corresponding $U(1)$ gauge field in AdS$_4$. Since a conserved current has no anomalous dimension, the dimension of $\mathcal{U}$ is uncorrected in going from the classical description to supergravity \cite{Klebanov:1999tb}. According to the general AdS/CFT prescription \cite{Klebanov:1999tb}, the VEV of the operator $\mathcal{U}$ is dual to the sub-leading correction to the warp factor. For large $r$ we can write
\begin{equation}
h(r,\theta_3)\sim\sum_{l=0}^\infty\, \mathcal{C}_{l}\, \Big(\frac{8b}{3r^2}\Big)^{3(1+\beta)/2}\, P_{l}(\cos\theta_3)~.
\end{equation}
Expanding the sum we then have
\begin{equation}
h(r,\theta_3)\sim \frac{R^6}{r^6}\left(1+\frac{18b\, \cos\theta_3}{5r^2}+\cdots\right)~.
\end{equation}
In terms of the AdS$_4$ coordinate $z= r^{-2}$ we have that the leading correction is of order $z$, which indicates that the dual operator $\mathcal{U}$ is dimension one. This is precisely the expected result, since this operator sits in the same supermultiplet as the broken baryonic current, and thus has a protected dimension of one. Furthermore, its VEV is proportional to $b$, the metric resolution parameter, which reflects the fact that in the conical (AdS) limit in which $b=0$ this baryonic current is not broken, and as such $\langle \mathcal{U}\rangle=0$.

The moduli space of the field theory in the new IR is equivalent to the geometry close to the branes. 
Recall that we placed the $N$  M2 branes at the north pole $\{\theta_3=0\}$ of the exceptional
$(\theta_3,\phi_3)$ sphere at $r=0$. Defining $\tilde{\psi}=\psi+\phi_3$ and introducing the new radial variables $\tilde{r}=\frac{b}{2}\,\theta_3$, 
$\rho=\frac{\sqrt{3}}{2}\, r$,  the geometry close to the branes becomes to leading order
\begin{eqnarray}
\dd s^2&=&\dd\rho^2+\frac{\rho^2}{9}\Big(\dd\tilde{\psi}+\sum_{i=1}^2 \cos\theta_i\dd\phi_i\Big)^2+\frac{\rho^2}{6}\Big(\dd\theta_2^2+\sin^2\theta_2\dd\phi_2^2\Big)\nonumber \\ &&+\frac{\rho^2}{6}\Big(\dd\theta_1^2+\sin^2\theta_1\dd\phi_1^2\Big)+\Big(\dd\tilde{r}^2+\tilde{r}^2\dd\phi_3^2\Big)~,
\label{c_x_conifold}
\end{eqnarray}
which is precisely the Ricci-flat K\"ahler metric of $\mathcal{C}(T^{11})\times \mathbb{C}$, in accordance with the discussion in the previous subsection.

\subsection{Higgsing the $Q^{111}$ field theory}\label{sec:HiggsQ111}

We have argued that the warped resolved supergravity solutions described in the previous section are dual to 
spontaneous symmetry-breaking in the SCFT in which the M5 brane states appear as baryonic-type operators. 
Let us study this in more detail in the field theory described in \secref{sec:3}. 
In this SCFT the symmetries $U(1)_I$ and $U(1)_{II}$ in (\ref{gaugedU(1)charges}) are global, rather than gauge, symmetries, 
with the corresponding conserved currents coupling to the baryonic $U(1)$ gauge fields in AdS$_4$. By inspection 
of this charge matrix we conclude that it is possible to give a VEV to the $A_i$, $B_i$ and $C_i$ fields. 
These VEVs then break the corresponding baryonic $U(1)$ symmetries. In particular, by giving a VEV to 
any pair of fields ($A$s, $B$s or $C$s) we break only one particular $U(1)$ baryonic symmetry, leaving another combination
unbroken. In this section we will examine the resulting Higgsings of the gauge theory obtained by giving VEVs to 
different pairs of fields, and compare with the gravity results of the previous section. 

\subsubsection{Higgsing $\mathcal{C}(Q^{111})$}
As explained in \secref{sec:2}, at each of the two poles for each copy of $\mathbb{CP}^1$ in the K\"ahler-Einstein base of $Q^{111}$, 
there is a supersymmetric five-cycle that may be wrapped by an M5 brane. Altogether these are six M5 brane states, corresponding to the 
toric divisors of $\mathcal{C}(Q^{111})$. Each pair are acted on by one of the $SU(2)$ factors in the isometry group 
$SU(2)^3\times U(1)_R$, rotating one into the other. Quantizing the BPS particles in AdS$_4$ one obtains dual baryonic-type operators
given by (\ref{b-operators}). In particular, consider the M5 branes that sit at a point on the copy of $\mathbb{CP}^1=S^2$ with coordinates 
$(\theta_3,\phi_3)$. In the next section we will compute the VEV of these M5 brane operators in the partially resolved gravity background 
described by (\ref{Q111-warp-factor}), 
showing that the baryonic operator dual to the M5 brane at $\theta_3=0$ vanishes, while that at the opposite pole $\theta_3=\pi$ 
is non-zero and proportional to the resolution parameter $b$ (see equation (\ref{vev})). Considering the $A$ fields in the field theory 
this corresponds to the fact that, after breaking the baryonic symmetry by giving diagonal VEVs to these fields, it is possible to use the $SU(2)$ flavour symmetry to find one combination of $A$ fields with zero VEV, and an orthogonal combination with non-vanishing VEV. Let us assume for example that $\langle A_1\rangle=0$ and $\langle A_2\rangle=b\, I_{N\times N}$. Thus only one baryonic operator in (\ref{b-operators}) has non-vanishing VEV, 
namely\footnote{As anticipated in \secref{sec:3}, at the IR superconformal fixed point the dimensions of the chiral fields are expected to be different from the free field fixed point. That is why generically the VEV of the baryonic operator is $\langle\mathscr{B}_{A_2}\rangle = b^{N\, \Delta_{A_2}}$.} $\langle\mathscr{B}_{A_2}\rangle = b^{N\, \Delta_{A_2}}$. This situation was analyzed in \cite{Franco:2009sp}, where it was shown that the effective field theory in the IR has CS quiver
\begin{center}
\includegraphics[scale=1.2]{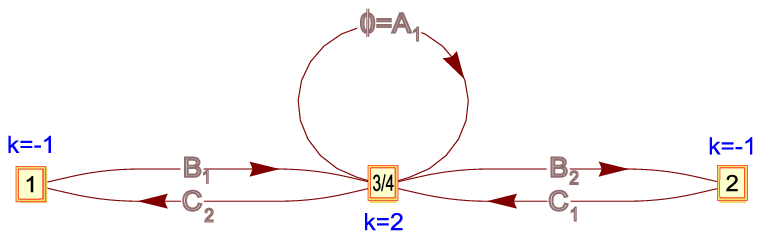}
\end{center}
with superpotential
\begin{equation}\label{con1}
W\,=\, {\rm Tr}\,\Phi\, \Big(\,C_2\, B_1\, B_2\, C_1\,-\,B_2\, C_1\, C_2\, B_1\,\Big)~.
\end{equation}
As shown in \cite{Franco:2009sp}, the moduli space of this theory is indeed $\mathcal{C}(T^{11})\times\C$. This is of course expected from the gravity analysis of equation (\ref{c_x_conifold}). Any other VEV for the $A$ fields corresponds to placing the M2 branes on $SU(2)$-equivalent points on the blown-up $\mathbb{CP}^1$, and therefore results in the same near horizon geometry.

The manifest symmetry exhibited by the Lagrangian of the $Q^{111}$ field theory is $SU(2) \times U(1)_R$, which is only a subgroup of the full $SU(2)^3 \times U(1)_R$ symmetry which is expected from the isometry of the $Q^{111}$ manifold. Therefore, in contrast to the situation with the 
$A$ fields, we see that different VEVs for the pairs of $B$ or $C$ fields result in different theories in the IR. Giving a non-vanishing VEV to $C_1$ and $C_2$, for example, results in the CS quiver
\begin{center}
\includegraphics[scale=1.0]{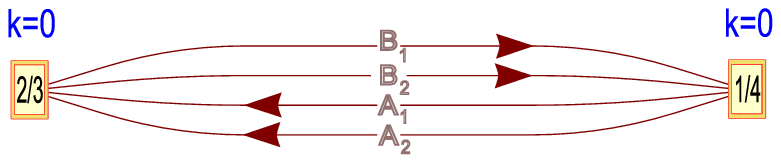}
\end{center}
with superpotential
\begin{equation}\label{zeroCS}
W\,=\, {\rm Tr}\,\Big(\,A_2\, B_1\, B_2\, A_1\,-\,B_2\, A_2\, A_1\, B_1\,\Big)~,
\end{equation}
while when the VEV of only one field is non-vanishing we instead obtain the CS quiver
\begin{center}
\includegraphics[scale=1.2]{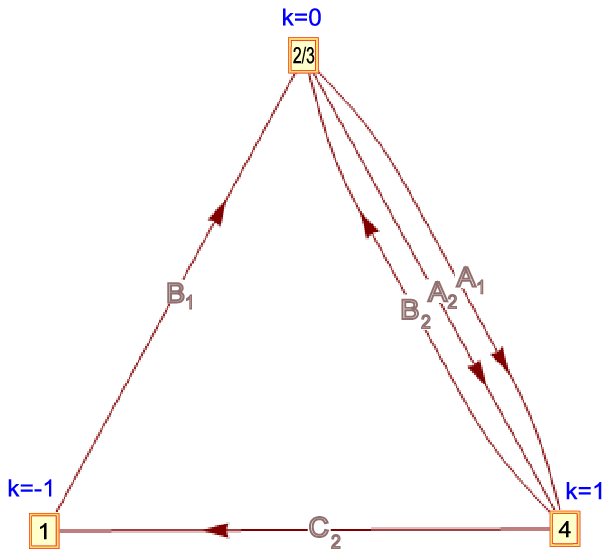}
\end{center}
with superpotential
\begin{equation}\label{con2}
W\,=\, {\rm Tr}\,C_2\, \Big(\,A_2\, B_1\, B_2\, A_1\,-\, A_2\, A_1\, B_1\,B_2\,\Big)~.
\end{equation}
Of course, both cases correspond geometrically to blowing up the same $\mathbb{CP}^1$, as can be seen from the explicit construction of the field 
theory moduli space. However, recall that the position of the M2 branes depends on the VEVs of $C_1$ and $C_2$. While in the gravity picture all  locations on the exceptional $\mathbb{CP}^1$ are $SU(2)$-equivalent, in the field theory since only part of the global symmetry is manifest we obtain different theories for different VEVs. The supergravity analysis hence suggests that the theories obtained in the IR above are dual. Indeed, one can check that the moduli spaces of these theories are the same, the details appearing in \cite{Davey:2009qx}.

The QCS theory for $\mathcal{C}(T^{11})\times\C$ in (\ref{zeroCS}) has zero CS levels, and there is therefore no tunable coupling parameter in this theory. The reason for that can be seen from the brane construction. After blowing up the cone to the partial resolution and placing the stack of M2 branes at a residual singular point on the exceptional $\mathbb{CP}^1$, the tangent cone geometry at this point is $\mathcal{C}(T^{11})\times\C$. In the field theory the $S^1$ that rotates the copy of $\C$ corresponds to the M-theory circle. Shrinking the size of this circle in the M-theory supergravity solution corresponds to a Type IIA limit describing a black D2 brane solution with no smooth near horizon. This Type IIA background is not an appropriate description of the system as the string coupling close to the branes diverges and therefore one has to use the M-theory description. The lack of tunable coupling in the field theory is due to the fact that no such coupling exists in M-theory. On the other hand, in the remaining two field theories (\ref{con1}), (\ref{con2}) for $\mathcal{C}(T^{11})\times\C$ described in this section 
the M-theory circle involves a $U(1)$ that acts also on $\mathcal{C}(T^{11})$\footnote{The M-theory circle can be deduced by computing the $\Z_k$ orbifold action on the moduli space in each case.} and a weak coupling limit in the field theory does exist. In this limit the field theory can be treated classically and therefore a Lagrangian description can be found. These issues will be discussed in more detail in \secref{sec:exo}.

As a final comment in this section, notice that in general we will have an infinite set of CFTs dual to the $Q^{111}$ geometry, with different boundary conditions on the baryonic gauge fields (\ref{3-form-to-global}) in AdS$_4$. From our earlier discussion, these will have the same QCS theories in the $SU(N)$ sector, but different $U(1)$ sectors. In particular, in general different combinations of the diagonal $U(1)$s in $U(N)$ may be gauged, and with different $U(1)$ CS levels 
from the levels $k_a$ of the $SU(N)$ factors. These $U(1)$ sectors will in general behave differently under Higgsing. 
In particular, a global $U(1)$ can be spontaneously broken, while a gauged $U(1)$ can be Higgsed. It will be interesting 
to investigate this general structure in both the field theory and dual supergravity solutions, although we leave this 
for future work, focusing instead on the $U(1)^2\times SU(N)^4$ theory. 

\subsection{Baryonic condensates and M5 branes in the $Q^{111}$ solution}\label{sec:M5Q111}

In the previous section we discussed the RG flow triggered by giving a VEV to one of the fields, and hence baryonic operators,
in the $Q^{111}$ theory with gauge group $U(1)^2\times SU(N)^4$. In this section we calculate the VEV of this baryonic operator
in the corresponding gravity solution described by (\ref{Q111-warp-factor}). In order to do this we follow the analogous calculation in 
the Type IIB context, discussed in \cite{Martelli:2008cm, Klebanov:2007us}. In this prescription, to determine the VEV of a 
baryonic operator dual to an M5 brane wrapped on a five-submanifold $\Sigma_5\subset Y_7$ in the (partially) resolved 
supergravity background, we compute the Euclidean action of an M5 brane which wraps a minimal six-dimensional 
manifold in $X$ which at large $r$ asymptotes to the cone over $\Sigma_5$. In this section we 
present an explicit example, although later we will present a general formula for such VEVs.

Again, we focus on the partially resolved background described by (\ref{Q111-warp-factor}). We are interested in computing the 
VEV of the operator that was carrying charge under the baryonic symmetry before it was broken. As described in \secref{sec:2}, 
this symmetry originates on the supergravity side in a mode (\ref{3-form-to-global}) of the six-form potential $C_6$ along 
a five-cycle in the Sasaki-Einstein manifold $Q^{111}$. Consider a Euclidean M5 brane that is wrapped on the six-manifold 
at a fixed point in the $(\theta_3,\phi_3)$ copy of $\mathbb{CP}^1$, and is coordinatized by 
 $\{r,\psi,\theta_1,\theta_2,\phi_1,\phi_2\}$ in the partial resolution of $\mathcal{C}(Q^{111})$. This six-manifold is 
a divisor in the partially resolved background, and hence this wrapped M5 brane world-volume is a calibrated submanifold. 
The M5 brane carries charge 
under the $U(1)$ gauge field in $AdS_4$ that descends from the corresponding harmonic five-form in $Q^{111}$. 
We calculate the Euclidean action of this wrapped M5 brane 
 up to a radial cut-off $r=r_c$, identifying 
 $\me^{-S(r_c)}$ with the classical field dual to the baryonic operator, as in \cite{Klebanov:2007us}. Explicitly, the action is given by
\bea
S(r_c)=T_{5}\int_{r\leq r_c} h\,\sqrt{\det g_6}\, \dd^6x~,
\label{m5-action}
\eea
where $T_{5}=2\pi/(2\pi\ell_p)^6$ is the M5 brane tension, the warp factor $h$ is given by (\ref{Q111-warp-factor}), and $\det g_6$ is the determinant of the metric induced on the M5 brane world-volume. This induced metric is
\begin{eqnarray}
 \dd s_6^2&=&\kappa^{-1}\dd r^2+\kappa\frac{r^2}{16}\Big(\dd\psi+\sum_{i=1}^2 \cos\theta_i\dd\phi_i\Big)^2+\frac{r^2}{8}\Big(\dd\theta_2^2+\sin^2\theta_2\dd\phi_2^2\Big)\nonumber \\ &&+\frac{r^2}{8}\Big(\dd\theta_1^2+\sin^2\theta_1\dd\phi_1^2\Big)~,
\end{eqnarray}
where recall that $\kappa(r)$ is given by (\ref{kappa}).
A straightforward calculation hence gives
\bea
\sqrt{\det g_6}=\frac{\,r^5 \sin\theta_1 \sin\theta_2}{256}~.
\label{detg6}
\eea
Substituting these results into (\ref{m5-action}) then gives
\bea
\nn
S(r_c)&=&\frac{T_{5}}{256}\int \dd\phi_1\, \dd\phi_2\, \dd\theta_1\, \dd\theta_2\, \dd\psi\, \sin \theta_1\, \sin\theta_2\, \int_0^{r_c} \dd r\, r^5 \sum_{l=0}^\infty H_{l}(r) P_{l}(\cos\theta_3) \nn \\
&=&\frac{\pi^3 T_{5}}{4} \left[ \int_0^{r_c} \dd r\, r^5 H_0 (r) + \sum_{l=1}^{\infty} \int_0^{r_c} \dd r\, r^5 H_{l}(r) P_{l}(\cos\theta_3) \right]~.
\label{mem-action}
\eea
Let us evaluate the integrals separately. The first diverges in the absence of the cut-off $r_c$, since
\bea
\int_0^{r_c} \dd r\, r^5 H_0 (r) \simeq \frac{R^6}{2} \Big[\frac{1}{3}+\log\left(1+\frac{3r_c^2}{8b}\right)\Big]~.
\eea
The second integral is finite and can be calculated straightforwardly:
\bea
\nn
\int_0^{\infty} \dd r\, r^5 \sum_{l=1}^{\infty}H_{l}(r) P_{l}(\cos\theta_3)&=&\frac{3\,R^6}{4}\sum_{l=1}^{\infty}\frac{(2l+1)}{l(l+1)}P_{l}(\cos\theta_3) \nn \\
&=& -\frac{3\,R^6}{2}\Big(\frac{1}{2}+\log \sin\frac{\theta_3}{2}\Big)~.
\eea
Recall here that $R$ is given by (\ref{anothereqn}). Substituting these results into \eqref{mem-action} then gives
\bea
\me^{-S(r_c)}=\me^{{7N}/{18}} \left(\frac{8b}{3\,r_c^2}\right)^{\frac{N}{3}}\left(\sin \frac{\theta_3}{2}\right)^{N}~.
\label{vev}
\eea
The interpretation of this result is along the same lines as the discussion in the case of the conifold  in \cite{Klebanov:2007us}. We will therefore keep our discussion brief and refer the reader to \cite{Klebanov:2007us} for further details. 
Since the AdS$_4$ radial coordinate is related to $r$ as $z=r^{-2}$, we see that the operator dimension is $\Delta = \frac{N}{3}$, as anticipated by our  field theory discussion. Indeed, this provides a non-trivial check of the R-charge assignment required for the theory to have an IR superconformal fixed point, supporting the conjecture that the theory is indeed dual to M2 branes moving in $\mathcal{C}(Q^{111})$.

For the remaining M5 branes, wrapped on Euclidean six-submanifolds at a point in either the $(\theta_1,\phi_1)$ or $(\theta_2,\phi_2)$ 
copies of $\mathbb{CP}^1$, we note that $\me^{-S(r_c)}=0$. This is simply because the M5 brane world-volumes intersect 
the M2 brane stack on the exceptional $\mathbb{CP}^1$ parametrized by $(\theta_3,\phi_3)$ in these cases, and hence 
the world-volume action (\ref{m5-action}) is logarithmically divergent near to the M2 branes -- for
further discussion of this in the D3 brane case, see \cite{Martelli:2008cm}. Thus the dual gravity analysis of the 
partially resolved supergravity solution is in perfect agreement with the proposed $Q^{111}$ field theory, at least 
with the boundary conditions we study in this section.

\subsection{Wrapped branes and the phase of the condensate}\label{sec:wrapped}

In the resolved, or partially resolved, geometry one can consider various different kinds of stable wrapped branes in the $r\sim 0$ region. 
These shed further light on the physical interpretation of the supergravity solutions. 

In the fully resolved case with $a>0$ and $b>0$ one could consider an M5 brane wrapping the 
exceptional $\mathbb{CP}^1\times\mathbb{CP}^1$ at $r=0$ and filling the spacetime directions $x_0$, $x_1$. This is a domain wall in the Minkowski 
three-space in (\ref{metricagain}). Its tension, given by the energy of the probe brane, is $abT_5\pi^2/16$.
Notice that the warp factor cancels out in this computation, and the brane remains of finite tension even if the stack of M2 
branes is placed at $r=0$.

Of more interest for us is to consider an M2 brane wrapping an exceptional $\mathbb{CP}^1$, either in the resolved or 
partially resolved background. In the former case notice there are homologically two choices of such $\mathbb{CP}^1$ inside $\mathbb{CP}^1\times\mathbb{CP}^1$ 
at $r=0$. Again, in either case this is a calibrated cycle. This leads to 
a point particle in the Minkowski three-space, moving along the time direction $x_0$, whose mass is
$bT_2\pi/4$.
Again, its energy remains finite even when the M2 brane stack is at $r=0$. 

This M2 brane sources three-form fluctuations of the form
\begin{equation} \label{fluctuateC3}
\delta C_3=A\wedge \beta~,
\end{equation}
where $\beta$ is a two-form on the Ricci-flat K\"ahler manifold $X$. Demanding that $A$ is a massless gauge field in the Minkowski three-space leads to the equations
\begin{equation}\label{betaeqn}
\dd\beta=0\ ,\qquad \dd(\, h\star_8\beta)=0~,
\end{equation}
where $\star_8$ denotes the Hodge dual on $X$. In particular $\beta$ is closed and hence defines a cohomology class in $H^2(X,\R)$; we shall be interested in the case where this 
class is Poincar\'e dual to the two-cycle wrapped by the particle-like M2 brane.
In three dimensions the gauge field $A$ is dual to a periodic scalar, which can be identified as the Goldstone boson of the spontaneous symmetry-breaking.
Indeed, the M2 particle is a magnetic source for this pseudoscalar. The pseudoscalar modes will therefore have unit monodromy around a circle linking the M2 particle worldline. As in \cite{Klebanov:2007cx}, this Goldstone boson is expected to appear as a phase, through the Wess-Zumino action of the Euclidean M5 brane, in the baryonic condensate. We shall see that this is indeed the case. This wrapped M2 brane is the analogue of the global string that was discussed in \cite{Klebanov:2007cx} for the conifold theory in Type IIB. 

It thus remains to construct appropriate two-forms $\beta$ satisfying (\ref{betaeqn}) in the warped resolved backgrounds for $Q^{111}$. 
This will occupy us for the remainder of this subsection.
It will turn out to be simpler to use the metric in the form given in \cite{Benishti:2010jn}, which we reproduce here for convenience:
\begin{eqnarray}
\dd s^2&=&{U}^{-1}\dd\varrho^2+{U}\varrho\Big(\dd\psi+\sum_{i=1}^3 \cos\theta_i\dd\phi_i\Big)^2+(l_2^2+\varrho)\,\Big(\dd\theta_2^2+\sin^2\theta_2\dd\phi_2^2\Big)\nonumber \\ &&+(l_3^2+\varrho)\,\Big(\dd\theta_3^2+\sin^2\theta_3\dd\phi_3^2\Big)+\varrho\, \Big(\dd\theta_1^2+\sin^2\theta_1\dd\phi_1^2\Big)~,
\end{eqnarray}
where
\begin{equation}
 U(\varrho)= \frac{3\, \varrho^3+4\, \varrho^2\, (l_2^2+l_3^2)+6\, l_2^2\, l_3^2\varrho}{6\,(l_2^2+ \varrho)\, (l_3^2+ \varrho)}~.
\end{equation}
The constants $l_2$, $l_3$ are related to the constants $a,b$ in (\ref{resolvedQ111}) via 
$a=4l_2^2$, $b=4l_3^2$.

\subsubsection{Harmonic forms}

It is convenient to introduce the following vielbein
\bea
e_{\theta_i}=\diff \theta_i ,\quad e_{\phi_i}=\sin{\theta_i}\diff \phi_i ,\quad g_5=\diff \psi + \sum_{i=1}^3 \cos{\theta_i} \diff \phi_i~.
\eea
Since with the M2 brane stack at $\varrho=0$ the warp factor is then a function of $(\varrho, \theta_2, \theta_3)$, it is natural to consider the following ansatz for the two-form $\beta$:
\begin{equation}
\beta=e_{\theta_2}\wedge e_{\phi_2}+\dd\left(\, f_0\, g_5+\sum_{i=2}^3 f_i\, e_{\phi_i}\, \right)~,
\end{equation}
where now $f_{\mu}=f_{\mu}(\varrho,\,\theta_2,\,\theta_3)$, and $\mu=0,2$ or $3$. Since the corresponding $\delta C_3$ fluctuation in \eqref{fluctuateC3} couples to the M2 brane wrapped
over the $\mathbb{CP}^1$ at $\varrho=0$, $\beta$ should approach here the volume form of the finite sized $\mathbb{CP}^1$.
On the other hand,
as we will see later, in the large $\varrho$ region $\beta$ should asymptote to a harmonic two-form
$\omega_2$ on the singular cone.

The second equation in (\ref{betaeqn}) implies that the $f_{\mu}$ must satisfy
\begin{eqnarray}
&&\partial_{\rho}\Big(\frac{h\,\sqrt{\det g}\,U(\rho)}{\ell_j^2+\rho}\partial_{\rho} f_j \Big)+\sum_{i=2}^3 \frac{1}{\sin{\theta_i}} \partial_{\theta_i}\Big( \frac{h\,\sqrt{\det g}\,\sin{\theta_i}}{(\ell_j^2+\varrho)(\ell_i^2+\varrho)} \partial_{\theta_i}f_j \Big)-\\ \nn
&&-\partial_{\theta_j}\Big( \frac{h\,\sqrt{\det g}\,f_0}{(\ell_j^2+\varrho)^2}\Big)+\partial_{\theta_j}\Big( \frac{h\,\sqrt{\det g}\cot{\theta_j}}{(\ell_j^2+\varrho)^2}\Big)f_j +\partial_{\theta_j}\Big( \frac{h\,\sqrt{\det g}}{(\ell_j^2+\varrho)^2}\Big)\delta_{j,2}=0~,
\end{eqnarray}
for $\mu=j$, and
\begin{eqnarray}
&&\partial_{\varrho}\Big(h\, \sqrt{\det g}\, \partial_{\varrho}f_0\Big)+\sum_{i=1}^3\frac{1}{\sin\theta_i}\partial_{\theta_i}\Big(\frac{U^{-1}(\varrho)\, \sqrt{\det g}\, h}{(\varrho+l_i^2)}\, \sin\theta_i\,\partial_{\theta_i}\, f_0\Big) \\ \nonumber
&& +\sum_{i=2}^3\frac{\sqrt{ \det g}\, h}{(\varrho+l_i^2)^2\, \sin\theta_i}\partial_{\theta_i}\Big(\sin\theta_i\, f_i\Big)-\sum_{i=1}^3 \frac{\sqrt{\det g}\, h}{(\varrho+l_i^2)^2}f_0+\frac{h\, \sqrt{\det g}}{(\varrho+l_2^2)^2}=0~,
\end{eqnarray}
for $\mu=0$.
Note that $l_1\equiv 0$ and $\sqrt{\det g}=\varrho\, (\varrho+l_2^2)\, (\varrho+l_3^2)$.

We have three equations for three functions $f_{\mu}$, so we expect this system to contain the desired solution for $\beta$.  We take the boundary conditions to be $f_{0,2,3}(\varrho=0)=0$, $f_0(\varrho\rightarrow \infty)=1/3$ and $f_{2,3}(\varrho\rightarrow \infty)=0$. Furthermore, these equations can be seen to arise from minimizing the action
\bea
\nn
&&I=\int_X \beta\wedge\star_8h\,\beta = \int_0^{\pi} \diff \theta_2 \int_0^{\pi} \diff \theta_3 \int_0^{\infty} \diff \varrho\, h \, \sqrt{\det g} \sin{\theta_2} \sin{\theta_3} \Big[(\partial_{\varrho}\,f_0)^2 + \\ \nn 
&& +\frac{(\partial_{\theta_3} f_0)^2}{U\,(l_3^2 + \varrho)} + \frac{(\partial_{\theta_2} f_0)^2}{U\,(l_2^2 + \varrho)} + \frac{(f_0 - \partial_{\theta_3} f_3- \cot{\theta_3} f_3)^2}{(l_3^2 + \varrho)^2} + \frac{(\partial_{\theta_3} f_2)^2 + (\partial_{\theta_2} f_3 )^2}{(l_2^2 + \varrho) (l_3^2 + \varrho)} + \\ \nn
&& + \frac{(-1 + f_0 - \cot{\theta_2} f_2 -\partial_{\theta_2} f_2)^2}{(l_2^2 + \varrho)^2} + \frac{U}{(l_2^2 + \varrho)} (\partial_{\varrho} f_2)^2 
+ \frac{U}{(l_3^2 + \varrho)} (\partial_{\varrho} f_3)^2 + \frac{f_0^2}{\varrho^2} \Big]~.
\eea
With the boundary conditions above, one can check that $I$ is finite. We shall give a more general argument for existence of 
this two-form later in this section.

Going back to the original physical motivation, the warped harmonic form $\beta$ allows one to construct a three-form fluctuation $\delta C=A\wedge \beta$ that satisfies the linearized supergravity equations. The Hodge dual can be written as
\begin{equation}
\delta G_{7}=\star_3 \dd A\wedge h\star_8\beta=\dd p\,\wedge h\star_8\beta=\dd\left(p\, h\star_8\beta\right)~,
\end{equation}
thus obtaining a local form for the six-form fluctuation. At very large $\varrho$, $\beta$ can be approximated to leading order by
\begin{equation}
\beta\sim \frac{2}{3}\, e_{\theta_2}\wedge e_{\phi_2} - \frac{1}{3}\, e_{\theta_1}\wedge e_{\phi_1}- \frac{1}{3}\, e_{\theta_3}\wedge e_{\phi_3}~,
\end{equation}
so that
\begin{equation}
\delta C_{6}\supset p \,\,\frac{\dd\varrho}{\varrho^2}\wedge \dd\psi \wedge e_{\theta_1}\wedge e_{\phi_1}\wedge e_{\theta_2}\wedge e_{\phi_2}~.
\end{equation}
This form of the local potential couples to the baryonic M5 brane through the Wess-Zumino term, thus reinforcing the identification of the scalar $p$ as the phase of the baryonic condensate and Goldstone boson. Furthermore, at large $\varrho$ we can also write
\begin{equation}
\delta G_{7}\supset \dd(\, \varrho^{-1}\, \dd p\wedge \dd\psi\wedge e_{\theta_1}\wedge e_{\phi_1}\wedge e_{\theta_2}\wedge e_{\phi_2})~.
\end{equation}
We note the appearance of the volume form of the sub-manifold wrapped by the baryonic M5 brane. By reducing the corresponding asymptotic $\delta C_{6}$ over the five-cycle one obtains the gauge field in the AdS$_4$ space. From the discussion in \secref{sec:Vec-boun}, and noticing that the AdS$_4$ radial coordinate is related to $\varrho$ as $z \sim \varrho^{-1}$, we obtain the appropriate decay for a conserved current in three dimensions, and see that
\begin{equation}
\langle J^B_{\mu}\rangle \sim \partial_{\mu}p~.
\end{equation}

\section{General warped resolved CY backgrounds}\label{sec:generalwarped}

Much of our discussion of the resolved $Q^{111}$ backgrounds can be extended to general warped resolutions of 
CY cones. In this section we describe what is known about such generalizations. 
In particular we present a novel method for computing M5 brane condensates in such 
backgrounds, or more generally the world-volume actions of branes in warped CY geometries.

\subsection{Gravity backgrounds}

As for the $Q^{111}$ case we are interested in M-theory backgrounds of the form
\bea\label{background2} 
\dd s^2_{11} &=& h^{-2/3} \dd s^2(\R^{1,2}) + h^{1/3} \dd s^2(X)~, \\ \nn 
G &=& \dd^3 x \wedge \dd h^{-1}~,
\eea
where $X$ is a Ricci-flat K\"ahler eight-manifold that is asymptotic to a cone metric over some Sasaki-Einstein seven-manifold $Y_7$.  
Placing $N$ spacetime-filling M2 branes at a point $y\in X$ leads to the warp factor equation
\bea\label{green}
\Delta_x h[y] = \frac{(2\pi \ell_p)^6N}{\sqrt{\det g_X}} \delta^8(x-y)~.
\eea
Here $\Delta h = \diff^* \diff h = - \nabla^i\nabla_i h$ is the scalar Laplacian of $X$ acting on $h$. 
Thus $h[y](x)$ is simply the Green's function on $X$. More generally, one could pick different points $y_i\in X$, with $N_i$ M2 branes 
at $y_i$, such that $\sum_i N_i=N$. Then $h[\{y_i,N_i\}](x)$ will be a sum of Green's functions, weighted by $N_i$. We shall regard this as an obvious generalization. There are thus two steps involved in constructing such a solution: choose a CY metric 
on $X$, and then solve for the Green's function. If the latter is chosen so that it vanishes at infinity then the 
supergravity solution (\ref{background2}) will be asymptotically AdS$_4 \times Y_7$ with $N$ units of $G_7$ flux through $Y_7$.

If $Y_7$ is a Sasaki-Einstein manifold then $\mathcal{C}(Y_7)$ defines an isolated singularity at the tip of the cone $r=0$. 
We may then take a resolution  $\pi:X\rightarrow \mathcal{C}(Y_7)$, which defines our manifold $X$ as a complex manifold. 
The map $\pi$ is a biholomorphism 
of complex manifolds on the complement of the singular point $\{r=0\}$, so that in $X$ the singular point is effectively replaced by a higher-dimensional 
locus, which is called the exceptional set. We require that the holomorphic $(4,0)$-form on $\mathcal{C}(Y_7)$ extends to a smooth holomorphic $(4,0)$-form on 
$X$. Such resolutions are said to be crepant, and they are not always guaranteed to exist, even for toric singularities. 
In the latter case one can typically only partially resolve so that $X$ has at worst orbifold singularities. Having chosen such an $X$ we must then 
find a CY metric on $X$ that approaches the given cone metric asymptotically. Fortunately, mathematicians have very recently proved that 
one can \emph{always} find such a metric. Essentially, this is a non-compact version of Yau's theorem with a ``Dirichlet'' boundary condition, where we have a fixed Sasaki-Einstein metric 
at infinity on $\partial X=Y_7$ and ask to fill it in with a Ricci-flat K\"ahler metric. There are a number of papers that have developed this subject 
in recent years \cite{crep1,crep2,crep3,crep4,crep5, crep6}, but the most recent \cite{goto, vanC} prove the strongest possible result: that in each K\"ahler class in $H^2(X,\R)\cong H^{1,1}(X,\R)$ (see \cite{crep6}) there is a unique 
CY metric that is asymptotic to a fixed given Sasaki-Einstein metric on $Y_7=\partial X$. Note this result \emph{assumes} the existence of the 
Sasaki-Einstein metric -- it does not prove it.

The crepant (partial) resolutions of toric singularities are well understood, being described by toric geometry and hence fans of polyhedral cones. 
The extended K\"ahler cone for such resolutions is known as the GKZ fan, or secondary fan. The fan is a collection of polyhedral cones living in $\R^{b_2(X)}$, glued together along their boundaries, such that each cone corresponds to a particular choice of topology for $X$. Implicit here is the fact that 
$b_2(X)$ is independent of which topology for $X$ we choose.
A point inside the polyhedral cone corresponding to a given $X$ is a K\"ahler class 
on $X$. The boundaries between cones correspond to partial resolutions, where there are further residual singularities, and there is a topology change as 
one crosses a boundary from one cone into another. The GKZ fan for $Y_7=Q^{111}$ was described in Figure \ref{fig:GKZ1}. If we combine this description with the above 
existence results, we see that the GKZ fan is in fact classifying the space of resolved asymptotically conical CY metrics. 

Having chosen a particular resolution and K\"ahler class, hence metric, 
we must then find the warp factor $h$ satisfying (\ref{green}). This amounts to finding the Green's function on $X$, and this always exists 
and is unique using very general results in Riemannian geometry. A general discussion in the Type IIB context may be found in 
 \cite{Martelli:2007mk}, and the comments there apply equally to the M-theory setting. 
 
\begin{figure}[ht]
\center
\includegraphics[scale=0.6]{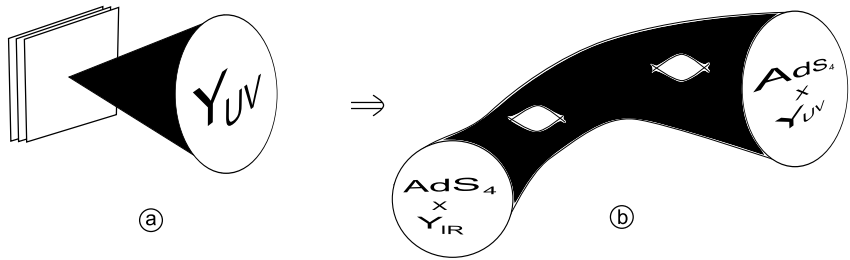}
\caption{{(a) A stack of $N$ M2 branes transverse to the CY cone singularity $\mathcal{C}(Y_{\mathrm{UV}})$; (b) the supergravity geometry
describing an RG flow dual to a diagonal Higgsing.}
\label{fig:UVIR}}
\end{figure}

In the warped metric (\ref{background2}) the point $y\in X$ is effectively sent to infinity, and 
the geometry has two asymptotically AdS$_4$ regions: one near $r=\infty$ that is asymptotically AdS$_4\times Y_7$, and one near 
to the point $y$, which is asymptotically AdS$_4\times Y_{\mathrm{IR}}$. Here the tangent space at $y$ is the cone $\mathcal{C}(Y_{\mathrm{IR}})$. Thus if $y$ is a smooth point, $Y_{\mathrm{IR}}=S^7$.
For further discussion, see \cite{Martelli:2007mk, Martelli:2008cm}. Denoting the original Sasaki-Einstein seven-manifold as $Y_7=Y_{\mathrm{UV}}$, the spacetime is shown in Figure \ref{fig:UVIR}.

The above discussion implies that we expect a supergravity solution to exist for \emph{any} choice
of Higgsing in the field theory. Conversely, since for any partial resolution we have a supergravity solution, there should exist a field theory dual to this given by an appropriate Higgsing pattern. It is important to notice that this discussion applies for \emph{zero} $G$-flux on $Y_{\mathrm{UV}}$ and $Y_{\mathrm{IR}}$. We will discuss the more involved case in which $G$-flux is included in the next section.

If one places the $N$ M2 branes at the same position $y\in X$, the moduli space is naturally a copy of $X$. 
The $b_2(X)$ K\"ahler moduli are naturally complexified by noting that $H^6(X,Y_7,\R)\cong H_2(X,\R)\cong \R^{b_2(X)}$ by Poincar\'e 
duality, and that this group classifies the periods of $C_6$ through six-cycles in $X$, which may either be closed or have a 
boundary five-cycle on $Y_7=\partial X$. More precisely, taking into account large gauge transformations leads to 
the torus $H^2(X,Y_7,\R)/H^2(X,Y_7,\Z)\cong U(1)^{b_2(X)}$. Altogether this moduli space of supergravity solutions 
should be matched to the full moduli space of the dual SCFT. At least for toric 
$X$ one can prove quite generally via an exact sequence that $b_2(X)=b_2(Y_7)+b_6(X)$, where $b_6(X)$ is also the number 
of irreducible exceptional divisors in the resolution. In toric language, this is the number of internal lattice points 
in the toric diagram. We shall discuss such examples in \secref{sec:5}: the presence of calibrated six-cycles 
is expected to lead to M5 brane instanton corrections in these backgrounds.

\subsection{Harmonic two-forms}

Recall we are also interested in fluctuations of the form
\begin{equation}
\delta C_3=A\wedge \beta~,
\end{equation}
where $A$ leads to a massless gauge field in the Minkowski three-space if
\begin{equation}\label{betaagain}
\dd\beta=0\ ,\qquad \dd(\, h\star_8\beta)=0~.
\end{equation}
For trivial warp factor $h\equiv 1$ this just says that $\beta$ is harmonic. It is a general result that 
if we also impose that $\beta$ is $L^2$ normalizable, or equivalently that $A$ has finite kinetic energy 
in three dimensions, then such forms are guaranteed to exist and are in 1-1 correspondence with 
$H^2(X,Y_7,\R)\cong H_6(X,\R)$ \cite{hausel}. Thus there are always $b_6(X)$ $L^2$ normalizable harmonic two-forms $\beta$ 
in the unwarped case. 

However, this case is not the physical case for applications to AdS/CFT. Instead we should look for 
solutions to (\ref{betaagain}) where $h$ is the Green's function on $X$. Again, fortunately 
there are mathematical results that we may appeal to to guarantee existence of such forms. 
These are again described in the Type IIB context in \cite{Martelli:2008cm}. In the warped 
case $\beta$ is harmonic with respect to the metric $h^{1/2} \diff s^2(X)$. This manifold 
has an asymptotically conical end with boundary $S^7$ (or more generally $Y_{\mathrm{IR}}$ if the M2 brane stack 
is placed at a singular point with horizon $Y_{\mathrm{IR}}$), and an isolated conical singularity with 
horizon metric $\tfrac{1}{4}\diff s^2(Y_7)$. The number of $L^2$ harmonic two-forms on such a space 
is in fact known, and is $b_2(X)$. To see this requires combining a number of mathematical results 
that are described in \cite{Martelli:2008cm}. In particular, since $b_2(X)=b_2(Y_7)+b_6(X)$ there is a 
corresponding harmonic form, and hence Goldstone mode, for each of the $b_2(Y_7)$ baryonic $U(1)$ 
symmetries. Indeed, these $b_2(Y_7)$ harmonic forms can be seen to asymptote to the harmonic two-forms
on $Y_7$ at $r=\infty$. Thus the analysis at the end of \secref{sec:wrapped} carries over 
in much more general backgrounds.
We shall analyse the asymptotics of the $b_6(X)$ unwarped $L^2$ harmonic forms 
in more detail in \secref{sec:5}, where they will be given a very different interpretation.

\subsection{Baryonic condensates: M5 branes in general warped geometries}\label{sec:M5general}

As discussed in \secref{sec:baryons}, M5 branes wrapped on five-manifolds $\Sigma_5\subset Y_7$ lead, with 
appropriate choice of quantization of the gauge fields in AdS$_4$, 
to scalar operators 
in the dual SCFT that are charged under the $U(1)^{b_2(Y_7)}$ baryonic symmetry group. We have already described 
how to compute the VEV of such an operator in the (partially) resolved $Q^{111}$ background. More generally 
one should compute the action of a Euclidean M5 brane which is wrapped on a minimal six-submanifold $D\subset X$ with boundary $\partial D=\Sigma_5$. 
Similar computations, in some specific examples, have been performed in \cite{Baumann:2006th}. In this section we explain how 
this Euclidean action may be computed exactly, in general, in the case where $D$ is a divisor. This is essentially a technical computation that may be skipped if the reader is not interested in the details: the final formula is (\ref{thefinalcountdown}).

Let suppose that we are given a warped background (\ref{background2}), where $X$ has an asymptotically conical CY metric and 
the warp factor satisfying (\ref{green}) is given, with a specific choice of point $y\in X$ where the stack of $N$ M2 branes are located. 
We would like to compute 
\bea
\exp\left(-T_5\int_D \sqrt{\det g_D}\, h[y]\, \diff^6x\right)~.
\eea
Here $T_5=2\pi/(2\pi \ell_p)^6$ is the M5 brane tension, and the integrand is the world-volume action (in the absence of a self-dual two-form). Thus $g_{D}$ denotes the pull-back of the unwarped metric to the world-volume $D$. 
We also assume that $D$ is a divisor, in order to preserve supersymmetry. Then $D$ is also minimal and the integral is
\bea
\int_D \sqrt{\det g_D}\, h[y] \, \diff^6 x = \int_D \frac{\omega^{3}}{3!}\, h[y]~,
\eea
where $\omega$ denotes the K\"ahler form of the unwarped metric, pulled back to $D$. 

Before beginning our computation, we note that on a K\"ahler manifold the scalar 
Laplacian $\Delta h = \dd^* \dd h$ can be written as
\bea\label{KahlerLaplace}
\Delta h = -\omega\lrcorner \dd \dd^c h = -2\ii \omega\lrcorner\partial \bar{\partial}h~.
\eea
Here
$\dd^c \equiv I\circ \dd = \ii (\bar{\partial}-\partial)$, 
where $I$ is the complex structure tensor. The contraction 
$\omega\lrcorner \dd \dd^c h$ is then in local complex coordinates
\bea
\omega\lrcorner\dd \dd^c h = 4\omega^{i\bar{j}}\frac{\partial^2h}{\partial z^i\partial\bar{z}^{\bar{j}}} = -\Delta h~ \quad
\text{where} \quad
\omega=\frac{\ii}{2}\omega_{i\bar{j}}\diff z^i\wedge \diff\bar{z}^{\bar{j}}~.
\eea

For simplicity we shall study the case in which $D$ is described globally by the equation 
$D=\{f=0\}$, where $f$ is a global holomorphic function on $X$. This means 
that the homology class of $D$ is trivial, and hence in fact the wrapped 
M5 brane carries zero charge under $U(1)^{b_2(Y_7)}$. The cases studied in 
\cite{Baumann:2006th} are of this form. More generally, since $D$ is a complex divisor it defines 
an associated holomorphic line bundle $\mathcal{L}_D$ over $X$. Then we may take $D$ to be the zero 
set of a holomorphic section of $\mathcal{L}_D$, with a simple zero along $D$. 
To extend the computation below to this case would require combining the argument we give here 
with the arguments in \secref{sec:5}.

Thus, suppose that $f$ is a holomorphic function with a simple zero along $D$, and introduce the
two-form
\bea
\eta_D \equiv \frac{1}{2\pi} \dd \dd^c \log |f| = -\frac{1}{2\pi \ii} \partial\bar{\partial} \log |f|^2=-\frac{1}{2\pi \ii} \partial\bar{\partial}(\log f + \log \bar{f})~.
\eea
This shows that away from the locus $f=0$, which is the divisor $D$, 
in fact $\eta_D=0$. On the other hand, locally along $D$ we can write 
$f = zg$ 
where $z$ is a local coordinate normal to $D$, and $g=g(z,w_1,w_2,w_3)$, where 
$w_1,w_2,w_3$ are local complex coordinates along $D$ and $g$ has no zero in this local chart. 
We may then write $z=r \me^{\ii\theta}$, and note that 
\bea
\frac{1}{2\pi}\dd \dd^c \log r = \delta^2(0) \frac{\ii}{2}\dd z\wedge\dd \bar{z}~.
\eea
This is just the elementary statement that $(1/2\pi)\log r$ is the unit 
Green's function in dimension two (the local transverse space to $D$). 
Thus we have shown that $\eta_D$ is zero away from $D$, and is  a unit 
delta function supported along $D$. 

Using these properties of $\eta_D$ we may hence write
\bea
V\equiv T_5\int_D \frac{\omega^3}{3!}\, h[y] = \int_X \frac{T_5 h[y]}{2\pi} \frac{\omega^3}{3!} \wedge \dd \dd^c \log |f|~.
\eea
Note in particular that $T_5 h[y]/2\pi$ is $N$ times the unit Green's function (with unit delta function source), {\it i.e.}
\bea
\Delta_x \left(\frac{T_5 h[y]}{2\pi}\right) = \frac{N}{\sqrt{\det g_X}}\delta^8(x-y)~.
\eea
We then integrate by parts
\bea\label{eq1}
V = \int_{\partial X=Y_7} \frac{T_5 h[y]}{2\pi} \frac{\omega^3}{3!} 
\wedge \dd^c \log |f| - \int_X \dd\left(\frac{T_5 h[y]}{2\pi}\right)\wedge\frac{\omega^3}{3!} \wedge \dd^c \log |f|~.
\eea
We will deal with the boundary terms later, focusing for now on the integrals over $X$. First note that
\bea
\gamma\wedge \frac{\omega^3}{3!} = - \star I(\gamma)~.
\eea
holds for any one-form $\gamma$. Using this we may write
\bea
V &=& -\int_X \dd\left(\frac{T_5 h[y]}{2\pi}\right)\wedge\frac{\omega^3}{3!} \wedge \dd^c \log |f| + \mbox{boundary term}\nn\\
  &=& \int_X \diff^c\left(\frac{T_5 h[y]}{2\pi}\right)\wedge\frac{\omega^3}{3!} \wedge \dd \log |f| + \mbox{boundary term}~.
\eea
We then again integrate by parts
\bea\label{eq2}
V = \int_X \dd \dd^c \left(\frac{T_5 h[y]}{2\pi}\right)\wedge\frac{\omega^3}{3!}\log |f| + \mbox{boundary terms}~,
\eea
where explicitly now
\bea
\mbox{boundary terms} = \int_{Y_{7}} \left[\frac{T_5 h[y]}{2\pi}\dd^c \log |f|  - \log |f| \dd^c \left(\frac{T_5 h[y]}{2\pi}\right)\right]\wedge \frac{\omega^3}{3!}~.
\eea
In fact this boundary integral is divergent -- a key physical point in interpreting it holographically. For now 
let us deal with the integral over $X$ in (\ref{eq2}). Using (\ref{KahlerLaplace}) we may write
\bea
-\dd\dd^c \left(\frac{T_5 h[y]}{2\pi}\right)\wedge \frac{\omega^3}{3!} = \Delta \left(\frac{T_5 h[y]}{2\pi}\right) \frac{\omega^4}{4!} = 
\frac{N}{\sqrt{\det g_X}}\delta^8(x-y) \frac{\omega^4}{4!}~,
\eea
and thus
\bea
V = -\int_X N \delta^8(x-y)\, \log |f|\, \diff^8 x = - N \log |f(y)|+\mbox{boundary terms}~.
\eea

Let us turn now to the boundary terms. In order to render this finite, we cut off the integral at some large $r=r_c$, and 
write the boundary integral as
\bea\label{bterms}
\mbox{boundary terms} = \int_{Y_{r_c}} \left[\frac{T_5 h[y]}{2\pi}\dd^c \log |f|  - \log |f| \dd^c \left(\frac{T_5 h[y]}{2\pi}\right)\right]\wedge \frac{\omega^3}{3!}~.
\eea
We require that $D$ is asymptotically conical, so that at large $r_c$ it approaches a cone over a compact five-manifold $\Sigma_5\subset Y_7$. 
In the cone geometry, a conical divisor with trivial homology class is specified as the zero set of a homogeneous function under $r\partial_r$. Thus we take
$
|f| = Ar^{\lambda}\left(1+\ldots\right)
$, where $A$ is homogeneous degree zero ({\it i.e.} a function on $Y_7$) and the $\ldots$ are terms that go to zero as $r\rightarrow\infty$. Thus $f$ has asymptotic homogeneous degree $\lambda>0$. Now, the volume form on $Y_7$ is 
\bea
\diff\vol(Y_7)=\eta\wedge \frac{(\diff\eta)^3}{2^3\cdot 3!}~,
\eea
where $\eta=\dd^c \log r$ and
$\omega_{\mathrm{cone}} = \frac{1}{2}\diff (r^2 \eta)$ is the K\"ahler form on the cone over $Y_7$. Asymptotically,
$\omega = \omega_{\mathrm{cone}}(1 + O(r^{-2}))$.
The $O(r^{-2})$ follows since the leading correction to the cone metric is a harmonic two-form on $Y_7$, which is down by a factor 
of $r^{-2}$ relative to the cone metric. We also have
\bea
\frac{T_5 h[y]}{2\pi} = \frac{N}{6\vol(Y_7) r^6}(1 + \ldots)~.
\eea
{\it Cf.} (\ref{warping}). Thus the first term in (\ref{bterms}) is \emph{convergent} and gives
\bea
\lim_{r_c\rightarrow \infty} \int_{Y_{r_c}} \frac{T_5 h[y]}{2\pi}\dd^c \log |f| \wedge \frac{\omega^3}{3!} = \frac{N}{6\vol(Y_7)}\cdot \lambda \cdot \vol(Y_7) = \frac{N\lambda}{6}~.
\eea
Note here that the function $A$ does not contribute to the integral as it is independent of $r$, and thus $J(r\partial_r)\lrcorner \dd^c \log A=0$. On the other hand, the second term in (\ref{bterms}) is divergent, the leading divergent piece being
$N\lambda \log r_c~.$
Provided the $\ldots$ terms in $|f|$ and $h$ fall off as $o(r^{-\epsilon})$, for some $\epsilon>0$, then in fact this is the only divergence (since 
any positive power of $r$ grows faster than $\log r$). There is also a finite part, namely $N\int_{Y_7} \log A/\vol(Y_7)$. However, the important point is that this 
depends only on asymptotic data. 

Let us interpret this divergence. Suppose that the Sasaki-Einstein manifold $Y_7$ is quasi-regular, meaning that it is a $U(1)$ 
bundle over a K\"ahler-Einstein orbifold $Z$.\footnote{In fact the irregular case can be approximated arbitrarily closely by the quasi-regular case.} 
Then asymptotically $f$ is, in its dependence on $Z$, a holomorphic section of $L^k$ for some integer $k\in\Z_{>0}$, where $L=K_Z^{-1/I}$ with $I=I(Z)\in\Z_{>0}$ being the orbifold Fano index of $Z$. Here $K_Z$ denotes the orbifold canonical bundle of $Z$, and $I$ is by definition 
the largest integer so that the root $L$ is defined. It follows that $\lambda = 4k/I$,
where $4=\dim_\C X$. The five-manifold $\Sigma_5\subset Y_7$ is then the total space of a $U(1)$ bundle over an orbifold surface $S\subset Z$, with 
the Poincar\'e dual to $S$ being represented by $c_1(L^k)=kc_1(Z)/I$. The K\"ahler-Einstein condition on $Z$ gives
\bea\label{KE}
8[\omega_Z] = 2\pi c_1(Z)\in H^{1,1}(Z,\R)~,
\eea
where $\omega_Z=(1/2)\diff\eta$ denotes the K\"ahler form of $Z$. 

Now, the conformal dimension of the operator dual to an M5 brane wrapped on $\Sigma_5$ is given by the general formula
$
\Delta(\mathcal{O}[\Sigma_5]) = \pi N \vol(\Sigma_5)/6\vol(Y_7)
$.
In the quasi-regular case at hand, the length of the $U(1)$ circle cancels in the numerator and denominator and we can write this as 
\bea
\Delta(\mathcal{O}[\Sigma_5]) = \frac{\pi N \int_S \frac{\omega_Z^2}{2!}}{6 \int_Z \frac{\omega_Z^3}{3!}} = \frac{\pi N \frac{k}{I}\int_Z \frac{\omega_Z^2}{2!}\wedge 
c_1(V)}{2\int_Z \frac{\omega_Z^3}{2!}} = \frac{2Nk}{I}~.
\eea
Here in the last step we used the K\"ahler-Einstein condition (\ref{KE}). Thus we have the general result that the divergent part of the 
integral is
\bea
N\lambda \log r_c = 2\Delta(\mathcal{O}[\Sigma_5]) \log r_c = -\Delta(\mathcal{O}[\Sigma_5]) \log z~,
\eea
where we have changed to the usual AdS$_4$ coordinate $r_c^2 = 1/z$. Thus we see that the divergent part of the 
integral is precisely such that we can interpret its coefficient as the VEV of the operator $\mathcal{O}[\Sigma_5]$ in this background. 
This coefficient is, from the above computations, 
\bea\label{thefinalcountdown}
\exp(-V_{\mathrm{reg}}) = |f(y)|^N\exp\left(-\frac{N\lambda}{6}-\frac{N}{\vol(Y_7)}\int_{Y_7} \log (|f|/r^\lambda)\right)~.
\eea
This is an exact result for the VEV, or regularized exponential of the M5 brane action, in terms of the defining function $f$ of the divisor $D$. The integral is understood in the limit
$Y_7=Y_{r_c}$ as $r=r_c\rightarrow \infty$, which is convergent as the integrand is independent of $r$ in this limit. Notice, in particular, that 
if one multiplies $f$ by a constant, this constant drops out of the formula, as it should.

\section{Torsion $G$-flux in warped resolved Calabi-Yau backgrounds}\label{sec:Gflux}

In this section we discuss the effect of adding \emph{torsion $G$-flux} to the AdS background AdS$_4\times Y_7$, where $Y_7$ is a Sasaki-Einstein seven-manifold. We show that this has important implications on the existence of warped resolved CY backgrounds.

\subsection{$G$-flux and the supergravity dual of Higgsing}\label{sec:generalGflux}
As was discussed in \secref{sec:Cfield}, for seven-dimensional Sasaki-Einstein manifolds with non-trivial $H^{4}_{\mathrm{tor}}(Y_7,\Z)$ a flat torsion $G$-flux can be turned on without affecting the supergravity equations of motion, or the supersymmetry of the background. This was first discussed in the context of QCS theories for the ABJM theory in \cite{Aharony:2008gk}, although here the torsion in $H^4(S^7/\Z_k,\Z)\cong\Z_k$ is due to the $\Z_k$ quotient, giving $\pi_1(S^7/\Z_k)\cong\Z_k$. More generally there are examples in which the torsion $G$-flux is \emph{not} associated to the CS level quotient by $\Z_k$ -- for example, the $Y^{p,q}$ geometries discussed in detail in \cite{Martelli:2008rt}.

Consider adding $G$-flux to this background. As we discussed, to preserve the AdS$_4$ symmetry $SO(3,2)$ and supersymmetry, $G$ has to be flat, and thus the different choices of $G$-flux in the AdS background are classified by the torsion cohomology class $[G]\in H^{4}_{\mathrm{tor}}(Y_7,\Z)$. 

We wish to discuss the implications of turning on such torsion $G$-flux on the Higgsing of the field theory. As was explained in the previous section, the dual supergravity solution to the RG flow induced by the Higgsing has two boundaries: AdS$_4\times Y_{\mathrm{UV}}$ and AdS$_4\times Y_{\mathrm{IR}}$. For \emph{zero} $G$-flux on $Y_7=Y_{\mathrm{UV}}$ we expect a supergravity solution to exist for \emph{any} choice of Higgsing in the field theory. However, when we turn on torsion $G$-flux on $Y_{\mathrm{UV}}$ the existence of such solution is not guaranteed. For such solutions to exist we must be able to extend $G$ over the (partial) resolution $\hat X$ of the original cone $X$ in a way that satisfies the appropriate supersymmetric equations of motion.

The key point here is that when $G=0$ on $Y_{UV}$, we may obviously extend this as $G=0$ on the partial resolution,
while for non-trivial torsion $G$ the process of completing the supergravity solution is much more involved.
There are two steps: first, it must be possible to extend the cohomology class
$[G]\in H^4(Y_{\mathrm{UV}},\Z)$ to a cohomology class\footnote{We assume here that the membrane anomaly
on $\mathcal{X}$ is zero. This will be true in the example that we shall study. The membrane anomaly on
$Y_7$ is automatically zero, as it is zero on any oriented spin seven-manifold.} in $H^4(\mathcal{X},\Z)$, where we have defined
$\mathcal{X}=\hat{X}\setminus\{y\}$, and $y$ is the isolated singular point -- {\it a priori} there might be topological obstructions to this; second, if this is possible, we must choose a flux in this cohomology class to satisfy the supersymmetry
conditions
\cite{Becker:1996gj}, which require
that $G$ must be primitive, so $G\wedge \omega=0$ where $\omega$ is the K\"ahler form, and have Hodge type $(2,2)$ with respect to the complex structure (which implies it is self-dual). 

This leads to two issues: (i) if the choice of $G$-flux on
$Y_{\mathrm{UV}}$ cannot be so extended then the supergravity solution \emph{does not exist}, and therefore
the SCFT dual to $Y_{\mathrm{UV}}$ with this $G$-flux cannot be Higgsed to the partial resolution
corresponding to $\hat{X}$, (ii) the choice of $G$-flux may not be unique, meaning that
the SCFT should be Higgsable to the partial resolution but with
potentially more than one choice of torsion $G$-flux in $H^4(Y_{\mathrm{IR}},\Z)$.
Indeed, notice that choosing an extension of $[G]$ over $\mathcal{X}$ immediately leads by restriction
to a choice of $G$-flux in $H^{4}(Y_{\mathrm{IR}},\Z)$, and thus a torsion $G$-flux in the IR theory dual to
AdS$_4\times Y_{\mathrm{IR}}$.

To conclude, one expects M2 brane QCS theories dual to torsion $G$-flux backgrounds to display different behaviour to those without $G$-flux --
namely, one should see obstructions to Higgsings to certain partial resolutions in theories with $G$-flux. We shall investigate
this in detail in the remainder of this section for a particular example, and show that this behaviour is indeed realized.

\subsection{$Y^{1,2}(\mathbb{CP}^2)$: Gravity results}\label{sec:gravity}
We now investigate the above discussion in detail in a particular example:
the toric CY cone
$X=\mathcal{C}(Y^{1,2}(\mathbb{CP}^2))$ \cite{Martelli:2008rt}, the toric diagram of which is presented in Figure~\ref{f:phase1}. 
Here the Sasaki-Einstein metric on
$Y_{\mathrm{UV}}=Y^{1,2}(\mathbb{CP}^2)$ is known explicitly,
and was constructed in \cite{Gauntlett:2004hh}.
The complex structure of the cone singularity $X$ may be described
as the affine holomorphic quotient of $\C^5$
by $\C^*$ with charges $(1,2,-1,-1,-1)$.
This is of course the same complex structure induced by the
K\"ahler quotient, at moment map level zero, of $\C^5$ by $U(1)$ with the same charges.
There are precisely two (partial) CY resolutions of this singularity,
given by taking the moment map level $\zeta<0$ or $\zeta>0$.

To describe these partial resolutions, let $z_1,\ldots,z_5$ denote coordinates on $\C^5$.
The moment map/GLSM D-term equation is then
\bea\label{momentum}
|z_1|^2+2|z_2|^2 - |z_3|^2 -|z_4|^2-|z_5|^2 = \zeta~.
\eea
For $\zeta<0$ this describes the smooth CY four-fold
$\hat{X}_-\equiv$ total space of $\mathcal{O}(-1)\oplus\mathcal{O}(-2)\rightarrow
\mathbb{CP}^2$. The zero-section, which is a copy
of $\mathbb{CP}^2$, is at $\{z_1=z_2=0\}$, while the boundary $
\partial \hat{X}_-=Y_{\mathrm{UV}}=Y^{1,2}(\mathbb{CP}^2)$. In fact, note that an \emph{explicit}
Ricci-flat K\"ahler metric on this manifold was constructed in
\cite{metrics}.
Since $\hat{X}_-$ is contractible to $\mathbb{CP}^2$, it follows
that $H^0(\hat{X}_-,\Z)\cong H^2(\hat{X}_-,\Z)\cong H^4(\hat{X}_-,\Z)\cong\Z$, with
all other cohomology vanishing. Moreover, since
$\hat{X}_-$ is the total space of a rank four real vector bundle over
$\mathbb{CP}^2$, the Thom isomorphism implies that
$H^4(\hat{X}_-,Y_{\mathrm{UV}},\Z)\cong H^6(\hat{X}_-,Y_{\mathrm{UV}},\Z)\cong H^8(\hat{X}_-,Y_{\mathrm{UV}},\Z)\cong\Z$, where
the generator of $H^4(\hat{X}_-,Y_{\mathrm{UV}},\Z)$ is the Thom class.
It follows that the image of the generator
in the map
$H^4(\hat{X}_-,Y_{\mathrm{UV}},\Z)\rightarrow H^4(\hat{X}_-,\Z)\cong H^4(\mathbb{CP}^2,\Z)$ is the Euler class of the bundle $\mathcal{O}(-1)\oplus\mathcal{O}(-2)$.
Denoting $H$ the hyperplane class that generates $H^2(\mathbb{CP}^2,\Z)\cong\Z$, we have
\bea
e(\mathcal{O}(-1)\oplus\mathcal{O}(-2))&=&c_2(\mathcal{O}(-1)\oplus\mathcal{O}(-2))=c_1(\mathcal{O}(-1))\cup c_1(\mathcal{O}(-2))\nonumber\\
&=& (-H) \cup (-2H) = 2\in H^4(\mathbb{CP}^2,\Z)\cong\Z~.
\eea
Recall here that $H\cup H$ generates $H^4(\mathbb{CP}^2,\Z)\cong \Z$.
Thus the long exact sequence
\bea
H^4(\hat{X}_-,Y_{\mathrm{UV}},\Z)\stackrel{f}{\rightarrow} H^4(\hat{X}_-,\Z)\rightarrow H^4(Y_{\mathrm{UV}},\Z)\rightarrow
H^5(\hat{X}_-,Y_{\mathrm{UV}},\Z)\cong 0
\eea
implies, since the first ``forgetful'' map $f$ is multiplication
by the Euler number $e=2$, that $H^4(Y_{\mathrm{UV}},\Z)\cong \Z_2$.
Thus we may turn on precisely one non-trivial
torsion $G$-flux on $Y_{\mathrm{UV}}=Y^{1,2}(\mathbb{CP}^2)$.
It is similarly straightforward to show that the only
other non-trivial cohomology groups of $Y_{\mathrm{UV}}$ are $H^2(Y_{\mathrm{UV}},\Z)\cong H^5(Y_{\mathrm{UV}},\Z)\cong\Z$. Notice this agrees with
\cite{Martelli:2008rt}, where the
cohomology groups of $Y_{\mathrm{UV}}$ were computed
via a completely different method.

Now consider the other partial resolution, with $\zeta>0$ in (\ref{momentum}).
This may be described as $\hat{X}_+\equiv$ total space of $\mathcal{O}(-1)^3\rightarrow
\mathbb{WCP}^1_{[1,2]}$, where the zero-section weighted
projective space $\mathbb{WCP}^1_{[1,2]}$ is now at
$\{z_3=z_4=z_5=0\}$. $\hat{X}_+$ has a single, isolated singular
point at $y=\{z_1=z_3=z_4=z_5=0\}$, which has tangent cone
$\C^4/\Z_2$ where the $\Z_2$ generator acts with equal charge on
each coordinate of $\C^4$; thus this is the ABJM $k=2$
quotient. If we remove $y$ from $\hat{X}_+$, we obtain
a smooth eight-manifold $\mathcal{X}_+$ with boundaries $Y_{\mathrm{UV}}=Y^{1,2}(\mathbb{CP}^2)$
and $Y_{\mathrm{IR}}=S^7/\Z_2$, as in Figure~\ref{fig:UVIR}.

From our general discussion in \secref{sec:generalGflux}, one
thus expects to be able to
Higgs the field theory dual to
AdS$_4\times Y_{\mathrm{UV}}$ with zero $G$-flux to
field theories dual to AdS$_4\times S^7$ and AdS$_4\times S^7/\Z_2$ with zero $G$-flux.
Here the latter corresponds to putting the $N$ M2 branes
at the singular point $y$ of $\hat{X}_+$, while putting the
M2 branes anywhere else on $\hat{X}_+$, or at any point on
$\hat{X}_-$, should have near horizon limit given by the AdS$_4\times S^7$ theory.
To investigate what happens \emph{with}
$G$-flux, we must extend the non-zero $G$ over either $\mathcal{X}_\pm$, satisfying the
appropriate supersymmetry equations for the flux. Analysing this is
in fact quite technical, although for this relatively simple example
we will be able to provide a complete answer to the problem.
\subsection*{The geometry $\mathcal{X}_+$}
To see whether or not we can extend the non-zero torsion flux in $H^4(Y_{\mathrm{UV}},\Z)\cong\Z_2$, we need to know something about the cohomology of the smooth eight-manifold
$\mathcal{X}_+\equiv \hat{X}_+\setminus \{y\}$.
This has boundary $\partial \mathcal{X}_+=Y_{\mathrm{UV}}\amalg Y_{\mathrm{IR}}$
with two connected components
$Y_{\mathrm{UV}}=Y^{1,2}(\mathbb{CP}^2)$, and $Y_{\mathrm{IR}}=S^7/\Z_2$.
To extend the non-trivial element of $H^4(Y_{\mathrm{UV}},\Z)\cong\Z_2$
over $\mathcal{X}_+$ we need to examine the exact sequence
\bea
H^4(\mathcal{X}_+,\Z)\rightarrow H^4(\partial \mathcal{X}_+,\Z)\rightarrow
H^5(\mathcal{X}_+,\partial \mathcal{X}_+,\Z)~.
\eea
This says we may extend an element of
$H^4(\partial \mathcal{X}_+,\Z)$ over $\mathcal{X}_+$ if and only if
it maps to zero in $H^5(\mathcal{X}_+,\partial \mathcal{X}_+,\Z)$.
Thus we need to compute the latter group, and
also the map. By Poincar\'e-Lefschetz duality, notice that
$H^5(\mathcal{X}_+,\partial \mathcal{X}_+,\Z)\cong H_3(\mathcal{X}_+,\Z)$.

We compute by covering $\mathcal{X}_+\subset \hat{X}_+$ with two
open sets, and then using the resulting Mayer-Vietoris
sequence. We first define
$V_1=\{z_1\neq 0\}\cong\C^*\times\C^4\subset\C^5$, with $\C^5$ having coordinates $(z_1,\ldots,z_5)$.
The invariants under the $\C^*$ action on $\C^5$ with charges
$(1,2,-1,-1,-1)$ are
spanned by $x_1=z_2/z_1^2$,
$w_1=z_3z_1$, $w_2=z_4z_1$ and $w_3=z_5z_1$.
Thus $V_1/\C^*\equiv U_1\cong \C^4$, with coordinate functions
$x_1,w_1,w_2,w_3$. We similarly
define $V_2=\{z_2\neq 0\}\subset \C^5$.
The invariants are now
spanned by the 10 functions $x_2=z_1^2/z_2$,
$y_1=z_3z_1$, $y_2=z_4z_1$, $y_3=z_5z_1$,
$y_4=z_3^2z_2$, $y_5=z_4^2z_2$, $y_6=z_5^2z_2$,
$y_7=z_3z_4z_2$, $y_8=z_4z_5z_2$, $y_9=z_3z_5z_2$.
These satisfy the 6 relations
\bea \label{patch}
&& y_1y_2=y_7x_2~, \qquad y_2y_3=y_8x_2~, \qquad y_1y_3=y_9x_2~,\nonumber\\
&& y_1y_7=y_2y_4~, \qquad y_2y_8=y_3y_5~, \qquad y_3y_9=y_1y_6~.
\eea
This precisely defines the affine variety $\C^4/\Z_2$,
and thus $V_2/\C^*\equiv U_2\cong \C^4/\Z_2$.
Indeed, if $u_1,u_2,u_3,u_4$ denote standard
coordinates on $\C^4$, with the $\Z_2$ action multiplication by
$-1$ on all coordinates, then the invariants
are $u_1^2$, $u_2^2$, $u_3^2$, $u_4^2$, $u_1u_2$, $u_1u_3$, $u_1u_4$,
$u_2u_3$, $u_2u_4$, $u_3u_4$. We may identify
$x_2=u_1^2$, $y_1=u_1u_2$, $y_2=u_1u_3$, $y_3=u_1u_4$,
$y_4=u_2^2$, $y_5=u_3^2$, $y_6=u_4^2$,
$y_7=u_2u_3$, $y_8=u_3u_4$, $y_9=u_2u_4$.

The two coordinate patches $U_1\cong\C^4$, $U_2\cong\C^4/\Z_2$ in fact now cover
$\hat{X}_+$, since one cannot have both $z_1=0$ \emph{and}
$z_2=0$ -- such points violate the moment map equation
(\ref{momentum}) for $\zeta>0$.
Hence $\mathcal{X}_+=\hat{X}_+\setminus\{y\}$ is covered
by $A_1\equiv U_1\cong\C^4$ and $A_2\equiv U_2\setminus\{y\}\cong \R\times S^7/\Z_2$.
The coordinate patch $U_1$ overlaps
$U_2$ where $z_2\neq 0$. In $U_1$, this is the subset
$\{x_1\neq 0\}$. Thus $U_1\cap U_2\cong A_1\cap A_2\cong
\C^*\times\C^3\cong S^1\times \R^7$,
where the first $\C^*$ coordinate is $x_1$.

Consider now the Mayer-Vietoris sequence:
\bea\label{MVA}
\nn
0\cong H_3(A_1\cap A_2,\Z)\rightarrow
H_3(A_1,\Z)\oplus H_3(A_2,\Z)\rightarrow \\
\rightarrow H_3(\mathcal{X}_+,\Z)\rightarrow H_2(A_1\cap A_2,\Z)\cong0~.
\eea
Since $H_3(A_2,\Z)\cong H_3(S^7/\Z_2,\Z)\cong\Z_2$,
it thus follows that $H_3(\mathcal{X}_+,\Z)\cong\Z_2$, which
is the homology group of interest.
Moreover, $U_2\cong\C^4/\Z_2$ is the tangent cone to
the singular point $y$, whose link is thus
$Y_{\mathrm{IR}}=S^7/\Z_2$. The generator of
$H_3(S^7/\Z_2,\Z)\cong\Z_2$ thus trivially
maps to the generator of $H_3(A_2,\Z)$, whose image via inclusion we have
shown generates $H_3(\mathcal{X}_+,\Z)\cong\Z_2$.
The Poincar\'e-Lefschetz dual of this is thus
that we have shown that the map
\bea\label{seq}
\Z_2\oplus\Z_2\cong H^4(S^7/\Z_2,\Z)\oplus H^4(Y_{\mathrm{UV}},\Z)
\rightarrow H_3(\mathcal{X}_+,\Z)\cong \Z_2
\eea
takes $(1,0)\in \Z_2\oplus\Z_2$ to $1\in\Z_2$.
To determine the map completely, we need to also
know the image of $(0,1)$. This is the image
of $H_3(Y_{\mathrm{UV}},\Z)$ in $H_3(\mathcal{X}_+,\Z)$ under inclusion. We may compute this
with a slight modification of the above argument.

Let $X_0=\R_+\times Y_{\mathrm{UV}}$. This is $\hat{X}_+$ minus the
$\mathbb{WCP}^1_{[1,2]}$ zero-section, which recall
is $\{z_3=z_4=z_5=0\}$. We would like to remove
these points from $\hat{X}_+$ to obtain $X_0$.
In terms of the coordinate
patches, this gives
$B_1\equiv U_1\setminus\{w_1=w_2=w_3=0\}\cong
\C\times \R\times S^5\cong \R^3\times S^5$
and $B_2\equiv U_2\setminus \{y_i=0, i=1,\ldots,9\}\cong
(\C\times \R\times S^5)/\Z_2$, where the $\Z_2$
acts as $-1$ on $\C$, and is the $\Z_2\subset U(1)$ in the Hopf $U(1)$ action on $S^5$
(and thus acts freely of course).
Now $B_1\cap B_2$ is still $\{x_1\neq 0\}\subset B_1$,
which gives $B_1\cap B_2\cong \C^*\times \R\times
S^5\cong S^1\times S^5\times \R^2$.
The Mayer-Vietoris sequence for
$X_0=B_1\cup B_2$ is hence
\bea
\nn
0\cong H_3(B_1\cap B_2,\Z)\rightarrow
H_3(B_1,\Z)\oplus H_3(B_2,\Z)\rightarrow \\
\rightarrow H_3(X_0,\Z)\rightarrow H_2(B_1\cap B_2,\Z)\cong0~.\nonumber
\eea
Again, $H_3(B_2,\Z)\cong\Z_2$. If we coordinatize
$B_2$ with the coordinates $u_1,u_2,u_3,u_4$,
the generator may be taken to be
$u_1=u_2=0$, $|u_3|^2+|u_4|^2=1$,
which is a copy of $S^3/\Z_2\subset B_2$.
The above sequence thus proves
that $H_3(X_0,\Z)\cong H_3(Y_{\mathrm{UV}},\Z)\cong \Z_2$, which
of course we already knew. However,
the key point is that this shows that the
generator of $H_3(Y_{\mathrm{UV}},\Z)$ is represented by the above copy of $S^3/\Z_2$. But this is also contained in $A_2\supset B_2$,
and similarly generates $H_3(A_2,\Z)\cong\Z_2$, and the
Mayer-Vietoris sequence (\ref{MVA}) thus proves that the generator
of $H_3(Y_{\mathrm{UV}},\Z)\cong\Z_2$ maps to the generator of $H_3(\mathcal{X}_+,\Z)\cong\Z_2$ under inclusion. Hence $(0,1)\in\Z_2\oplus\Z_2$ in (\ref{seq}) also maps to $1\in\Z_2$, and thus
the map (\ref{seq}) is simply addition of the two
factors.

All this rather abstract algebraic topology thus shows that
zero $G$-flux on the UV boundary $Y_{\mathrm{UV}}\cong Y^{1,2}(\mathbb{CP}^2)$ lifts to a (necessarily torsion) $G$-flux on the RG flow manifold $\mathcal{X}_+$ only if there is \emph{zero} $G$-flux on the IR boundary $Y_{\mathrm{IR}}$.
On the other hand, non-trivial torsion $G$-flux on the UV
boundary, where $H^4(Y_{\mathrm{UV}},\Z)\cong\Z_2$,
lifts to a $G$-flux on the RG flow manifold only if there is
\emph{non-trivial} $G$-flux on the IR boundary $Y_{\mathrm{IR}}$,
where $H^4(Y_{\mathrm{IR}},\Z)\cong\Z_2$. In the field
theory, it follows that the dual to
$Y_{\mathrm{UV}}$ with/without torsion $G$-flux can be Higgsed to
the field theory dual to $S^7/\Z_2$
with/without torsion $G$-flux, \emph{respectively}. 
\subsection*{The geometry $\mathcal{X}_-$}
Finally, we consider the resolution $\hat{X}_-\cong \mathcal{O}(-1)\oplus\mathcal{O}(-2)\rightarrow\mathbb{CP}^2$. In this case zero $G$-flux
on $Y_{\mathrm{UV}}\cong \partial \hat{X}_-$ clearly extends
as zero $G$-flux over $\hat{X}_-$, but for non-zero flux we must necessarily
turn on a \emph{non-torsion} $G$-flux in $H^4(\hat{X}_-,\Z)\cong\Z$. More precisely, we should pick 
a point $y\in \hat{X}_-$ and extend $G$ in $H^4(\mathcal{X}_-,\Z)$ where $\mathcal{X}_-=\hat{X}_-\setminus\{y\}$, although the difference
between $\hat{X}_-$ and $\mathcal{X}_-$ will not affect our discussion of flux, since removing $y$ does not affect the cohomology of interest.
The flux in turn must be primitive and type $(2,2)$ in order
to satisfy the supersymmetry equations (and hence equations of motion).

To see the existence of such a flux, we may appeal to the
results of \cite{hausel}. The latter reference proves that
for a complete asymptotically
conical manifold $(X,g_{X})$ of real dimension $m$, we have
\bea
\label{tamas}
\mathcal{H}^k_{L^2}(X,g_{X}) \cong \left\{
\begin{array}{ll} H^k(X,\partial X,\R), & k<m/2 \\
f(H^{m/2}( X,\partial X,\R))\subset H^{m/2}( X,\R), & k=m/2 \\
H^k( X,\R), & k>m/2\end{array}\right.~,
\eea
where $f$ denotes the ``forgetful'' map, forgetting that a class is relative. Thus the space of $L^2$ harmonic forms $\mathcal{H}^k_{L^2}(X,g_{X})$ is topological.
In particular, we showed earlier that $H^4(\hat{X}_-,\partial \hat{X}_-,\R)\cong H^4(\hat{X}_-,\R)\cong \R$
under the forgetful map (which is multiplication by 2),
and we thus learn that there is a unique
$L^2$ harmonic four-form $G$ on $\hat{X}_-$, up to scale.
If we normalize $(2\pi \ell_p)^{-3}\int_{\mathbb{CP}^2} G = 2M+1$ to be
odd, then this maps under reduction modulo 2 to the generator of $H^4(Y_{\mathrm{UV}},\Z)\cong\Z_2$, for any $M\in\Z$.
Next note that $\omega\wedge G=0$, where $\omega$ is the K\"ahler form on $\hat{X}_-$,  follows since if $\omega\wedge G$ were not zero
it would be an $L^2$ normalizable harmonic six-form on $\hat{X}_-$, and
there are not any of these by (\ref{tamas}).
Thus the $L^2$ harmonic four-form on $\hat{X}_-$ is necessarily primitive. Next, all of the cohomology
on a toric CY four-fold is of Hodge type $(2,2)$.
Each Hodge type is separately harmonic, and thus
again we see that $G$ has to be purely type $(2,2)$
(any other type would be topologically trivial and harmonic,
and thus zero by (\ref{tamas})). Recall that the explicit asymptotically conical Ricci-flat
K\"ahler metric on $\hat{X}_-$ is known \cite{metrics}, and so in principle
one should be able to construct this harmonic four-form $G$ explicitly.
\subsection{$Y^{1,2}(\mathbb{CP}^2)$: Field theory results} \label{ftresults}
The expectations from the supergravity analysis derived above have been shown to match the field theory analysis in the original paper that this manuscript is based on \cite{Benishti:2009ky}. Such Higgsings were later studied with the aid of Hanany-Witten brane constructions, and generalized to other geometries, in \cite{Cremonesi:2010ae}.

\section{Exotic renormalization group flows} \label{sec:exo}

In the previous section we studied the warped resolved backgrounds dual to RG flows induced by Higgsing of the $\mathcal{C}(Y^{1,2}(\mathbb{CP}^2))$ theory. The examples that were discussed in the end of that section were field theories for which the RG flows can be described by a Lagrangian. However, in general, the reduction to Type IIA over the M-theory circle, that is encoded by the field theory CS levels, suggests that a Lagrangian description of the RG flow cannot always be obtained. In this section we want to show how such exotic RG flows can occur. 

\subsection{M-theory circle and Higgsing}\label{sec:extra}

\begin{figure}[ht]
\centering
\includegraphics[scale=0.8]{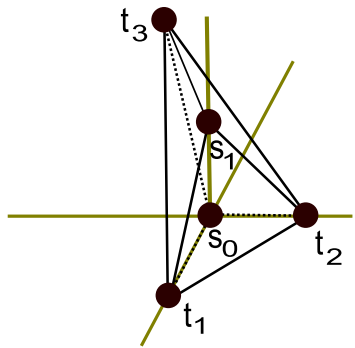}
\caption{Toric diagram for the $\mathcal{C}(Y^{1,2}(\mathbb{CP}^2))$ geometry.}
\label{f:phase1}
\end{figure}

In this subsection we want to pay closer attention to the importance of the choice of the M-theory circle in gravity solutions dual to RG flows. We consider the cone over $Y^{1,2}(\mathbb{CP}^2)$, the toric diagram of which is presented in Figure~\ref{f:phase1} and plotted in such a way that the z-axis corresponds to the M-theory circle. This M-theory circle, as we explain later on, is different from the one that is picked by the field theory mentioned in \secref{ftresults}. Theories that differ by the choice of M-theory circle should be connected by duality. It will be interesting to see that due to the different reduction to Type IIA the description of the dual theories changes drastically. As will become clear later, the generalization of our discussion to other geometries is straightforward.

To start, note that the choice of M-theory circle can be added to the GLSM that describes the geometry, as discussed in the previous section. It will be useful to follow the notations of \cite{Benini:2011cm}, as then our discussion can be easily applied to the theories discussed there. For the cones over $Y^{1,2}(\mathbb{CP}^2)$, the GLSM charge matrix can be written as
\be
\label{full GLSM dP0}
\begin{array}{c|cccccccccc|c}
\text{CY}_4 & t_1 & t_2 & t_3 & s_0 & s_1 & \text{FI} \\
\hline
U(1) & 1 & 1 & 1 & -1 & -2 & \xi^c \\
\hline
U(1)_M & 0 & 0 & 0 & 1 & -1 & r_0
\end{array}~,
\ee
where the last line was added to define the $U(1)_M$ symmetry acting on the M-theory circle (the role of $r_0$ will be explained later on). As should be clear from the discussion in the previous section, the first and second lines denote the GLSM fields and their charges under the $U(1)$ gauge group respectively. The FI parameter, $\xi^c$, controls resolutions of the geometry.

One can consider the seven-manifold $X_7 = \text{CY}_4 / U(1)_M$ which is obtained after KK reduction to Type IIA along the isometry circle $U(1)_M$. Using the GLSM description, the charges of the combinations of fields that parametrize the seven-manifold carry zero charge under $U(1)_M$ \cite{Aganagic:2009zk,HananyTalk}. As opposed to the K\"ahler quotient $\text{CY}_3 = \text{CY}_4 // U(1)_M$, that is diagrammatically described by the projection of the three-dimensional toric diagram on the $xy$ plane, the moment map $r_0$ in our quotient is not fixed. In Type IIA we then get $N$ D2 branes probing a fibration of $\text{CY}_3$ over the real line, parametrized by the moment map $r_0$ \cite{Aganagic:2009zk,HananyTalk}.

To obtain the exact form of this fibration notice that, depending on the sign of $r_0$, one of the $s_i$ fields can be eliminated. More specifically, we can define
\be
\label{s_coordinate}
t_0 = s_0  \quad \mathrm{if} \quad r_0 \leq 0~, \quad \text{or} \quad t_0 = s_1  \quad \mathrm{if} \quad r_0 > 0~, 
\ee
and then rewrite the CY$_3$ GLSM as
\be
\label{generic_GLSM_of_CY3}
\begin{array}{c|cccc|c}
\text{CY}_3 &  t_1 & t_2 & t_3 & t_0 & \text{FI} \\
\hline
U(1) & 1 & 1 & 1 & -3 & \chi(r_0)
\end{array}~.
\ee
The FI parameter is
\be
\label{resol_param_CY3}
\chi(r_0) = \xi^c - 2 \, r_0 +3 r_0\, \Theta(r_0) \;,
\ee
where $\Theta(x)$ is the Heaviside step function. 

Whenever $\chi(r_0)\geqslant 0$ this describes a resolved $\C^3/\Z_3$ with a blown-up $\mathbb{CP}^2$ of size $\chi(r_0)$. This is always the case when $\xi^c\geqslant 0$, which is the $\hat{X}_-$ resolution that we studied in the last section. The K\"ahler parameter $\chi(r_0)$ is continuous in $r_0$, while its first derivative jumps by 3 at $r_0 = 0$. This is due to the presence of a $\overline{\text{D6}}$ brane wrapping the toric divisor $\mathbb{CP}^2$ in $\C^3/\Z_3$ \cite{Benini:2011cm}. 

However, considering the second resolution $\hat{X}_+$ by taking $\xi^c < 0$, one can see that for some region of $r_0$ the K\"ahler parameter $\chi(r_0)$ is negative and the interpretation of the reduction as $\C^3/\Z_3$ fibered over the real line is not valid anymore. Moreover, the $U(1)_M$ is rotating the exceptional $\mathbb{WCP}_{[1,2]}^1$, so it fixes its north pole $\{t_{1,2,3}=0,s_0=0\}$ and south pole $\{t_{1,2,3}=0,s_1=0\}$. So in M-theory there are two co-dimension 8 fixed points. In addition, we see that $\Z_3\subset U(1)_M$ fixes the $\{t_{1,2,3}=0,s_0\neq 0,s_1\neq 0\}$ locus. There is no simple interpretation of the resulting singularity in the dilaton in Type IIA string theory. Thus, we expect the baryonic branch that describes the $\hat{X}_+$ resolution in the field theory to be described by a strongly coupled sector.

We can also consider the RG flow that is dual to the near-horizon of the $N$ M2 branes probing the residual singularity in the $\hat{X}_+$ resolution. We are interested in warped resolved CY backgrounds in which the IR and UV boundaries are isolated four-fold singularities. The case we just analysed is one concrete example, as the tangent cones to the stack of $N$ M2 branes in the UV and IR are $\mathcal{C}(Y^{1,2}(\mathbb{CP}^2))$ and $\C^4/\Z_2$ respectively. The reduction on the M-theory circle results in a Type IIA background with two boundaries as well. The discussion above suggests that a Lagrangian description of the RG flow exists only when this Type IIA background is smooth. One should of course reduce the two boundaries on the same M-theory circle. This can be inferred from the toric diagram after deleting the corresponding points that give the desired resolution. 

In the $\mathcal{C}(Y^{1,2}(\mathbb{CP}^2))$ example the resolution of interest corresponds to deleting $s_1$ from the toric diagram. The tangent cone at the residual singular point is simply $\C^4/\Z_2$, and the M-theory circle still corresponds to the z-axis. Notice that this is a different M-theory circle from the one that is described by the ABJM theory with $k=2$. The corresponding reduction of the near-horizon geometry to Type IIA is singular, and thus, following the discussion in \cite{Jafferis:2009th}, we do not expect to find a field theory with Lagrangian description of this RG flow. To see explicitly that the reduction is singular we use the coordinate patch defined in the last section around \eqref{patch}. In the notations of the last section, $z_1$ and $z_2$ are the coordinates that are charged under $U(1)_M$. Thus, for the $\C^4/\Z_2$ that is parametrized by $\{u_1,u_2,u_3,u_4\}$ with the identification $\{u_1,u_2,u_3,u_4\}\sim  \{-u_1,-u_2,-u_3,-u_4\}$, the $U(1)_M$ action on the coordinates is $\{u_1,u_2,u_3,u_4\}\rightarrow  \{\lambda\,u_1,\lambda\,u_2,\lambda\,u_3,\lambda^{-3}\,u_4\}$, where $|\lambda|=1$. The quotient by $U(1)_M$ preserves the holomorphic $(4,0)$-form (hence Killing spinors), but introduces $\Z_3$ orbifold singularities that correspond to the fixed $\{0,0,0,u_4\}$ locus.

Notice that similar considerations show that deleting $s_0$ results in $\C^4$ as a tangent cone with singular reduction over $U(1)_M$. Although a VEV for $s_0$ corresponds to the $\hat{X}_+$ resolution, the branes are not sitting on the singular point this time, and thus the tangent cone is flat space. The same remarks from the last paragraph apply also here. 

As an aside, notice that different VEVs that result in $\C^4$ as tangent cone may be given. Especially interesting is the $\hat{X}_-$ resolution that is obtained by giving VEVs to the $t_i$ fields. In fact, using the $SU(3)$ symmetry, we may without loss of generality give the only non-vanishing VEV to $t_1$. To obtain the tangent cone we delete $t_1$ from the toric diagram. We then see that the tangent cone is just $\C^4$ and the $U(1)_M$ fibre degenerates over a non-compact four-cycle in $\C^3$. The latter is the base of the fibration, as can be seen from the projection of the toric diagram after the deletion. Thus, the Type IIA reduction contains a D6 brane wrapping this cycle. The field theory which is dual to such a background was studied in \cite{Benini:2009qs, Jafferis:2009th}, and a candidate Lagrangian description was proposed there. Recall that the $\hat{X}_-$ resolution involves a $\overline{\text{D6}}$ brane, which is wrapped over the $\mathbb{CP}^2$ in the Type IIA reduction. It is interesting to see that this $\overline{\text{D6}}$ brane is connected to the one in the tangent cone. After giving a VEV to $t_1$ the stack of D2 branes sits on the blown-up $\mathbb{CP}^2$ at $r_0=0$. Thus, close to the D2 branes, the $\overline{\text{D6}}$ brane that was wrapped on the $\mathbb{CP}^2$ is now wrapped on the non-compact four-cycles in the tangent cone.

We want now to study in more detail the M5 branes that are dual to the exotic baryonic operators. A better understanding of such M5 branes and the identification of the dual baryonic operators in the field theory might shed light on the exotic field theories in the far IR. This can be done by studying the RG flows induced by giving VEVs to such operators. We will consider the Euclidean M5 instantons wrapped on holomorphic six-cycles of the CY in the presence of the stack of M2 branes. In the spirit of \cite{Forcella:2008au} the M5 brane instantons lead to insertions of baryonic operators in the stack of M2 branes. 
The mapping of the M5 branes wrapping the different non-compact divisors to branes in Type IIA after the reduction can be obtained from the GLSM equations \eqref{full GLSM dP0}. We want to start with the $s_0=0$ six-cycle. The VEV of the corresponding operator parametrizes the $\hat{X}_+$ resolution. From the third line in \eqref{full GLSM dP0} we can see that $r_0 \leq 0$, so that $s_1$ can be eliminated. We hence find that the six-cycle is mapped to a five-manifold spanned by $\{r_0 \leq 0, t_1, t_2, t_3\}$ with the constraint in \eqref{generic_GLSM_of_CY3} and $t_0=0$. Thus, the M5 brane wrapping the $s_0=0$ six-cycle maps to D4 brane wrapping the $\mathbb{CP}^2$ with $\chi(r_0)$ as the K\"ahler modulus fibered over $r_0 \leq 0$. Similar considerations show that the $s_1=0$ six-cycle maps to the $\mathbb{CP}^2$ with $\chi(r_0)$ as the K\"ahler modulus fibered over $r_0 \geq 0$. The reduction of the $t_i=0$, where $i=1,2,3$, six-cycles can be read more easily from \eqref{generic_GLSM_of_CY3}. We see that these are mapped to the non-compact divisors in $\C^3/\Z_3$ fibred over the unconstrained $r_0$ coordinate. By T-dualizing to Type IIB along the real line (after compactification), one can see that the M5 wrapping the $t_i=0$ six-cycles are mapped to D3 branes wrapping the non-compact divisors of the $\C^3/\Z_3$ cone. These are the familiar instantons that correspond to the insertions of baryonic operators in $(3+1)$d theories. It is immediately obvious that the two other M5 branes, wrapping $s_i=0$, will not map to familiar objects in Type IIB. This is partly due to the fact that the wrapped cycles in $\C^3/\Z_3$ are compact. Moreover, it is not clear how this object should be T-dualized. More work should be done in order to understand what type of insertions those D4 instantons correspond to.

Having addressed field theories with exotic M-theory circles, we want now to make additional comments on the $\mathcal{C}(Y^{1,2}(\mathbb{CP}^2))$ field theory that was mentioned in the last section. The field theory discussed in \secref{ftresults} corresponds to a choice of M-theory circle for which the reduction to Type IIA of the resolved CY four-folds, and the corresponding warped resolved backgrounds, are always smooth. The resolutions can be described by blowing-up vanishing cycles and moving the D2 branes along the $r_0$ coordinate, in the Type IIA picture. Moreover, in this case the UV and IR boundaries of the warped resolved solutions always admit smooth reductions to Type IIA. This explains why, in this example, we have a full Lagrangian description of the resolutions in the field theory, and of the RG flow of the corresponding Higgsings.

The above discussion can easily be applied to more general geometries, especially the $Y^{p,q}(\mathbb{CP}^2)$ examples discussed recently in \cite{Benini:2011cm}. In the latter case, if we restrict ourselves to isolated singularities for which $0<q<3\,p$, it is easy to see that whenever $q \neq 3$, $\chi(r_0)$ in (3.15) of \cite{Benini:2011cm} can take negative values in a certain range of the $r_0$ coordinate, for some choices of the FI parameters. For negative values of $\chi(r_0)$ the corresponding resolutions of the CY$_4$ singularity cannot be described by resolutions of the CY$_3$, together with displacement of the D2 and D6 branes along the $r_0$ coordinate, in the Type IIA picture. Therefore a similar discussion to that above shows that the corresponding baryonic operators and RG flows are expected to be exotic.

\section{Six-cycles and non-perturbative superpotentials} \label{sec:5}

An isolated toric CY four-fold singularity will typically have exceptional divisors when it is resolved.
The irreducible components of these divisors are in 1-1 correspondence with the internal lattice points in the toric diagram, and a simple homology calculation (see \eqref{generate}) shows that these generate the group $H_6(X,\R)$ of six-cycles. This immediately 
raises the question of what is the AdS/CFT interpretation of such six-cycles, as we mentioned briefly at the end of \secref{sec:generalwarped}. Again, in order to make our discussion concrete we will begin by focusing on a simple example, namely the cone over $Q^{222}$. 

\begin{figure}[ht]
\begin{center}
\includegraphics[scale=0.8]{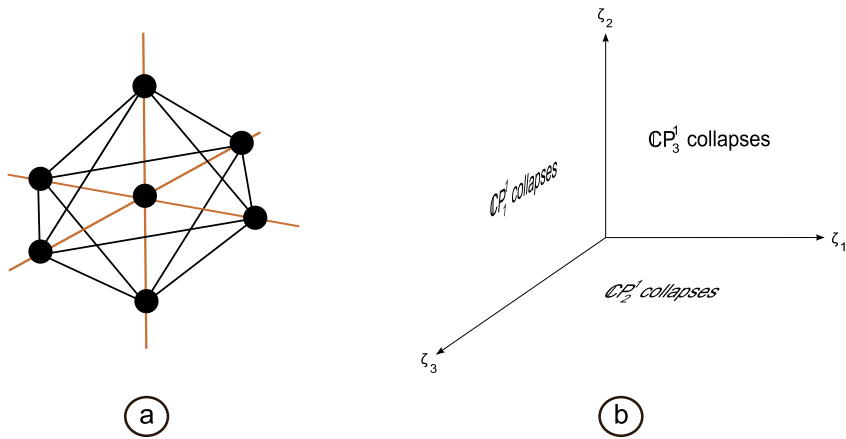}
\end{center}
\caption{(a) The toric diagram for $\mathcal{C}(Q^{222})$. (b) The GKZ fan for $Q^{222}$ is $(\R_{\geq 0})^3$. The axes $\zeta_a$, $a=1,2,3$, may be identified with the K\"ahler classes 
of each factor in $\mathbb{CP}^1\times\mathbb{CP}^1\times\mathbb{CP}^1$, or equivalently the FI parameters in the GLSM.}
\label{fig:toricdiagramQ222}
\end{figure}

Inspection of the $\mathcal{C}(Q^{222})$ toric diagram in Figure \ref{fig:toricdiagramQ222}
shows an interior lattice point, signalling the possibility of blowing up a six-cycle. We shall discuss the geometry of such resolutions 
in more detail later in this section. For more details on the notations that will be used we refer the reader to appendix C of \cite{Benishti:2010jn}. As we discuss in the next subsection, such six-cycles have been shown to be responsible for non-perturbative superpotentials in CY compactifications via wrapped Euclidean M5 branes. We are interested in such contributions to a non-perturbative superpotential 
in warped CY backgrounds. The warping here is induced by the back-reaction of point-like M2 brane sources on the CY\footnote{It might be possible to generalize this to the case with SUSY $G$-flux, in which the flux also sources the warp factor.}.

\subsection{Non-perturbative superpotentials}

The toric geometries under consideration do not have three-cycles, either in the boundary Sasaki-Einstein manifold 
$Y_7$, or in the CY resolution $X$ of $\mathcal{C}(Y_7)$. Thus there are no cycles on which to wrap Euclidean M2 branes. However, EM5 branes may be wrapped on the irreducible 
components of the exceptional divisor in the CY resolution. Such cycles, being complex sub-manifolds, are automatically supersymmetric. 

A very similar situation was considered in \cite{Witten:1996bn}, where compactifications of M-theory to three dimensions on a CY four-fold were discussed. In that reference the role of Euclidean instantonic M5 branes, and their possible contribution to non-perturbative superpotentials, was studied in detail. In order to generate such contributions, the number of zero modes, which includes the Goldstinos of the SUSYs broken by the brane, must be appropriate to saturate the superspace measure. In particular, such instantons must wrap cycles without infinitesimal holomorphic deformations, since the superpartners of the deformation moduli would provide additional fermionic zero modes. In \cite{Witten:1996bn} it was shown that the appropriate zero mode counting in the case of an M5 brane wrapping a divisor $D$ in this set-up requires the necessary condition that
\bea
\label{Todd}
\chi(D,\mathcal{O}_D)\equiv \sum_{i=0}^3 (-1)^i \dim H^i(D,\mathcal{O}_D)=1~.
\label{arithmetic-genus}
\eea
This is the \emph{arithmetic genus} of $D$. 
Assuming this necessary condition is satisfied, the structure of the non-perturbative superpotential generated by an EM5 brane also requires one to understand its dependence on the K\"ahler moduli. As explained in \cite{Witten:1996bn}, the dependence on these K\"ahler moduli is known exactly, being encoded entirely in the semi-classical term $\me^{-V + \ii\phi}$. Here $V$ denotes the volume of the six-cycle, while $\phi$ is the expected linear multiplet superpartner to $V$ and is given by the period of $C_6$ through $D$. 
This latter structure is determined from holomorphy of the superpotential.

It is well-known that for \emph{any} smooth compact toric manifold the arithmetic genus in (\ref{Todd}) is indeed equal to 1. However, at this point we should recall that the situation in \cite{Witten:1996bn} is slightly different from the one at hand. Firstly, our CY four-fold is 
non-compact, so that gravity is decoupled from the point of view of reduction to Minkowski three-space. Secondly, 
 our set-up contains also point-like M2 branes, and moreover in the warped solutions the back-reaction of these M2 branes is also included. This leads to the asymptotically 
AdS$_4$ backgrounds discussed in \secref{sec:FR}. It should however be possible to extend the analysis of  \cite{Witten:1996bn} to such warped cases. 

Let us first briefly discuss a similar situation in the more controlled Type IIB scenario.  In that case one can consider a CY three-fold singularity with colour and fractional D branes wrapping the collapsed cycles at the singularity, leading 
to a four-dimensional $\mathcal{N}=1$ SUSY field theory at the singularity.
In addition one can consider instantonic Euclidean Ep branes. The various types of strings stretching between these branes can give rise to non-perturbative contributions to the superpotential 
of the field theory at the CY singularity.
Exactly as for the M-theory case, in order for this non-perturbative superpotential to be generated at all the right number of zero modes must be present. An important remark here is that the Ep-Ep sector sees the full $\mathcal{N}=2$ CY three-fold background, thus generically leading to too many zero modes to saturate the $\mathcal{N}=1$ superspace measure. Therefore in order for a non-perturbative superpotential to be generated, some method of eliminating these extra zero modes is required. On the other hand, the situation in M-theory is very different since the colour M2 branes do not break 
any further the SUSYs of the CY four-fold background. From this point of view, we then expect EM5 instantons to generically contribute to non-perturbative corrections to the superpotential in three 
dimensions.

Nevertheless, the structure and interpretation of such corrections is far from clear. 
We can think of the gravitational background as a warped CY four-fold compactification, albeit one which is asymptotically an AdS$_4$ background. As such, we expect to be able to promote all moduli, both K\"ahler and those related to the positions of the M2 branes in the colour stack, to  spacetime fields in the Minkowski three-space. These will be dynamical fields provided the fluctuations are normalizable in the warped metric; 
at least for the K\"ahler moduli this is expected to be the case, as discussed in the Type IIB context in \cite{Martelli:2008cm}.
However, finding an ansatz for such a reduction is far from trivial, and at the time of writing there is no complete proposal for such an ansatz which would allow one 
to compute the precise form of the non-pertubative superpotential for these modes. The most recent paper on this subject is \cite{Frey:2008xw}, 
where the authors consider only the universal K\"ahler modulus in a warped compactification. 
 On the other hand, following the more controlled Type IIB case, one might expect that computing the warped volume of the Euclidean brane is the dual ``closed membrane channel'' of the picture above, described in terms of M2 branes in the blown up CY four-fold in the presence of the EM5 branes. In the IIB case it has been explicitly checked \cite{Baumann:2006th} in some simple situations how the computation of open string diagrams in the relevant sector \cite{Berg:2004ek, Berg:2005ja} can be reproduced though the computation of warped volumes, which are then interpreted as a resummation of such open string diagrams. Of course, it should be stressed that in the M-theory scenario at hand this can be taken only as a heuristic picture. In any case, one would expect that a general \emph{holographic} interpretation of the superpotential should also be available, since the warped 
background is asymptotically AdS$_4$. One natural suggestion is that this might come from considering the boundary behaviour of the  six-form fluctuation 
sourced by the EM5 branes. We leave a more complete investigation of these issues for further work. Instead in this thesis we focus 
on computing the warped volumes of the EM5 branes as a function of the moduli. Understanding precisely how this is related to non-perturbative 
corrections in these warped resolved geometries will require further work.


\subsection{Instantons in the resolved $Q^{222}$ background} \label{s:GKZ}

There is a unique resolution of $\mathcal{C}(Q^{222})$ where one blows up the six-cycle corresponding to the internal lattice point in 
Figure \ref{fig:toricdiagramQ222}. This gives a CY four-fold $X$ which is the total space of the canonical bundle 
$\mathcal{O}(-2,-2,-2)\rightarrow\mathbb{CP}^1\times\mathbb{CP}^1\times\mathbb{CP}^1$. There is a K\"ahler class 
for each factor in the zero section exceptional divisor $(\mathbb{CP}^1)^3$, leading to a GKZ fan which is 
$(\R_{\geq 0})^3$, as shown in Figure \ref{fig:toricdiagramQ222}. 

By the general theorem mentioned earlier we know that there will be a unique Ricci-flat K\"ahler metric, which is asymptotic to the cone metric over $Q^{222}$, for each choice of K\"ahler class. Again, in this case one can write these metrics explicitly:
\begin{eqnarray}
\diff s^2&=&\kappa(r)^{-1}\diff r^2+\kappa(r)\frac{r^2}{16}\Big(\diff \psi+\sum_{i=1}^3 \cos\theta_i \diff \phi_i\Big)^2+\frac{(2a+r^2)}{8}\Big(\diff \theta_2^2+\sin^2\theta_2 \diff \phi_2^2\Big)\nonumber \\ &&+\frac{(2b+r^2)}{8}\Big(\diff \theta_3^2+\sin^2\theta_3 \diff \phi_3^2\Big)+\frac{r^2}{8}\Big(\diff \theta_1^2+\sin^2\theta_1 \diff \phi_1^2\Big)~,
\end{eqnarray}
where now
\begin{equation} \label{kappaQ222}
\kappa(r)=\frac{r^8+\frac{8}{3}\, (a+b)\, r^6+8\, a\,b \,r^4-16\, c}{r^4\, (2\,a+r^2)\,(2\,b+r^2)} \ .
\end{equation}
Here $a$, $b$ and $c$ are arbitrary constants, and correspond to the choice of K\"ahler classes $\zeta_i$, $i=1,2,3$. 
In particular, setting $a=b=0$ implies that all three $\mathbb{CP}^1$s have the same volume, and the metric 
simplifies considerably. In this case it is convenient to define $r_\star^8=16c$, so that the exceptional 
divisor is at the radial position $r=r_\star$. For further details about these metrics we refer the reader to \secref{sec:D}.

Again, it is also possible to solve explicitly for the warp factor for these metrics. In the simplified case 
with $a=b=0$, one can place the stack of $N$ M2 branes at an arbitrary radial position $r=r_0\geq r_\star$ and solve 
for the Green's function. Again, we refer the reader to \secref{sec:D} for details of this warp factor. 
From now on we focus exclusively on the case $a=b=0$, where $\zeta_1=\zeta_2=\zeta_3$ is parametrized by the radius $r_\star>0$.

\subsubsection{Warped volumes}

We are interested in computing the warped volume of the compact exceptional divisor $(\mathbb{CP}^1)^3$ at $r=r_{\star}$, with the stack of $N$ M2 branes at the position $y=(r_0,\xi_0)$. Here $\xi_0$ denotes the point in the copy of $Q^{222}$ at radius $r=r_0$. Thus we define
\begin{equation}
S=T_5\, \int_D\, \sqrt{\det g_D}\, h\, \diff^6 x
\end{equation}
where $D$ is the exceptional divisor and $h$ is the (pull back of the) warp factor. Here the latter is given by the expression (\ref{general-warp}) 
in terms of the mode expansion discussed in \secref{sec:D}. The determinant of the metric pulled back to the divisor is simply
\begin{equation}
\sqrt{\det g_D}=\frac{r_{\star}^6}{8^3}\, \sin\theta_1\, \sin\theta_2\, \sin\theta_3~.
\end{equation}
After substituting these results into the world-volume action one obtains
\begin{equation}
S=\frac{{T_5}\,r_{\star}^6}{8^3}\, \sum_I\,  Y_I(\xi_0)^*\,\psi_I(r_{\star})\,\int_D\, \sin\theta_1\, \sin\theta_2\, \sin\theta_3 \, Y_I(\xi)\, \diff^6 x~.
\end{equation}
Explicitly, the integral reads
\bea
\nn
\int_D\, \sin\theta_1\, \sin\theta_2\, \sin\theta_3 \, Y_I(\xi)\, \diff^6 x=\mathcal{C}_{I}\,\prod_{i=1}^3  \int_{\theta_i=0}^\pi\int_{\phi_i=0}^{2\pi} \sin\theta_i\, \me^{\ii\, m_i\,\phi_i}J_{0,\, l_i,\, m_i}(\theta_i)\, \diff\theta_i\, \diff\phi_i~.\\
\eea
Here $\mathcal{C}_I$ is a normalization constant that ensures the mode $Y_I$ has unit norm.
The $\phi_i$ integrals vanish unless $m_i=0$. Then $J_{0,\,l_i,\, 0}(\theta_i)$ reduces to a Legendre polynomial ${\rm P}_{l_i}(\cos\theta_i)$, so that
\begin{equation}
\int_{\theta_i=0}^\pi \sin\theta_i\, {\rm P}_{l_i}(\cos\theta_i)\, \diff\theta_i=2\, \delta_{l_i,\,0}~.
\end{equation}
Therefore
\begin{equation}
\int_D\, \sin\theta_1\, \sin\theta_2\, \sin\theta_3 \, Y_I(\xi)\, \diff^6 x=2^6\, \pi^3\, \mathcal{C}_{0}\,\prod_{i=1}^3 \delta_{l_i,\,0}\,\delta_{m_i,\,0}~.
\end{equation}
Substituting this back into the expression for $S$ one finds
\begin{equation}
S=\frac{2^6\, \pi^3\,{T}_5\,r_\star^6}{8^3}\, |\mathcal{C}_0|^2 \,\psi_{0,0,0}(r_{\star})~.
\end{equation}
From \eqref{warp-Q222-I} and (\ref{warp-Q222-II}) we have
\begin{equation}
\psi_{0,0,0}(r_{\star})=\frac{1}{r_0^2}\, _2F_1\Big(\frac{3}{4},\, 1,\, \frac{7}{4},\frac{r_{\star}^8}{r_0^8}\Big)\, , \quad\, |\mathcal{C}_0|^2=\frac{2^9\, \pi^2\, N\, \ell_p^6}{3}~.
\end{equation}
Finally, substituting the explicit M5 brane tension results in the warped volume
\begin{equation}
S=\frac{2\,N}{3}\,\frac{r_{\star}^6}{r_0^6}\, _2F_1\Big(\frac{3}{4},\, 1,\, \frac{7}{4},\frac{r_{\star}^8}{r_0^8}\Big)
\label{Q222-action}~.
\end{equation}
Note that for $r_0 \rightarrow r_{\star}$ we get $S\sim -\frac{N}{2}\,\log(r_0-r_{\star})$.
\subsubsection{The $L^2$ harmonic two-form}

The key observation in this subsection is that the warped volume (\ref{Q222-action}) of the exceptional divisor 
is closely related to the $L^2$ normalizable two-form $\beta$ which is Poincar\'e dual to the six-cycle. 
The claim is that the precise relation between the two is (for $N=1$)
\bea
\beta=\frac{1}{\pi \, \ii}\,\partial \bar{\partial} S~.
\label{beta-S}
\eea
Here the derivatives are regarded  as acting on the coordinates of the point $y=(r_0,\xi_0)$, which recall is the location of the stack 
of M2 branes. 
We shall prove this claim in full generality in \secref{s:general-V}. Here we first prove it in the current explicit example, where it 
is convenient to use the coordinate system in section B.1 of \cite{Benishti:2010jn}. Setting $N=1$ and after some algebra it can be shown that 
\bea
\nn
(\bar{\partial}-\partial)S=\ii \,\frac{r_{\star}^6}{r_0^6}\,g_5~,
\quad \text{where} \quad
g_5=\dd\psi+\sum_{i=1}^3\cos\theta_i\, \diff\phi_i~,
\eea
and therefore (\ref{beta-S}) reads
\begin{equation}
\beta=\partial \bar{\partial}\,\left(\frac{1}{\pi \, \ii}\,S\right)=\frac{1}{2}\dd(\bar{\partial}-\partial)\left(\frac{1}{\pi \, \ii}\,S\right)=\frac{1}{2 \pi}\dd\left(\frac{r_{\star}^6}{r_0^6}\,g_5\right)~.
\end{equation}
This is a harmonic two-form with respect to the unwarped metric, and also $L^2$ with respect to this metric. To see that $\beta$ is indeed Poincar\'e dual to the six-cycle, we choose the following closed form on $X=\mathcal{O}(-2,-2,-2)\rightarrow (\mathbb{CP}^1)^3$
\begin{equation}
\mu = \frac{\sin\,\theta_1\,\sin\,\theta_2\,\sin\,\theta_3}{2^6 \, \pi^3} \dd \theta_1 \wedge \dd \phi_1 \wedge \dd \theta_2 \wedge \dd \phi_2 \wedge \dd \theta_3 \wedge \dd \phi_3~.
\end{equation}
It is easy to see that 
\begin{equation}
\int_D\,\mu = 1 \, , \quad \, \int_X\,\beta\,\wedge\,\mu = 1~.
\end{equation}
Hence $\beta$ is $L^2$ normalizable and Poincar\'e dual to the divisor, as claimed.

\subsubsection{Critical points}

Formally, the superpotential that is induced by the instanton action that we have calculated in the previous subsections is given by
\bea\label{supQ222}
W=\me^{-S}=\exp\left[\frac{2\,N}{3}\,\frac{r_{\star}^6}{r_0^6}\, _2F_1\Big(\frac{3}{4},\, 1,\, \frac{7}{4},\Big(\frac{r_{\star}}{r_0}\Big)^8\Big)\right]~,
\eea
where $S$ is given in (\ref{Q222-action}). This is something of a formal statement, since in reality what we have computed is the 
on-shell Euclidean action of the wrapped M5 brane as a function of the moduli of the supergravity background. Here essentially $r_\star$ is a K\"ahler 
modulus, while $r_0$ is a modulus associated to the position of the stack of M2 branes. On the other hand, the superpotential 
should be a function of the corresponding spacetime fields in Minkowski space, obtained by promoting these moduli to dynamical fields. 
In the unwarped case there is essentially no distinction between the two, but in the warped case it is not known how to do this at present, and the situation is much less clear. Note also that we have only computed the real part of $S$, and hence absolute value of $W$. Thus the best we can do is to examine the critical points of $\me^{-S}$ interpreted directly as a 
superpotential on the supergravity moduli space. 
It is straightforward to compute
\bea
\partial_{r_0}\,S=\frac{4\,N\,r_{\star}^6\,r_0}{r_{\star}^8-r_0^8}~.
\eea
For $r_{\star}>0$ there are no critical points of $S$. In order for $\diff W=0$ we must then necessarily have $r_0=r_{\star}$, which gives $S=+\infty$ and $W=0$ on this locus. In this case the branes move only on the six-cycle. Clearly, this is always a solution since the ``superpotential'' (\ref{supQ222}) is identically zero if the branes are moving on the divisor $D$. For $r_{\star}=0$ there are formally no contributions of instantons to the superpotential. However, notice this is a singular limit of the supergravity solution in which the six-cycle is blown down. 
In the absence of instantons of course the M2 branes are free to propagate on the cone $\mathcal{C}(Q^{222})$ in which the six-cycle is blown down. 

\subsection{EM5 brane instantons: general discussion} \label{s:general-V}

In this section we describe how the above calculations for $Q^{222}$ generalize to more general CYs. 

\subsubsection{Geometric set-up}

Throughout this section we assume we are given an asymptotically conical K\"ahler manifold $X$ of complex dimension $n$, with metric $g=g_X$ and K\"ahler two-form $\omega$. This means that the manifold $X$ is non-compact, and that the metric $g$ asymptotically approaches the cone metric $\diff r^2 + r^2 g_Y$,
where $Y_7=\partial X$ is the compact base of the cone. The metric $g_Y$ is then Sasakian, by definition. 
In fact this is slightly too general for the situation we are interested in. More precisely, we want $X$ to be a \emph{resolution} of the singularity at $r=0$ of the cone $\mathcal{C}(Y_7)\cong (\R_{\geq 0}\times Y_7)\cup \{r=0\}$. This means that there is a proper birational map $\pi:X\rightarrow \mathcal{C}(Y_7)$ which is a biholomorphism on the restriction $X\setminus E\rightarrow \{r>0\}\subset \mathcal{C}(Y_7)$. Less formally, the resolution $X$ replaces the singular point $\{r=0\}$ of the cone by the \emph{exceptional set} $E=\pi^{-1}(r=0)$. 
For physical applications we require the metric $g$ on $X$ to be Ricci-flat and hence CY. Then by definition $X$ above is a \emph{crepant} resolution of the cone singularity. In fact this won't really affect the computations that follow.

Of particular interest for us in this section are the exceptional divisors in 
$X$. These are the irreducible (prime) divisors of $E=\pi^{-1}(r=0)$. Call these irreducible components $E_i$. Since $X$ is contractable onto $E$ we have
\bea \label{generate}
H_{2n-2}(X,\R)\cong H_{2n-2}(E,\R)\cong \bigoplus_i H_{2n-2}(E_i,\R)\cong \R^{b_{2n-2}(X)}~,
\eea
so that $i=1,\ldots,b_{2n-2}(X)=\dim H_{2n-2}(X,\R)$. Thus the 
exceptional divisors $E_i$ generate the homology of $X$ in codimension two. Notice that if $b_{2n-2}(X)=0$ then the resolution has no exceptional divisors and the resolution is said to be \emph{small}. For example, the resolved conifold is a small resolution of the cone over $T^{1,1}$, since the exceptional set is $E\cong \mathbb{CP}^1$; similarly, the resolutions of the geometries that were classified in \secref{sec:gen} are small.

\subsubsection{$L^2$ harmonic two-forms}

Another key result for us is that in the above situation
\bea
\mathcal{H}^2_{L^2}(X,g)\cong H^2(X,Y,\R)\cong H_{2n-2}(X,\R)~,
\eea
for $n>2$.
Here $\mathcal{H}^2_{L^2}(X,g)$ denotes the $L^2$ normalizable harmonic two-forms on $X$ (which thus depends on the metric $g$). 
That is, the codimension two cycles in 
$X$ are 1-1 with the prime exceptional divisors, and these are also 1-1 with the $L^2$ normalizable harmonic two-forms on $X$, as long as $n>2$. This result about $L^2$ harmonic forms holds in general for complete asymptotically conical manifolds, and was proven in 
\cite{hausel}. In dimension $n=2$, instead $\mathcal{H}^2_{L^2}(X) \cong \mathrm{Im}\left[H^2(X,Y,\R)\rightarrow 
H^2(X,\R)\right]$. For example, for the  Eguchi-Hanson manifold the map $H^2(X,Y,\Z)\rightarrow H^2(X,\Z)$ is multiplication by 2, so that 
there is a unique $L^2$ harmonic two-form, up to scale.

We shall need some more information about these harmonic forms. 
First, we note that the harmonic two-forms are of Hodge type $(1,1)$. This is because they are Poincar\'e dual to divisors in $X$, so if there 
was a $(0,2)$ part of the harmonic form it would be cohomologically trivial and hence\footnote{Notice this result definitely fails for spaces that are not asymptotically conical. A good example is the Taub-NUT space, which has no two-cycles but does have an $L^2$ harmonic two-form.} identically zero by the result of \cite{hausel}.

Pick a particular $D=E_i$ and normalize the associated $L^2$ harmonic two-form so that it is Poincar\'e dual to $D$. 
We may then 
think of the harmonic form as the curvature of a Hermitian line bundle $\mathcal{L}=\mathcal{L}_D$ -- the divisor bundle for 
$D$. If $s$ is a \emph{local} nowhere zero holomorphic section of $\mathcal{L}$ over an open set $U\subset X$, and $H$ is the Hermitian metric 
on $\mathcal{L}$, then we may write the harmonic form as
\bea\label{betaform}
\beta\mid_U\, = \frac{1}{2\pi\ii}\, \partial\bar{\partial} \log H(s,s)~.
\eea
This is a standard result. Notice that $\beta$ here does not depend on the choice of local holomorphic section $s$. 
We will be interested in the special case where we take $U$ as large as possible, which is $U = X\setminus D$. 
By construction, the line bundle $\mathcal{L}_D$ is trivial over $U$, with trivializing nowhere zero holomorphic section $s$. 
We pick an $s$, and write $H=H(s,s)$ for the corresponding real function on $X\setminus D$.

We next note that $\log H$ is itself a harmonic function on $X\setminus D$. 
To see this, suppose generally we have a harmonic $(1,1)$-form $\beta$. Because $(X,g)$ is complete, this means 
$\beta$ is both closed and co-closed. The co-closed condition involves computing the Hodge dual of $\beta$, which is $\star\beta = -\beta\wedge \omega^{n-2} + (\omega\lrcorner\beta)\omega^{n-1}$. Here we have simply used that $\vol = \omega^n/n!$ and that $\beta$ is type $(1,1)$. Thus 
$\beta^{ij}\omega_{im}\omega_{jn}=\beta_{mn}$. By definition, $\omega\lrcorner\beta=(1/2!)\omega_{ij}\beta^{ij}$. 
Thus if $\beta$ is closed, then $\star\beta$ is co-closed if and only if $\omega\lrcorner\beta$ is closed, {\it i.e.} 
constant. 

Thus we learn that for a harmonic $(1,1)$-form $\beta$, $\omega\lrcorner\beta$ is in fact constant. 
Now, for an $L^2$ form on an asymptotically conical manifold this constant is in fact necessarily zero. 
To see why, we must look at the asymptotics of $\beta$. This was studied in appendix A of 
\cite{Martelli:2008cm}. Here we have a two-form, so $p=2$ in Table 4 of that reference, and we are interested in 
$L^2_{\infty}$, so that the form is normalizable at infinity. For $n>2$, $p<n$ so the normalizable modes 
are of type II and type III$^-$. However, the type II modes are constructed from harmonic one-forms on the 
base $Y$, and there are not any of these as $(Y,g_Y)$ is a positive curvature Einstein manifold so $b_1(Y)=0$ by Myers' theorem \cite{Myers}.
Thus asymptotically, the two-form $\beta$ 
is to leading order of the form
$
\beta \sim \diff (r^{2-n-\nu}\beta_\mu)
$, where $\beta_\mu$ is a massive co-closed one-form along $Y$,
$
\Delta_Y \beta_\mu = \mu\beta
$,
and 
$
\nu = \sqrt{(n-2)^2+\mu}
$.
Consider now $\omega\lrcorner\beta$. Since asymptotically $\omega\sim (1/2)\diff (r^2\eta)$, where $\eta=\ii (\bar\partial-\partial)\log r$ is the contact one-form of the Sasakian manifold $(Y,g_Y)$, we have
\bea
\omega\lrcorner\beta \sim r^{-n-\nu}\left[(2-n-\nu)\beta_\mu\lrcorner\eta + \frac{1}{2}\diff\beta_\mu\lrcorner\diff\eta\right]~.
\eea
Since $\nu>0$ it follows that $\omega\lrcorner\beta\rightarrow 0$ at infinity. Since we have also shown that $\omega\lrcorner\beta$ is constant, it follows that 
this constant is zero.
If we write 
$\beta$ as in (\ref{betaform}) on $X\setminus D$, then $\omega\lrcorner\beta=0$ is equivalent to saying that $\log H$ is harmonic. This just follows from the form 
of the scalar Laplacian on a K\"ahler manifold: $\Delta f = -2\omega\lrcorner\ii\partial\bar{\partial} f$. 

As an aside comment that will be important below, the first non-zero eigenvalue $\mu$ is bounded below by $4(n-1)$. In fact, this is saturated precisely by \emph{Killing} one-forms -- 
see, for example, \cite{DNP}. Thus $\mu\geq 4(n-1)$ and correspondingly $\nu \geq \sqrt{(n-2)^2+4(n-1)} = n$.

We thus conclude that $\log H$ is a harmonic function. It will be crucial in what follows that $\log H$ is not in fact defined everywhere on $X$. 
By construction, it is defined only on $X\setminus D$. Along $D$ in fact $\log H$ is singular. This is simply because $D$ is the zero set of $s$, which has a simple zero along $D$ by assumption. Thus if $z=\rho\, \me^{\ii\theta}$ is a local complex coordinate normal to $D$, with $D$ at $\rho=0$, then $\log H$ blows up near to $D$ like $\log \rho^2 = 2 \log \rho$. 

\subsubsection{The instanton action}

An instantonic brane wrapped on an exceptional divisor $D=E_i$ is calibrated and supersymmetric -- for example, a D3 brane 
for $n=3$ in Type IIB string theory or an M5 brane for $n=4$ in M-theory. These are in 1-1 correspondence with the homology classes $H_{2n-2}(X,\R)$, and moreover there is a unique $L^2$ harmonic two-form associated to each irreducible exceptional divisor, which is Poincar\'e dual to the divisor, as discussed above. 

Let $G[y](x)$ denote the Green's function on $X$, with a fixed (Ricci-flat) K\"ahler metric, normalized so that 
\bea
\Delta_x G[y](x) = 2\pi \frac{1}{\sqrt{\det g_X}}\, \delta^{2n}(x-y)~. 
\eea
Consider the on-shell action of an instantonic brane, given by the following Green's function weighted volume of $D$:
\bea
V = \int_D G[y]\sqrt{\det g_D}\, \diff^{2n-2}x = \int_D G[y]\frac{\omega^{n-1}}{(n-1)!}~.
\eea
This is the relevant formula both for D3 branes and M5 branes, where the warp factor is $h= NG/T$, with $N$ the number of 
spacetime filling branes and $T$ the tension of the wrapped instanton brane.
The warped volume $V$ depends on the source point $y\in X$, so $V=V(y)$. Of course, it also depends on the choice of K\"ahler metric. If we consider the CY case, we know that there is a unique metric in each K\"ahler class, so we may think of $V$ also as a function of the K\"ahler class: $V=V(y;[\omega])$. Then we claim that
\bea
V(y) = -\frac{1}{2}\log H(y)~.
\eea
Of course, $\log H$ depends implicitly on the K\"ahler metric since the associated harmonic form does also. 
Notice this result actually provides a \emph{formula} for the $L^2$ harmonic two-forms on an asymptotically conical 
K\"ahler manifold, in terms of the Green's function on $X$. This is a new mathematical result, as far as we are aware\footnote{We thank Tamas Hausel 
for discussions on this.}.

The strategy for proving this claim involves three steps: (i) show that $V(y)$ is a harmonic function on $X\setminus D$, (ii) show that $V$ diverges as $-\log \rho$ along $D$, (iii) show that the two-form $\ii\partial\bar{\partial} V$ is $L^2$. These steps show that the latter two-form is an $L^2$ harmonic form that is Poincar\'e dual to $D$. We may then appeal to the uniqueness of such a form.

Step (i). This is straightforward. We want 
to compute $\Delta_y$ acting on $V$. Using that 
the Green's function $G[y](x)$ is symmetric in its arguments, so $G[y](x)=G[x](y)$, then 
provided $y\notin D$ 
\bea
\Delta_y V= \int_{D}(\Delta_y G[y](x))\sqrt{\det g_D}\, \diff^{2n-2}x = 0~.
\eea
The last step follows since $y\notin D$.
This shows that $V(y)$, interpreted as a function on $X\setminus D$, is indeed harmonic. 

Step (ii). Near to the source point $y$ we have
\bea\label{Gnearsource}
G[y](x) = \frac{2\pi}{(2n-2)\vol(S^{2n-1})\rho^{2n-2}} (1+{o}(1))~,
\eea
where $\rho$ denotes geodesic distance from $y$, where we regard the latter as fixed. Here $\vol (S^{2n-1})$ is the volume of a unit $(2n-1)$-sphere, which appears 
in this computation as a small sphere around the point $y$. 
The divergence is a local question, so this is really a question in Euclidean space. Let us take the local model $\C^{n-1}\times \C$, with complex coordinate $z$ on $\C$. We suppose that the divisor $D$ is locally $z=0$ in this patch, and that the point $y$ is $(0,z)$ in this coordinate system. The integral over $D$ is finite outside a ball $B_\delta$ of fixed radius $\delta>0$ in $\C^{n-1}\times\{0\}$, so we would like to analyse the divergence in the integral
\bea
\frac{2\pi}{(2n-2)\vol(S^{2n-1})}\int_{B_\delta} \frac{1}{(R^2 + |z|^2)^{n-1}}R^{2n-3}\diff R \, \diff\vol(S^{2n-3})~,
\eea
as $|z|\rightarrow 0$. This is easily seen to be 
\bea
\left(-\frac{1}{2}\log |z|^2\right) \cdot \frac{2\pi\vol(S^{2n-3})}{(2n-2)\vol(S^{2n-1})} = -\frac{1}{2}\log |z|^2 = - \log |z|~.
\eea
This shows that $V$ diverges as $-\log |z|=-\log \rho$ near to $D$, as claimed.

Step (iii). We consider the two-form $\diff \diff^c V(y)$ as the point $y$ tends to infinity. Since we have already shown this two-form is harmonic, it will have an asymptotic expansion as in Appendix A of \cite{Martelli:2008cm}. There are three types of modes, I, II and III. 
As already mentioned, there are in fact no modes of type II since $Y$ has no harmonic one-forms. The modes of type I are pull-backs of harmonic two-forms on $Y$, which are not $L^2$. The modes of type III$^{\pm}$ are of the form $
\diff\left(r^{2-n\pm\nu}\beta_\mu\right)$, with $\nu=\sqrt{(n-2)^2+\mu^2}$ as above. The III$^{-}$ modes are $L^2$, while III$^{+}$ are not. Thus we must show that $\diff\diff^c V$ has leading term of type III$^{-}$, so that it is normalizable at infinity and hence normalizable.

Since $V$ is globally defined on $X\setminus D$, we immediately see that there can be no mode of type I since near infinity $\diff\diff^c V$ is exact. Thus we are reduced to analysing the asymptotic $r$-dependence of the one-form $\diff^c V$. 

If we regard the point $x$ as fixed, then as $y$ tends to infinity we have
\bea
G[y](x) = \frac{2\pi}{(2n-2)\vol(Y)r^{2n-2}} (1+o(1)).
\eea
Then
\bea
\beta = \frac{1}{2\pi}\diff\diff^c \int_D G[y]\frac{\omega^{n-1}}{(n-1)!}~,
\eea
gives to leading order
\bea\label{asym}
\beta \sim -\frac{\mathrm{vol}(D)}{\mathrm{vol}(Y)}\diff\left(r^{2-2n}\eta\right)~.
\eea
We thus conclude that $2-n\pm\nu = 2-2n$. From the comment above, this means that we indeed have a normalizable mode III$^{-}$, and moveover
that $\nu=n$ and hence $\mu=4(n-1)$ saturates the lower bound on the eigenvalue. The Killing one-form $\eta$ is of course dual to the R-symmetry. This completes our proof.

\subsubsection{The superpotential} \label{s:W}

To conclude our results thus far, we have proven in general that $\me^{-V}(y,[\omega]) = \sqrt{H(y)}$, where $(1/2\pi \ii) \partial\bar{\partial} \log H$ is the unique $L^2$ harmonic two-form that is Poincar\'e dual to the divisor $D$ wrapped by the 
instanton. The on-shell action of this instanton is $V$. Notice that $\sqrt{H}$ has a simple zero along $D$, as expected on general grounds. In fact locally $H=H(s,s)=\me^{2g}|s|^2$, where $g$ is function and $s$ is a holomorphic section of the divisor bundle $\mathcal{L}_D$. Again, this was expected from arguments in \cite{Ganor:1996pe}, where the phase that pairs with $V$, coming from the Wess-Zumino term in the action, was studied. Thus the result 
presented here is rather complimentary to the discussion in reference \cite{Ganor:1996pe}.
Restoring the factor of $N$, our computation hence shows that, formally at least, we have the superpotential
\bea
W = \me^{-NV}(y;[\omega]) = \sqrt{H^N} = \me^{Ng} |s|^N~.
\eea
This is interpreted as a function of both the K\"ahler class, and also the position of the stack of M2 branes $y\in X$, and generalizes 
the result (\ref{supQ222}) we derived explicitly for $Q^{222}$.
A critical point of this $W$  requires either a critical point of $V$,  or else $V=\infty$. Since $V=-\frac{1}{2}\log H$, the first case requires a critical point of the harmonic function $\log H$. By the maximum principle, notice that such a critical point cannot be either a local maximum or a local minimum of $\log H$. 
\chapter{Conclusions}\label{sec:6}

We set out to study Abelian symmetries in the context of the AdS$_4$/CFT$_3$ correspondence. In particular, we considered gauge fields in AdS$_4$ arising from KK reduction of the supergravity potentials over the $b_2(Y_7)$ topologically non-trivial two-cycles. In contrast to its better-understood AdS$_5$/CFT$_4$ relative, the case at hand displays many more subtleties. The key difference resides in the fact that gauge fields in AdS$_4$ admit, in a consistent manner, quantizations with either of two possible fall-offs at the boundary, implying that the gauge field can be dual to either a global symmetry or to a dynamical gauge field in the boundary CFT. From the bulk perspective, electric-magnetic duality  in the four-dimensional electromagnetic theory in AdS$_4$ amounts to exchanging these two boundary fall-offs. In addition, from the bulk perspective one can shift the $\theta$-angle by $2\pi$. Following \cite{Witten:2003ya}, these two actions translate into particular operations on the boundary theory, the $\mathcal{T}$ and $\mathcal{S}$ operations reviewed in \secref{sec:baryons}, which then generate the group $SL(2,\mathbb{Z})$. As stressed in the main text, these actions exchange different boundary conditions for the gauge field in AdS$_4$. Correspondingly, the dual boundary CFTs are different. Indeed, the whole of $SL(2,\mathbb{Z})$ acts on the boundary conditions for the bulk gauge fields, leading in general to an infinite orbit of CFTs for each $U(1)$ gauge symmetry in AdS$_4$. Understanding the structure of such orbits is a very interesting problem which we set aside for further work. 

We have contented ourselves with studying these issues in the particular case of M2 branes moving in $\mathcal{C}(Q^{111})$\footnote{Although as described in \secref{sec:gen} we expect similar results to hold for other toric isolated four-fold singularities with no exceptional six-cycles.}. In \cite{Franco:2008um, Franco:2009sp} a $U(N)^4$ dual theory was proposed and further studied. In \secref{sec:Q111} we proposed a choice of quantization for the Abelian vector fields in the Betti multiplets such that precisely two $U(1)$s are ungauged, leading to the gauge group  $U(1)^2\times SU(N)^4$. This leaves precisely two global symmetries that may be identified with the two gauge fields coming from KK reduction of the supergravity six-form potential over five-cycles in $Q^{111}$. A key point in that identification is that the corresponding boundary conditions in the bulk AdS$_4$ allow for electric wrapped M5 brane states. These M5 branes can be easily identified in terms of the toric geometry of the variety. Since the field theory realizes the minimal GLSM, it is then straightforward to identify the relevant $U(1)$ symmetries in the QCS theory. In turn, this allows one to construct dual baryonic operators to such M5 branes. It is then natural to consider the spontaneous breaking of such symmetries, where the operator dual to such an M5 brane acquires a VEV.
We analyzed in detail such SSB in \secref{sec:4}. In particular, we have been able to compute the VEV of the baryonic condensate, with precise agreement with field theory expectations. We stress that this is a non-trivial check of the dual theory, as this suggests that it does admit an IR superconformal fixed point with the correct properties (specifically, R-charges) to be dual to M2 branes moving in $\mathcal{C}(Q^{111})$. Along the lines of \cite{Klebanov:2007us, Klebanov:2007cx}, we have also been able to identify the Goldstone boson of this SSB.  However, a comprehensive understanding of these resolutions in the context of the actions on the boundary conditions is still lacking. We postpone this for further work. An interesting by-product of our computation is the finding of general expressions for warped volumes in CY backgrounds, which are potentially of interest for other, similar computations.

In AdS$_4 \times Y_7$, as opposed to AdS$_5 \times Y_5$ backgrounds, it is possible to turn on different torsion $G$-flux, classified by $H^4_{\mathrm{tor}}(Y_7,\Z)$, and each such background corresponds to a physically distinct SCFT. We focused attention on the singularity $\mathcal{C}(Y^{1,2}(\mathbb{CP}^2))$, which has two toric (partial) resolutions. A key ingredient in our analysis was a careful consideration of the effect of \emph{torsion $G$-flux} in the dual AdS backgrounds, which we described for general four-fold singularities. In the case at hand,
the AdS solution AdS$_4\times Y^{1,2}(\mathbb{CP}^2)$ has two choices of torsion $G$-flux, labelled by $H^4(Y^{1,2}(\mathbb{CP}^2),\Z)\cong \Z_2$. 
This analysis can in principle be used for other theories, and should serve as a further constraint on M2 brane theories at CY four-fold singularities.
In general, we expect from the existence of a supergravity dual, and also from the examples studied, that {\it theories with no $G$-flux should be Higgsable to all toric sub-diagrams}. In \cite{Benishti:2009ky} we checked this statment for two non-trivial orbifold theories. First we studied the $\C^2/\Z_n \times \C^2/\Z_n$ orbifold , whose field theory may be obtained via a $\Z_n$ projection of the ABJM theory. We proved there that by Higgsing this theory one can indeed produce QCS theories for all toric sub-diagrams. We then proceeded to study another example, namely the orbifold $\C^4/\Z_2^3$. This has 18 inequivalent toric sub-diagrams, corresponding to 18 toric partial resolutions, including, of course, the space $\C^4$. Since a direct Douglas-Moore projection of the ABJM theory does not produce this theory, we constructed instead the dual theory using the Un-Higgsing algorithm which is derived in \secref{sec:unhiggsing}. This is a general algorithm for toric quiver-Chern-Simons theories. Sequential un-Higgsing starting from the
ABJM theory for $\C^4$ then leads to the $\C^4/\Z_2^3$ field theory. We have demonstrated in \cite{Benishti:2009ky} that this theory can be Higgsed to theories for all 18 toric partial resolutions.

In \secref{sec:exo} we studied yet another subtlety that does not arise in the better-understood AdS$_5$/CFT$_4$ relative. We showed that, in general, certain resolutions of the CY four-fold singularities do not admit a smooth reduction over the $U(1)$ isometry that acts on the M-theory circle. This observation suggests that the baryonic operators in the dual field theory that correspond to such resolutions cannot be described as usual as determinants of bifundamental fields. We expect such operators to reside in sectors of the field theory that do not admit a Lagrangian description. More work should be done in order to understand the nature of these operators. When considering the RG flow induced by giving VEVs to such operators one can see that in general the IR geometry does not admit a smooth reduction to Type IIA. When such a geometry corresponds to an isolated CY four-fold it is expected, as explained in the main text, that a Lagrangian description of the field theory at the IR fixed point does not exist, and thus a Lagrangian description of the RG flow should not exist either. A better understanding of such RG flows could lead to new insights on the field theories in the far IR. We plan to continue with the study of these issues in the future.

It is natural to extend our analysis and consider backgrounds with exceptional six-cycles, as we briefly considered in \secref{sec:5}. Upon resolution, Euclidean M5 branes can be wrapped on these exceptional divisors. As opposed to the Type IIB counterpart case for four-cycles, the  M2 branes sourcing the background do not break SUSY any further than that preserved by the Euclidean brane in the resolution of the cone. Thus, in very much the same spirit as in \cite{Witten:1996bn}, it is natural to expect that these Euclidean branes contribute as non-perturbative effects to the superpotential, even in the warped case. Nevertheless, a comprehensive understanding of these issues is lacking. We have however taken some first steps towards understanding this by computing the warped volume of such branes. In particular we studied in detail the example $\mathcal{C}(Q^{222})$, which is a certain $\mathbb{Z}_2$ orbifold of $\mathcal{C}(Q^{111})$. In extending our findings to more general geometries we have found expressions which might be of relevance in other contexts. It will be interesting to see how our findings can be realized in the dual field theory candidates that were proposed recently in \cite{Benini:2011cm,Davey:2011mz}.
\part{Symmetry Breaking in AdS$_5$/CFT$_4$} \label{part2}
\chapter{Overview} \label{overview-p2}

Having addressed symmetry-breaking in AdS$_4$/CFT$_3$, we now move on to some new results in the more traditional AdS$_5$/CFT$_4$ story. We will be interested in the richest examples of the AdS$_5$/CFT$_4$ correspondence which are both tractable and also non-trivial. These are given by the $(3+1)$d $\mathcal{N}=1$ gauge theories that arise on a stack of D3 branes probing singular toric CY three-folds. In the case of interest for us the CY singularity is a cone over a five-dimensional Sasaki-Einstein manifold $Y_5$, and as a result the gravity dual is of the form AdS$_5 \times Y_5$ \cite{Klebanov:1998hh,Acharya:1998db,Morrison:1998cs}. For example, one may take $Y_5 = T^{1,1}$ \cite{Klebanov:1998hh}, or the more recently discovered infinite families of Sasaki-Einstein manifolds, $Y^{p,q}$ \cite{Gauntlett:2004zh,Gauntlett:2004yd} and $L^{a,b,c}$ \cite{Cvetic:2005vk,Martelli:2005wy}. In all these cases, the dual field theories \cite{Martelli:2004wu,Bertolini:2004xf,Benvenuti:2004dy,Benvenuti:2005ja,Franco:2005sm,Butti:2005sw} are conjectured to be supersymmetric gauge theories, at an IR conformal fixed point. Over the past few years very powerful techniques have been developed \cite{Hanany:2005ve,Franco:2005rj,Franco:2005sm,Hanany:2005ss,Feng:2005gw} to describe these models in terms of quiver and dimer diagrams, that provide the relevant information concerning the spectrum and couplings of the corresponding gauge theory (for a review we refer the reader to \cite{Kennaway:2007tq}).

Similar to the QCS theories discussed in the previous part, the moduli space of the field theory should at least contain a branch which is dual to the position of the $N$ D3 branes on the geometry. This corresponds to VEVs of mesonic operators in the field theory, which are just traces of product of fields that form closed loops in the quiver. In addition, the moduli space captures all the resolutions of the cone that the branes probe. The supergravity backgrounds constructed from these resolved cones allow, in general, for turning on flat NS-NS B-field and R-R forms. Those that vanish at the boundaries are part of the moduli space of the supergravity background. The supergravity moduli just discussed correspond, in the moduli space of the field theory, to VEVs of baryonic operators. Similar to the QCS theories studied in \secref{part1}, these are gauge-invariant determinant-like operators. The moduli space of such theories was studied recently in \cite{Forcella:2008bb,Martelli:2008cm} and a remarkable match was shown with the supergravity expectations \cite{Martelli:2008cm}. One aim in this part of the thesis is to resolve a puzzle that has been raised in this paper. We aim to show that the directions in the moduli space of the field theory that correspond to the NS-NS B-field moduli in the supergravity are \emph{periodic}. This is obviously expected as the NS-NS B-field is periodic in string theory due to large gauge transformations.

Quiver gauge theories have in general both anomalous and non-anomalous baryonic symmetries. The non-anomalous symmetries are dual to massless gauge fields in the bulk. Such gauge fields originate from KK reduction of the R-R four-form over three-cycles in the Sasaki-Einstein manifold. Thus, the number of non-anomalous symmetries is the same as the number of three-cycles in the Sasaki-Einstein manifold. Breaking of non-anomalous baryonic symmetries in field theory results in the appearance of massless Goldstone bosons and global strings in the IR \cite{Vilenkin:1982ni}. In the AdS/CFT context, this has been studied in the conifold case in \cite{Klebanov:2007cx} and was generalized in \cite{Martelli:2008cm} to general toric CY backgrounds. It was shown that in the supergravity, R-R four-form fluctuations that are sourced by wrapped D3 branes contain these Goldstone bosons. These branes wrap blown-up two-cycles in the bottom of the resolved cone $X$ and form global strings in the Minkowski space directions around which the Goldstone boson has a monodromy. In addition, there are $2\,b_4(X)$ anomalous baryonic symmetries in the classical theory that are broken by quantum effects.

As was shown in \cite{Martelli:2008cm}, for fully resolved geometries, the number of massless modes coming from R-R four-form fluctuations is $b_2(X)=b_4(X)+b_3(Y_5)$. Thus, it exceeds the number of broken non-anomalous symmetries in the field theory. Moreover, there are massless modes that originate from fluctuations in the R-R two-form; these correspond to an additional $b_4(X)$ massless modes\footnote{The scalar partners of the R-R four-form and two-form fluctuations come from metric and B-field fluctuations, respectively.}. Hence, the number of extra massless pseudo-scalars is equal to the number of broken \emph{anomalous} baryonic symmetries. However, the breaking of anomalous baryonic symmetries should not result in Goldstone bosons as these are not true symmetries of the quantum theory. Therefore, the interpretation of the additional massless modes just described in the field theory side is not clear. In \cite{Martelli:2008cm} it was suggested that such modes should be lifted by non-perturbative corrections. In this thesis we want to show that the additional massless modes coming from R-R four-form fluctuations can in fact be interpreted as Goldstone bosons in some special cases. 

As a start we study the moduli space of the field theory living on D3 branes probing the $\C^3/\Z_3$ orbifold. We examine the appearance of non-conformal theories for specific Higgsings. Such phases, we claim, appear whenever the resolved geometry contains four-cycles. Such Higgsings were observed before in the brane tilling context \cite{Hanany:2005ve}, where the non-conformality translates to inconsistent tillings. This was also studied in \cite{Krishnan:2008kv} for the $\C^3/\Z_3$ orbifold. In this thesis, we identify systematically which VEVs lead to such non-conformal theories and suggest gravity interpretations of these RG flows. More specifically, we show that these VEVs correspond in the supergravity to specific values of the background compactly-supported B-field modes. These B-field values allow for D3 branes to wrap two-cycles in the resolved background. We suggest that these D3 branes are dual to global strings that appear in the field theory due to the breaking of the non-anomalous baryonic symmetry that emerges along the RG flow. This allows us to interpret the additional massless modes originating from R-R four-form fluctuations, which were discussed above, as Goldstone bosons.

The organization of this part is as follows. In \secref{sec:ads5} we review the quiver field theories and the dual AdS$_5 \times Y_5$ supergravity backgrounds of interest. In \secref{sec:B3} we discuss in detail the $\C^3/\Z_3$ orbifold. We compare its moduli space with that expected from supergravity. Then we turn to discuss the non-conformal theory and the VEVs that lead to it. We end this section with a supergravity analysis of the corresponding RG flow. A summary and concluding comments are given in \secref{sec:B3}. 
\chapter{AdS$_5$ backgrounds and field theories} \label{sec:ads5}
In this chapter we briefly review the $\mathcal{N}=1$ supersymmetric 
quiver field theories of interest, focusing in particular on their vacuum moduli spaces. 
For further details the reader is referred to \cite{Martelli:2008cm} and references therein. 
\section{Quiver gauge theories} \label{section2}
Our starting point is an $\mathcal{N}=1$ gauge theory in $(3+1)$d with 
product gauge group $\mathcal{G}\equiv\prod_{i=1}^G U(N_i)$, an arbitrary number of bifundamental $X_{ab}$ in the representation $(\square_a,\bar{\square}_b)$ under the $(a,b)$-th gauge groups, and a superpotential $W$. 

We want to show now that the $G$ central $U(1)$s in $\mathcal{G}$ become global symmetries in the IR. It is easy to see that the diagonal $U(1)$ does not couple to any matter field. The rest of the $U(1)$s, due to triangle anomalies in the quiver, are divided into anomalous and non-anomalous symmetries\footnote{Such anomalies were not discussed in \secref{part1} because they can occur only in even spacetime dimensions.}. The gauge coupling of the non-anomalous $U(1)$s vanishes in the IR and the anomalous $U(1)$ gauge fields become massive by mixing with closed
string fields (due to the cancellation of the anomaly via the Green-Schwarz mechanism). 
To see which symmetries are anomalous consider the $U(1)_{\vec{q}} \subset U(1)^G$ generated by $\mathcal{A}=\sum_{i}^G q_i\mathcal{A}_i$ where $\mathcal{A}_i$ are the generators of the central $U(1)_i\subset U(N_i)$. 
The condition for the cancellation of such triangle anomalies $Tr[U(1)_{\vec{q}}SU(N_k)^2]$ is just
\bea\label{anom}
\sum_{X_{i,j}|i=k}N_j\,q_j-\sum_{X_{i,j}|j=k}N_i\,q_i=0~.
\eea

Modding out by the global symmetries we see that in the IR the gauge group becomes
\bea\label{SUgroup}
\mathcal{SG}=\prod_{i=1}^G SU(N_i)~.
\eea

\subsection{Classical vacuum moduli space}

The classical VMS $\mathscr{M}$ is determined by the following equations 
\begin{eqnarray}\label{VMSeqns2}
\nn \partial_{X_{ab}} W &=& 0~,\\
\mu_a := -\sum\limits_{b=1}^G X_{ba}^{\dagger} {X_{ba}} + 
\sum\limits_{c=1}^G  {X_{ac}} X_{ac}^{\dagger} &=& 
0~
\end{eqnarray}
which are the F-term and D-term equations respectively. 

In the Abelian case the moduli space $\mathscr{M}$ is straightforward to describe. The first equation 
describes the space of F-term solutions, which is by construction an affine 
algebraic set. For the theories we study in this thesis, this is itself a toric variety, of dimension 
$4+(G-2) = G+2$. This is the same master space $\mathscr{F}_{G+2}$ that was mentioned in \secref{sec:3} and studied 
in detail in \cite{Forcella:2008bb}.

Finally, the combination of imposing the second equation in \eqref{VMSeqns2} and identifying by the gauge symmetries
may be described as a K\"ahler quotient of $\mathscr{F}_{G+2}$ by a subgroup\footnote{Notice that the subgroup by which we mod is the crucial difference between the computations of the moduli spaces of the $(2+1)$d quivers, studied in \secref{part1}, and the $(3+1)$d quivers that we discuss here.} 
$U(1)^{G-1}\subset U(1)^G$. This subgroup does not include the diagonal $U(1)$ that does not couple to the bifundamental fields. 
In particular, this K\"ahler quotient precisely sets the $\mu_i$ in \eqref{VMSeqns2}
equal to zero. To summarize, 
\begin{equation}\label{symp}
\mathscr{M} \cong \mathscr{F}_{G+2} \, //\, U(1)^{G-2}~,
\end{equation}
where the K\"ahler quotient is taken at level zero, implying that $\mathscr{M}$ is a 
K\"ahler cone. We will refer to this space as the Abelian mesonic moduli space. For a stack of $N$ coincident D3 branes transverse to a CY three-fold singularity, one expects the moduli space to be the $N$th symmetric product of the three-fold. 

The moment maps associated to the $G-1$ global $U(1)$s can take any non-vanishing values. In the physics literature this is often associated with turning on FI parameters $\eta_i$ that contribute to the D-terms\footnote{Strictly speaking, since the $U(1)$s are not gauged no FI parameters can be turned on. The FI-like contributions to the D-terms come from VEVs for fields.}. We will use this terminology throughout this thesis. The K\"ahler quotient in \eqref{symp} should then be taken with respect to 
\begin{eqnarray}\label{moment}
\mu_a := -\sum\limits_{b=1}^G X_{ba}^{\dagger} {X_{ba}} + 
\sum\limits_{c=1}^G  {X_{ac}} X_{ac}^{\dagger} &=& 
\eta_a~.
\end{eqnarray}
Setting the values of the FI parameters picks a point in $\mathscr{F}_{G+2}$ that solves \eqref{moment}. On the gravity side, as we explain in \secref{section3}, these FI parameters correspond to the K\"ahler class of the resolved CY geometry and the compactly-supported B-field periods. 

As opposed to the $(2+1)$d QCS theories discussed in \secref{part1}, the field theories that are dual to isolated CY three-fold singularities with vanishing four-cycles are not identified with the minimal GLSM. This is due to the fact that the moduli space in Type IIB is not completely geometrical and contains also directions coming from flat form-fields. Therefore, the matching between the field theory and the gravity moduli space is not as obvious as before.
It is then convenient to present the $\mathscr{F}_{G+2}$ moduli space of a field theory with the aid of a FI space. Since $\sum_a \eta_a=0$, as may be seen by summing over the moment maps in \eqref{moment}, this space is just $\R^{G-1}$. The additional phases, corresponding to the non-vanishing moment maps, survive the K\"ahler quotient and are fibered over the FI space to form $(\C^*)^{G-1}$. The structure of $\mathscr{F}_{G+2}$ corresponds (very loosely) to $\mathscr{M}(\eta)$ fibered over $(\C^*)^{G-1}$, where $\mathscr{M}(\eta)$ is obtained from \eqref{symp} with the corresponding moment map levels. As we will see later on, in the corresponding resolved CY three-fold $X$, $b_2(X)$ of the $G-1$ directions\footnote{Recall from \cite{Martelli:2008cm} that $G-1=b_2(X)+b_4(X)$.} in the FI space correspond to K\"ahler moduli, while the other $b_4(X)$ are dual to the compactly-supported B-field moduli. The fact that the latter are periodic, as expected from the periodicity of the B-field in string theory, is not obvious and we will show it later on in an example.

The FI space is expected to be divided into \textit{chambers}, where each one of them corresponds to the fully resolved geometry. These chambers are separated by \textit{walls}. Part of the walls correspond to singular $\mathscr{M}(\eta)$ spaces. When crossing such a wall the $\mathscr{M}(\eta)$ moduli space undergoes a form of small birational transformation called a flip \cite{flip}. Inside each such chamber the K\"ahler classes vary linearly with respect to the FI parameters and the $\mathscr{M}(\eta)$ spaces are isomorphic. These chambers, however, are further divided, as we will show later on in this thesis, by additional walls that correspond to critical values of the compactly-supported B-field periods. 

\section{Gravity backgrounds}\label{section3}


\subsection{AdS$_5$ backgrounds}

In this subsection we want to discuss the Type IIB supergravity solution obtained by placing $N$ D3 branes at the tip of CY three-fold $X$, which is a cone over a Sasaki-Einstein space $Y_5$. The corresponding metric and five-form are 
\bea\label{D3metric}
\diff s_{10}^2 &=& h^{-1/2}\diff s^2(\R^{1,3}) + h^{1/2} \diff s^2(X)~, \\ \label{metric}
G_5 & = & (1+*_{10})\diff h^{-1}\wedge \diff\vol_4~,
\eea
where the warp factor $h$ reads
\bea\label{Hfull}
h = 1+ \frac{L^4}{r^4}~.
\eea
Here $\diff s^2(\R^{1,3})$ is four-dimensional Minkowski space, with volume form $\diff\vol_4$, and
$L$ is a constant given by
\bea\label{Lrel}
L^4 = \frac{(2\pi)^4g_s(\alpha')^2N}{4\vol(Y_5)}~.
\eea
The near-horizon limit corresponds to the small $r$ limit; thus, the metric (\ref{D3metric}) becomes
\bea\label{AdSmetric}
\diff s_{10}^2\, =\, \frac{L^2}{r^2}\diff r^2 + \frac{r^2}{L^2} \diff s^2(\R^{1,3}) + L^2 \diff s^2(Y_5)~.
\eea
This is just the AdS$_5\times Y_5$ metric. To fix the background one needs to specify the flat NS-NS and R-R background fields. Turning on such fields corresponds to exactly marginal deformations in the $\mathcal{N}=1$ superconformal
field theories living on the boundary. For a comprehensive discussion on such deformations we refer the reader to \cite{Martelli:2008cm}. 

In the next subsection we want to review the supergravity backgrounds that are obtained by resolving the CY cones. These backgrounds correspond to non-vanishing VEVs of the scalar operators, which Higgses the field theory.

\subsection{Symmetry-breaking backgrounds}\label{defsection}

The symmetry-breaking backgrounds of interest for us are analogous to the backgrounds that were discussed in \secref{sec:generalwarped}, thus we keep our discussion in this subsection brief. The main emphasis will be placed on the important differences that arise in the analysis of the form-field moduli. 

The crepant (partial) resolutions of the toric three-fold singularities are described by the GKZ fan that lives in $\R^{b_2(X)}$. A point inside each of the polyhedral cones constructing the GKZ fan is a K\"ahler class on the resolved CY $X$. A CY metric on $X$ that approaches the given cone metric asymptotically can always be found \cite{crep4,crep5,crep6}. The ten-dimensional metric in the resolved background is
\bea
\diff s_{10}^2 &=& h^{-1/2}\diff s^2(\R^{1,3}) + h^{1/2} \diff s^2(X)~, \label{resolved-metric}
\eea
with $G_5$-flux still given by \eqref{D3metric}. Placing $N$ spacetime-filling D3 branes at a point $y\in X$ leads to the warp factor equation
\bea\label{green2}
\Delta_x h[y] = -\frac{(2\pi)^4g_s(\alpha^{'})^2N}{\sqrt{\det g_X}} \delta^6(x-y)~.
\eea
Here $\Delta h = \diff^* \diff h = - \nabla^i\nabla_i h$ is the scalar Laplacian of $X$ acting on $h$. 
Having chosen a particular resolution and K\"ahler class, hence metric, we must then find the warp factor $h$ satisfying (\ref{green2}) and this always exists and is unique. A general discussion in the Type IIB context may be found in \cite{Martelli:2007mk}. The warped metric \eqref{resolved-metric} describe a geometry with two asymptotically AdS$_5$ regions: one near $r=\infty$ that is asymptotically AdS$_5\times Y_5$, where $Y_5=Y_{\mathrm{UV}}$ is the base of the unresolved cone, and one near to the point $y$, which is asymptotically AdS$_5\times Y_{\mathrm{IR}}$, where $Y_{\mathrm{IR}}$ is the base of the tangent cone at the stack of D3 branes. 

Generally, the topology of $X$ allows one to turn on various topologically non-trivial flat form-fields. The forms of interest sit in compactly supported cohomology classes, and thus correspond to \emph{fixed} marginal couplings in the field theory in the UV. These form-field moduli are discussed in \cite{Martelli:2008cm}. In particular, we have the NS-NS B-field, as well as the R-R two-form $C_2$, which are harmonic two-forms that are $L^2$ normalizable with respect to the unwarped metric, and the four-form $C_4$ which is a harmonic $L^2$ normalizable four-form with respect to the warped metric. The B-field modes, which live in $H^2(X, Y_5 ;\R)/H^2(X, Y_5 ;\Z) = U(1)^{b_4(X)}$, are identified with $b_4(X)$ FI parameters in the field theory. The other $b_2(X)$ FI parameters correspond to $b_2(X)$ K\"ahler moduli. The GKZ fan, which was discussed above, should match the $\R^{b_2(X)} \subset \R^{G-1}$ sub-space that is obtained by projecting the FI space on the K\"ahler directions. Moreover, the K\"ahler moduli are naturally complexified by noting that $H^4(X,Y_5,\R)\cong H_2(X,\R)\cong \R^{b_2(X)}$ by Poincar\'e duality, and that this group classifies the periods of $C_4$ through four-cycles in $X$, which may either be closed or have a boundary three-cycle on $Y_5=\partial X$. More precisely, taking into account large gauge transformations leads to the torus $H^4(X,Y_5,\R)/H^4(X,Y_5,\Z)\cong U(1)^{b_2(X)}$. The pairing of the K\"ahler class with $C_4$, and the B-field with $C_2$, is reflected by the complexification $(\C^{\star})^{G-1}$ of the global baryonic symmetry group in the field theory. 

This supergravity background may be interpreted as a renormalization group flow from the initial $\mathcal{N}=1$ superconformal field theory to a new SCFT in the IR as was first shown in \cite{Klebanov:1999tb}. There may be additional light particles in the IR, namely Goldstone bosons associated to the spontaneous breaking of non-anomalous baryonic symmetries.
\chapter{The $\mathbb{C}^3/\mathbb{Z}_3$ orbifold}\label{sec:B3}

We wish to discuss field theories dual to toric CY three-folds with four-cycles in their resolutions. The simplest example of such a theory is the $\mathbb{C}^3/\mathbb{Z}_3$ orbifold theory. The space $\C^3/\Z_3$ is defined as the three-dimensional complex space $\C^3$ under the identification
\bea
\{x_1,x_2,x_2\} \sim \{w\,x_1,w\,x_2,w\,x_3\}, \ \, w^3=1~.
\eea
The only fixed point under this identification is the origin, therefore the near horizon geometry close to the point-like $N$ D3 branes is smooth. More specifically, from \eqref{AdSmetric} one sees that this is just the AdS$_5 \times S^5/\Z_3$ space. 

\section{Field theory description}
 
\begin{figure}[ht]
\begin{center} 
\includegraphics[scale=.8]{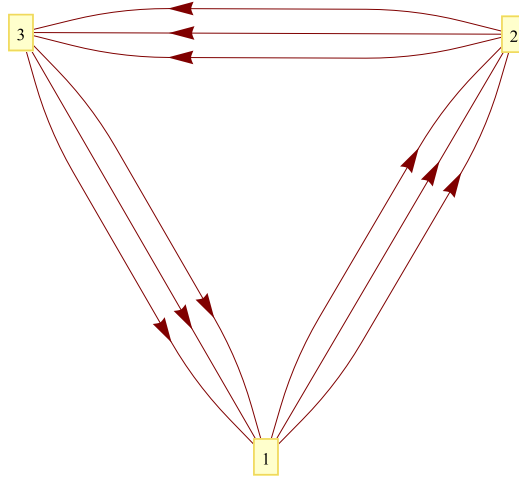} 
\end{center}
\caption{The quiver diagram for a dual of $\mathbb{C}^3/\mathbb{Z}_3$.}
\label{Z3-quiver}
\end{figure}

A stack of $N$ D3 branes propagating on the $\mathbb{C}^3/\mathbb{Z}_3$ orbifold is dual to a $(3+1)$d SCFT described by the quiver in Figure~\ref{Z3-quiver}.
The field theory has a superpotential that reads
\begin{equation}
W={\rm Tr}\Big(\, X_{12}^i\, X_{23}^j\, X_{31}^k\,\Big)\,\epsilon_{ijk}~.
\end{equation}
This theory was studied in detail in \cite{Gukov:1998kn}. In the UV the gauge group is just $U(N)^3$ containing a central $U(1)^3$ which become global symmetries in the IR. The diagonal $U(1)$ decouples as no matter field carries charge under it. In addition, any other combination of $U(1)$s is anomalous as can be seen from \eqref{anom}. One may choose the following orthogonal generators for these $U(1)$s 
\bea
\mathcal{A}_{B_1}=\mathcal{A}_1-\mathcal{A}_2 \ , \quad \mathcal{A}_{B_2}=\mathcal{A}_1+\mathcal{A}_2-\mathcal{A}_3 \ ,
\eea
where $\mathcal{A}_i$ are the generators of the corresponding $U(1)$s in the quiver.

\subsection{The GLSM description}

We can compute the moduli space in the usual way. At a generic point, by making use of gauge rotations, we can set all the fields to be diagonal. Then, the effective theory reduces to $N$ copies of the $U(1)$ theory. To compute the moduli space of the Abelian theory one can use the \textit{forward algorithm} of \cite{Douglas:1997de, Feng:2000mi}. The key point is that the algorithm takes the data of the matter content and the superpotential, and produces the GLSM charge matrix $Q_t$. The kernel of this is the $G_t$ matrix that encodes the toric diagram of the CY three-fold. We refer the reader to \cite{Feng:2000mi} for more details.

\begin{figure}[ht]
\begin{center} 
\includegraphics[scale=1.2]{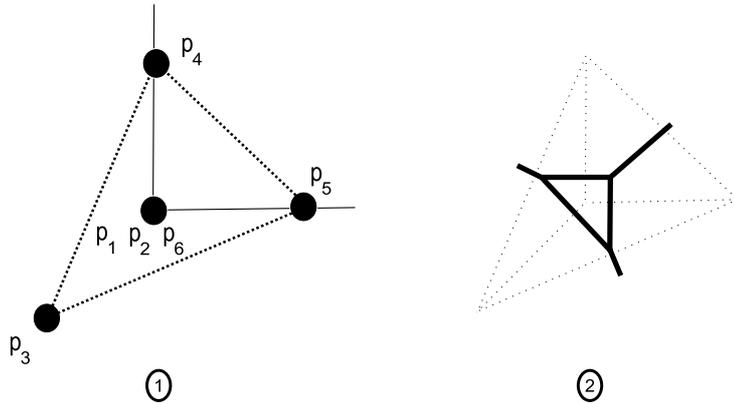} 
\end{center}
\caption{The $\C^3/\Z_3$ geometry. 1. The toric diagram of $\C^3/\Z_3$. 2. The pq-web is presented on top of the triangulation of the toric diagram that corresponds to the fully resolved $\C^3/\Z_3$ orbifold.}
\label{Z3-toric}
\end{figure}

Working in the GLSM description we may write the relations between the $X_{i\,j}^k$ fields and the GLSM fields, which are usually named $p$-fields, as follows
\bea
\nn \label{Z3-pfields}
&&X_{12}^1=p_1\,p_3 \ , \quad X_{12}^2=p_1\,p_4 \ , \quad X_{12}^3=p_1\,p_5 \ ,\\  \nn
&&X_{23}^1=p_2\,p_3 \ , \quad X_{23}^2=p_2\,p_4 \ , \quad X_{23}^3=p_2\,p_5 \ ,\\ 
&&X_{31}^1=p_6\,p_3 \ , \quad X_{31}^2=p_6\,p_4 \ , \quad X_{31}^3=p_6\,p_5~.
\eea
Using the forward algorithm one can show that 
\bea \label{Z3-Qt}
Q_t=\left(
\begin{array}{c c c c c c | c} 
 p_1 & p_2 & p_3 & p_4 & p_5 & p_6 & FI \\ \hline
 0 & 0 & 1 & 1 & 1 & -3 & a\\ 
 0 & -1 & 0 & 0 & 0 & 1 & b \\
 1 & 1 & -1 & -1 & -1 & 1 & 
\end{array}\right)~.
\eea
Now it is clear that, using the second and third rows in \eqref{Z3-Qt}, one can derive an expression for $p_1$ and $p_2$, which corresponds to internal points, in terms of the other $p$-fields. Explicitly one may write
\bea
|p_1|^2=|p_6|^2+a+b \ , \quad \ |p_2|^2=|p_6|^2-b \ .
\eea
In the singular cone, for which $a=b=0$, $|p_1|$ and $|p_2|$ are fixed by $|p_6|$. Moreover, each row $\ell$ in \eqref{Z3-Qt} encodes the charges of the $p$-fields under the GLSM $U(1)_{\ell}$ gauge transformation. One can use $U(1)_{2}$ and $U(1)_{3}$ to fix the phases of $p_1$ and $p_2$. We are then left with four degrees of freedom, $p_3$, $p_4$, $p_5$ and $p_6$, and the constraint coming from the first row in the charge matrix is
\bea
|p_3|^2+|p_4|^2+|p_5|^2=3\,|p_6|^2 \ ,
\eea
which after the corresponding gauge identification gives $\C^3/\Z_3$. Indeed, taking the FI parameters to zero one can compute the $G_t$ matrix by taking the null-space of $Q_t$ 
\bea
G_t=
\left(
\begin{array}{cccccc}
p_1 & p_2 & p_3 & p_4 & p_5 & p_6 \\ \hline
 1 & 1 & 1 & 1 & 1 & 1 \\
 0 & 0 & -1 & 0 & 1 & 0 \\
 0 & 0 & -1 & 1 & 0 & 0
\end{array}
\right)~.
\eea
The columns of the $G_t$ matrix are the coordinates in the toric diagram, which is presented in Figure~\ref{Z3-toric}. This is indeed the toric diagram that describes the $\C^3/\Z_3$ orbifold. Notice that the GLSM description is non-minimal, and this results in multiplicities of the lattice points in the toric diagram.
Note that each spacetime field is the product of an inner and an outer $p$-field.

\subsection{The non-conformal theory} \label{sec:noncon}

In \cite{Krishnan:2008kv} the Higgsing of the $\C^3/\Z_3$ field theory was studied, and two possible RG flows that end at $\mathcal{N}=4$ $SU(N)$ SYM theory were proposed. In the first scenario, the RG flow is induced by giving diagonal VEVs to two fields that connect different nodes in the quiver. In the second case, the RG flow is composed of two steps. In the first step, an interesting non-conformal theory appears after giving a VEV to one of the fields and RG flowing to the VEV scale. This corresponds in the GLSM language to giving a VEV to one external and one internal $p$-field as may be seen from \eqref{Z3-pfields}. One of the gauge nodes in this theory confines and by giving a VEV to a mesonic operator the theory seems to flow to the $\mathcal{N}=4$ $SU(N)$ SYM theory in the deep IR. 

\begin{figure}[ht]
\begin{center} 
\includegraphics[scale=.8]{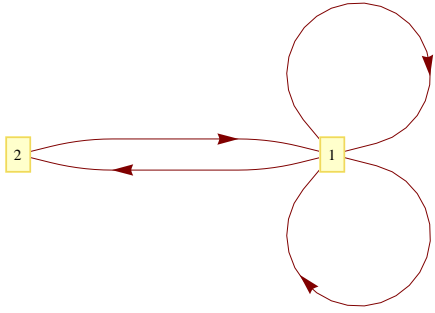} 
\end{center}
\caption{The quiver diagram for the non-conformal theory obtained by Higgsing the $\C^3/\Z_3$ theory. Both nodes are $SU(N)$.}
\label{HVZ-quiver}
\end{figure}

We want now to review how the non-conformal theory appears. For the sake of simplicity we will discuss just diagonal VEVs that correspond to keeping the D3 branes in one stack. Let us concentrate for the time being on the VEV $\langle X_{12}^1 \rangle=v\,I_{N \times N}$. This VEV breaks the anomalous global symmetry $\mathcal{A}_{B_1}$ and leaves $\mathcal{A}_{B_2}$ untouched. After giving such a VEV we see from the following part in the superpotential
\begin{equation}
W=\,v\,{\rm Tr}\Big(\, X_{23}^2\, X_{31}^3\,-\, X_{23}^3\, X_{31}^2\,\Big)\,+\,.\,.\,.
\end{equation}
that $X_{23}^2$, $X_{31}^3$, $X_{23}^3$ and $X_{31}^2$ become massive. At scales below $v$, these fields should be integrated out. The resulting quiver of the effective theory is presented in Figure~\ref{HVZ-quiver}.
The new effective superpotential reads
\begin{equation} \label{HVZ-W}
W={\rm Tr}\Big(\, [X_{11}^1,X_{11}^2]\, X_{12}\, X_{21}\,\Big)~.
\end{equation}

We may look, for example, at node 2. Since it has $N_f=N_c$, this theory cannot be in a conformal fixed point. 
In \cite{Krishnan:2008kv} it was argued that node 2 in this theory confines. This is indeed expected if the dynamical scale of node 2 is dominating. This will be consistent with the supergravity analysis that will be presented later on in this thesis.
In the IR one can treat the field theory effectively as a copy of $N_f=N_c$ SQCD coupled to a pair of singlets $X_{11}^i$ with the superpotential of the non-conformal theory. The IR of the SQCD theory is described in terms of mesons and baryons
\bea
\nn
&&\mathcal{M}^i_j=(X_{12})^j_{\alpha}\,(X_{21})^{\alpha}_i\, ,\qquad \mathcal{B}^{i_1\cdots i_N}=\epsilon^{\alpha_1\cdots \alpha_N}\, (X_{12})^{i_1}_{\alpha_1}\cdots (X_{12})_{\alpha_N}^{i_N}\, , \\ &&\tilde{\mathcal{B}}_{i_1\cdots i_N}=\epsilon_{\alpha_1\cdots \alpha_N}\, (X_{21})_{i_1}^{\alpha_1}\cdots (X_{21})^{\alpha_N}_{i_N}\, ,
\eea
where Latin indices stand for $SU(N)_1$ while Greek indices stand for $SU(N)_2$. Note that the superpotential is then written as
\begin{equation}
W={\rm Tr}\Big(\, \mathcal{M}\,[X_{11}^1,\, X_{11}^2]\,\Big)~.
\end{equation}
The baryons carry charge under the baryonic symmetry under which the fields transform as 
\begin{equation}
X_{12}\rightarrow \me^{i\theta}\, X_{12}\qquad X_{21}\rightarrow \me^{-i\theta}\, X_{21} ~.
\end{equation}
This, we claim, is a non-anomalous symmetry that was evolved from $\mathcal{A}_{B_2}$ during the RG flow.

As argued by Seiberg \cite{Seiberg:1994bz}, the theory has a quantum mechanically modified moduli space given by
\begin{equation}
{\rm det}\,\mathcal{M}-\mathcal{B}\, \tilde{\mathcal{B}}=\Lambda^{2N}~,
\end{equation}
where $\Lambda$ is the dynamically generated scale of the SQCD theory. In addition, we have to impose the F-terms coming from the superpotential. Thus, finally, the moduli space is given by the solutions to
\begin{equation}
{\rm det}\,\mathcal{M}-\mathcal{B}\, \tilde{\mathcal{B}}=\Lambda^{2N},\qquad [X_{11}^i,\, \mathcal{M}]=0,\qquad [X_{11}^1,\,X_{11}^2]=0~.
\end{equation}
Note that, since node 2 is confined, in the IR the relevant degrees of freedom are mesons and baryons. As such, the superpotential appears written in terms of the meson $\mathcal{M}$, and thus automatically generates an F-term ensuring that the $X_{11}^i$ fields commute. Let us introduce a Lagrange multiplier chiral field $\lambda$, so that we can write
\begin{equation}
W={\rm Tr}\Big(\, \mathcal{M}\,[X_{11}^1,\,X_{11}^2]\,\Big)+\lambda\, \Big({\rm det}\,\mathcal{M}-\mathcal{B}\, \tilde{\mathcal{B}}-\Lambda^{2N}\Big)~.
\end{equation}

Vanishing VEVs for the entire set of fields is not a point on the moduli space of this theory. As we explain now, a conformal fixed point can be reached by considering non-vanishing VEVs that induce continuation of the RG flow. In \cite{Krishnan:2008kv} the authors discussed the RG flow resulting from a non-vanishing VEV for the mesonic operator $\mathcal{M}$ only. This RG flow ends at the $\mathcal{N}=4$ $SU(N)$ SYM theory in the IR. It was then speculated there that such RG flows might be dual to RG flows that are induced by giving VEVs to two fields that connect different nodes in the quiver. This was justified by the fact that these flows also end at $\mathcal{N}=4$ $SU(N)$ SYM theory in the IR. This speculation will be falsified later on as it will be shown that the first RG flow corresponds to a choice of critical value for the compactly supported B-field in the dual supergravity background. 

In this thesis, however, we will be interested in the flow that results from giving VEVs to the baryons. After setting the only non-vanishing VEV as $\mathcal{B}\tilde{\mathcal{B}}=-\Lambda^{2N}$ and flowing to the new IR, the superpotential reduces to
\begin{equation}
W={\rm Tr}\Big(\, \mathcal{M}\,[X_{11}^1,\,X_{11}^2]\,\Big)~.
\end{equation}
We see that the theory in the IR is just $\mathcal{N}=4$ $SU(N)$ SYM. We expect to see a Goldstone boson and global string due to the fact that the non-anomalous baryonic symmetry was broken. 

It is interesting to notice, as was argued in \cite{Krishnan:2008kv}, that in the gravity there is just one scale, the size of the four-cycle. Thus, as expected, in the large $N$ limit in the field theory the scale in which the massive fields are integrated out is also the scale in which node 2 in the non-conformal theory confines. The strong coupling scale for the confining is related to the energy scale $E$ set by the VEV $v$ as:
\bea
\Lambda = E\,\me^{-\frac{8\,\pi^2}{2\,N\,g^2_{YM}(E)}}~,
\eea
where $E = \sqrt{v}$. For large t'Hooft coupling $\lambda \equiv N\,g_{YM}^2$ the scale $\Lambda$ and $E$ are of the same order and we cannot distinguish between them. Therefore, we cannot expect to see the moduli space of the non-conformal theory and the emerging baryonic symmetry in the dual supergravity. But in the far IR we can expect to see the Goldstone mode and global string that correspond to the broken global baryonic symmetry. In the next subsections we want to show that indeed these \emph{can} be observed in the dual supergravity background.

We want to start by describing the full moduli space of the $\C^3/\Z_3$ theory. It will prove useful to understand what are the moduli in the supergravity that correspond to the VEVs that lead to the non-conformal theory.
\subsection{$\C^3/\Z_3$ theory - moduli space}
From \eqref{Z3-Qt}, the VMS equations are just
\bea \label{vmsequ}
\nn
&&|p_3|^2\,+|p_4|^2\,+|p_5|^2\,-\,3\,|p_6|^2\,=\,a~, \\ \nn
&&|p_2|^2\,-\,|p_6|^2\,=\,-b~, \\ 
&&|p_1|^2\,+\,|p_2|^2\,-\,|p_3|^2\,+|p_4|^2\,+|p_5|^2\,+\,|p_6|^2\,=\,0~,
\eea
where one needs to mod out by the corresponding $U(1)$ transformations. We now show that these equations can be written in three different forms, in which two different internal $p$-fields can be eliminated. These correspond to three different chambers in the FI parameter space of the field theory. 

\subsection*{Chamber one}
We rearrange \eqref{vmsequ} as
\bea \label{vmsequ1}
\nn
&&|p_3|^2\,+|p_4|^2\,+|p_5|^2\,=3\,|p_2|^2\,+\,a\,+\,3\,b~, \\ \nn
&&|p_6|^2\,=|p_2|^2\,+\,b~, \\ 
&&|p_1|^2\,=\,|p_2|^2\,+\,a\,+2\,b~,
\eea
such that it is obvious that $p_1$ and $p_6$ can be eliminated with respect to the other $p$-fields, and their corresponding phases can be fixed. We have to take $b \geq 0$ and $a+2b \geq 0$ to guarantee that solutions to these fields always exist. We are left with the first line that describes branes on the resolved $\C^3/\Z_3$ where the K\"ahler class is proportional to $a+3b$. In every case we must necessarily find this geometry since the only CY resolution of the $\C^3/\Z_3$ singularity is $\mathcal{O}(-3)\rightarrow\mathbb{CP}^2$. This can be easily seen from the pq-web in Figure~\ref{Z3-toric} that represents the resolution of the CY. 

To obtain the pq-web one first needs to choose a triangulation of the toric diagram. This corresponds to a specific resolution of the CY space. For $\C^3/\Z_3$ there is just one choice, as can be seen from Figure~\ref{Z3-toric}, which also describes the fully resolved space. The pq-web is obtained by replacing faces by vertices, lines by orthogonal lines, and vertices by faces in the triangulation of the toric diagram. This allows one to map the topology of the resolved space into a two-dimensional diagram \cite{Aharony:1997bh,Leung:1997tw}. Indeed, the triangle in the pq-web represents the compact divisor $\mathbb{CP}^2$, which is in agreement with the GLSM picture. The $\mathbb{CP}^2$ zero section is at $p_2=0$ in this chamber of the gauge theory FI space.

If we take $|p_2| > 0$ we see from \eqref{vmsequ1} that some of the external $p$-fields and all the internal $p$-fields obtain a non-vanishing VEV. This corresponds to putting the D3 branes away from the four-cycle and in the field theory to a VEV of a closed loop of fields in the quiver that corresponds to a mesonic operator. This Higgses the $SU(N)^3$ gauge group to $SU(N)$ and the theory flows to $\mathcal{N}=4$ $SU(N)$ SYM in the IR. Notice that in order to obtain the non-conformal theory we were discussing, we need to blow-up the four-cycle by taking $a+3b>0$, put the D3 branes on the blown-up four-cycle by setting $p_2=0$, and take $b=0$ or $a+2b=0$. The latter condition guarantees that just one internal point has a non-vanishing VEV. On the other hand, setting $b>0$ and $a+2b>0$ results in an RG flow that ends at the $\mathcal{N}=4$ $SU(N)$ SYM in the IR. 

The position of the branes on the four-cycle is determined by $p_3$, $p_4$ and $p_5$, which are constrained by the first row in \eqref{vmsequ}. This is just a point on $\mathbb{CP}^2$ and all such points are equivalent, since $\mathbb{CP}^2$ is homogeneous. Thus, these directions can be suppressed when discussing the moduli space of the field theory. Therefore, the interesting information on the moduli space of the field theory is just the FI parameter space.

\subsection*{Chamber two}
A straightforward manipulation of \eqref{vmsequ} leads to
\bea
\nn
&&|p_3|^2\,+|p_4|^2\,+|p_5|^2\,=\,3\,|p_6|^2\,+\,a~, \\ \nn
&&|p_2|^2\,=\,|p_6|^2\,-b~, \\ 
&&|p_1|^2\,=\,|p_6|^2\,+a+b~.
\eea
One can see that $p_1$ and $p_2$ may be eliminated and their phases fixed. This can be achived if we restrict $a \geq 0$, $b \leq 0$ and $a+b \geq 0$. The discussion is then along the same lines as in the previous paragraph. This time, to obtain the non-conformal theory, we take $a>0$, $p_6=0$ and $b=0$ or $a+b=0$. Again, since $p_6=0$, the branes sit on the blown-up cycle with the resolved geometry being the same as before. 

\subsection*{Chamber three}
A straightforward manipulation of \eqref{vmsequ} leads to
\bea
\nn
&&|p_3|^2\,+|p_4|^2\,+|p_5|^2\,=\,3\,|p_1|^2\,-\,2a-3b~, \\ \nn
&&|p_2|^2\,=\,|p_1|^2\,-a-2b~, \\ 
&&|p_6|^2\,=\,|p_1|^2\,-a-b~.
\eea
One can see that $p_2$ and $p_6$ may be eliminated and their phases fixed. This can be achived if we restrict $a \leq - \frac{3}{2}\,b$, $a \leq -2b$ and $a \leq - b$. We see similar behaviour as for the other chambers that we just described. To obtain the non-conformal theory we have to take $-\,2a-3b>0$, $p_1=0$ and $a+2b=0$ or $a+b=0$, such that the branes again sit on the blown-up cycle of the resolved geometry. 
\begin{figure}[ht]
\begin{center}
\includegraphics[scale=0.7]{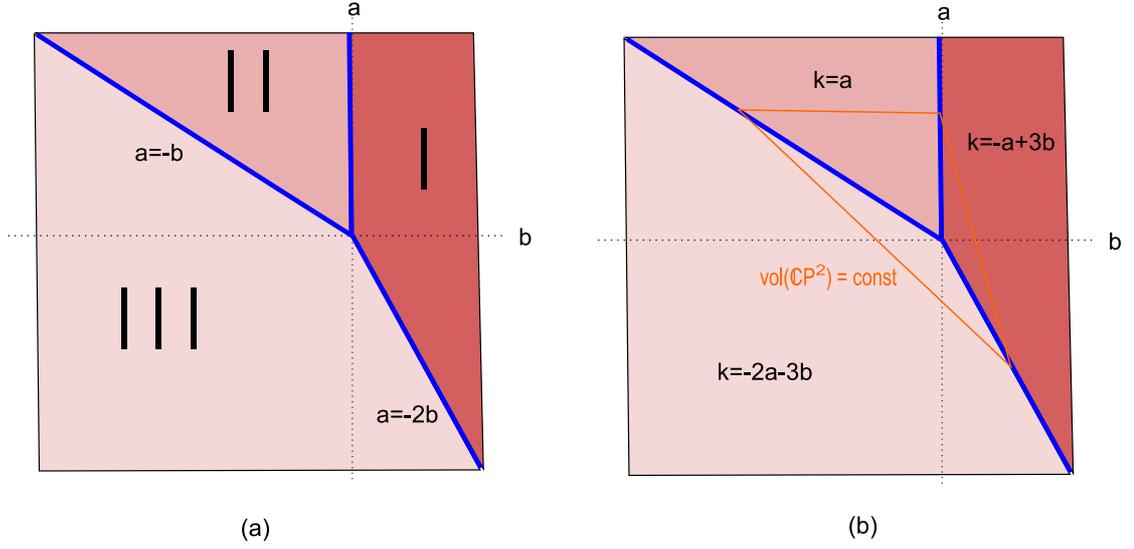}
\end{center}
\caption{The FI space of the $\C^3/\Z_3$ theory. (a) The three chambers of the moduli space. The blue lines correspond to the FI parameter values that result, after RG flow, in the non-conformal theory in Figure~\ref{HVZ-quiver}. (b) The FI path dual to the B-field period is plotted in orange for the fixed FI parameter $k$, which is the dual to the K\"ahler modulus in each chamber.}
\label{GKZ}
\end{figure}

To conclude, giving VEVs to fields, such that no closed loop of fields in the quiver has a non-vanishing VEV, results in a blown-up four-cycle, and the D3 branes sit on the exceptional cycle $\mathbb{CP}^2$ in $\mathcal{O}(-3)\rightarrow \mathbb{CP}^2$. Figure~\ref{GKZ} (a) describes this part of the moduli space. The blue lines, which are the borderlines of the different regions, correspond to the values of the FI parameters $a$ and $b$ that bring us to the non-conformal theory after RG flow to the VEV scale. Any other point in the diagram corresponds to VEVs that induce RG flows that end at $\mathcal{N}=4$ $SU(N)$ SYM in the IR.
In Figure~\ref{GKZ} (b) we show the path that corresponds to a constant K\"ahler class. Following this path it is easy to see that the FI parameter that is ``orthogonal'' to the one that represents the size of the cycle is \emph{periodic}. This will be consistent with interpreting this on the gravity side with the period of the B-field through the $\R^2$ fibre of $\mathcal{O}(-3)\rightarrow \mathbb{CP}^2$. We will also see that the non-conformal theory corresponds in the supergravity to a resolved background with one of the three critical values of the B-field period, as the diagram suggests. 
%
\section{Gravity description}
The gravity background corresponding to the resolved orbifold $\C^3/\Z_3$ is
\begin{equation}
\diff s^2(X)=\frac{\diff\rho^2}{\Big(1-\frac{a^6}{\rho^6}\Big)}+\frac{\rho^2}{9}\, \Big(1-\frac{a^6}{\rho^6}\Big)\, (\diff\psi-\mathcal{A})^2+\rho^2\, \diff s^2_{\mathbb{CP}^2}\, ,
\end{equation}
where
\bea
\mathcal{A}&=&\frac{3}{2}\,\sin^2\sigma\, (\diff\beta+\cos\theta\, \diff\phi)\, ,\nn\\
 \diff s^2_{\mathbb{CP}^2}&=&\diff\sigma^2+\frac{1}{4}\,\sin^2\sigma\,\Big(\diff\theta^2+\sin^2\theta\, \diff\phi^2+\cos^2\sigma\, (\diff\beta+\cos\theta\, \diff\phi)^2 \Big)~.
\eea
This is the usual Calabi ansatz, which works for the canonical line bundle over any K\"ahler-Einstein manifold. It is natural to introduce the vielbein
\bea
\nn
&&g_5=\diff\psi-\mathcal{A} \ , \quad e_{\beta}=\frac{1}{2}\, \sin\sigma\, \cos\sigma\, (\diff\beta+\cos\theta\, \diff\phi)\ , \\
&&e_{\theta}=\frac{1}{2}\,\sin\sigma\,\diff\theta\ , \quad e_{\phi}=\frac{1}{2}\,\sin\sigma\,\sin\theta\, \diff\phi~.
\eea
The metric then becomes
\begin{equation}
\diff s^2(X)=\frac{\diff\rho^2}{\Big(1-\frac{a^6}{\rho^6}\Big)}+\frac{\rho^2}{9}\, \Big(1-\frac{a^6}{\rho^6}\Big)\, g_5^2+\rho^2\, \diff\sigma^2+\rho^2\, e_{\beta}^2+\rho^2\,e_{\theta}^2+\rho^2\, e_{\phi}^2~.
\end{equation}
Substituting $\diff s^2(X)$ in \eqref{resolved-metric} one obtains the metric of the ten-dimensional Type IIB background. The warp factor $h$ is given by \eqref{green2}. As usual, we may solve \eqref{green2} by writing
\begin{equation}
\label{hI}
h=\sum\, h_I(\rho,\,\xi;\rho_p,\, \xi_p)\, Y_I^{\star}(\xi_p)\, Y_I(\xi)~,
\end{equation}
where the $Y_I$ are the relevant harmonics, $\xi$ denotes collectively the angular coordinates in the internal manifold, and $(\rho_p,\xi_p)$ stands for the particular point where the stack of D3 branes sits. We refer the reader to \cite{Krishnan:2008kv} for the explicit solution for this warp factor.

\subsection{Gravity moduli: B-field and $C_2$} \label{sec:2-form-moduli}

As discussed in \secref{section3}, the classical gauge theory has a large VMS. One thus expects to find massless scalar fields associated with these flat directions in field space. We are now interested in discussing the form-field moduli in the resolved background. As we reviewed, it is possible to turn on flat form-fields in the resolved background that restrict to trivial classes on the UV boundary. In this subsection we want to discuss the B-field and $C_2$ R-R two-form. The B-field moduli will become an important factor in the following discussion. The $C_4$ R-R four-form will not be discussed here; however fluctuations in this direction will become important later on in this section. 

The B-field and $C_2$ moduli correspond to harmonic two-forms that are $L^2$ normalizable with respect to the unwarped metric. For the resolved $\C^3/\Z_3$, $b_4=1$, thus we are looking for one such two-form. Let us define $J=d\mathcal{A}$
\begin{eqnarray}
J&=&3\sin\sigma\, \cos\sigma\, \diff\sigma\wedge \, (\diff\beta+\cos\theta\, \diff\phi)-\frac{3}{2}\,\sin^2\sigma\,\sin\theta\, \diff\theta\wedge \diff\phi\nonumber \\&=&6\, \diff\sigma\wedge e_{\beta}-6\, e_{\theta}\wedge e_{\phi}~.
\end{eqnarray}
We may consider the following closed four-form
\begin{equation}
\omega_4=J\wedge J + \diff\Big(f(\rho)\, g_5\wedge J\Big)~.
\end{equation}
The condition $\diff\star\omega_4=0$ boils down to
\begin{equation}
\partial_{\rho}\Big(\rho^{-1}\,\partial_{\rho}f\Big)+\frac{8}{\rho^3}=0~.
\end{equation}
The solution to this equation which vanishes at infinity is
\begin{equation}
f=1+\frac{A}{\rho^2}~,
\end{equation}
$A$ being an integration constant.

We may now compute the square $L^2$ norm of the form, which is finite
\begin{equation}
\int\omega_4\wedge \star\omega_4=864\,a^{-6}\,A^2\,\pi\,{\rm vol}(\mathbb{CP}^2)~.
\end{equation}
In turn, we have that $\omega_2=\star\omega_4$ may be written, as expected, as
\begin{equation} \label{2f-unwarped}
\omega_2=-6A\,\diff\Big(\rho^{-4}\,g_5\Big)~.
\end{equation}
We may now consider the following ansatz for the supergravity zero modes
\begin{equation}
B_2=b(x)\, \omega_2\, ,\qquad C_2=c(x)\, \omega_2~.
\end{equation}
From the supergravity equations of motion one can see that the scalar fields $b(x)$ and $c(x)$ are decoupled \cite{Martelli:2008cm}. Given the properties of $\omega_2$, it is straightforward to see that the $b(x)$ and $c(x)$ fields satisfy free field equations of motion in the Minkowski space directions.

It is convenient to choose the normalization factor to be $A = a^4/12\pi$. With this choice the above calculations show that 
\bea
\int_{\R^2_{\mathrm{fibre}}} \omega_2 = 1~.
\eea
Since $H^2_{\mathrm{cpt}}(X,\Z)\cong \Z$, this shows that $\omega_2$ is then an $L^2$ harmonic form 
representing the generator of this group. Here $X$ is the total space of $\mathcal{O}(-3)\rightarrow \mathbb{CP}^2$. 
An exact sequence shows that $\omega_2$ maps to $3$ times the generator of $H^2(\mathbb{CP}^2,\Z)\cong\Z$, where 
the $3$ is the same as in $\mathcal{O}(-3)$. Indeed, a simple calculation shows that
\bea \label{mult3} 
\int_{\mathbb{CP}^1} \omega_2 = 3~.
\eea

\subsection{The periodicity of $b(x)$ and $c(x)$}

Consider a Euclidean fundamental string running along $\{\rho,\, \psi\}$. Its (Euclidean) action is
\begin{equation}
S_E=\frac{T_F}{3}\int_{a}^{\rho_c} \diff\rho\, \int_0^{2\pi} \diff\psi\, \, h\, \rho+\ii\, T_F\, b
\end{equation}
where $\rho_c$ is some UV cut-off and $T_F$ is the string tension. In fact, this cut-off is not required since the integral
is convergent.

Under a shift
\begin{equation} \label{bperiod}
b\rightarrow b+\frac{2\pi n}{T_F}~, \qquad n\in\Z~,
\end{equation}
the partition function will not change. Thus, we have that $b$ is a periodic variable of period $2\pi/T_F$ (and so is $c$, had we considered a D1 string, upon changing $T_F\leftrightarrow T_1$).

We identify this B-field period, up to a constant factor, with the FI parameter in the field theory that corresponds to the closed path in Figure~\ref{GKZ}. Recall that supersymmetry pairs the B-field with $C_2$, and the latter corresponds to a fibre over the FI parameter space in the field theory.

\subsection{D3 branes on the resolved cone and global strings}
A spontaneously broken global symmetry in a field theory generally leads to Goldstone bosons and global strings. The first corresponds to the flat directions given by acting with a broken symmetry generator. In supergravity the Goldstones are fluctuations of R-R fields, and are hence pseudo-scalars. Their scalar supersymmetric partners come from metric and B-field fluctuations. 

In \cite{Martelli:2008cm} the authors have shown the existence of massless scalar fields on $\R^{1,3}$ associated with linearized deformations of the B-field and R-R field moduli, and argued that such modes can be obtained also from fluctuations of the metric. In general $b_4(X)$ such modes are coming from B-field and $C_2$ fluctuations and $b_2(X)$ from $C_4$ and metric fluctuations. Thus, it was argued that the dual field theory then includes corresponding massless particles. 

The linearized fluctuations just discussed may be associated with the Goldstone bosons and their supersymmetric partners. However, as discussed in \cite{Martelli:2008cm}, these modes cannot be interpreted as Goldstone bosons when the corresponding broken symmetries are anomalous, since these symmetries do not survive in the quantum theory. 

In the $\C^3/\Z_3$ field theory all the baryonic global symmetries are anomalous. The supergravity backgrounds constructed from resolved cones are dual to field theories in which these symmetries are ``broken''\footnote{They are really already broken by quantum effects.}. However, for three values of the B-field there is an emerging non-anomalous baryonic symmetry during the RG flow as we argued in the previous section. We claimed that the supergravity solution is dual to a field theory in which this symmetry is broken. We want to argue that the corresponding Goldstone boson and its supersymmetric partner originate from specific fluctuations in the R-R four-form and metric respectively. Since the fluctuations in the metric are somewhat involved, as explained in \cite{Martelli:2008cm}, we will discuss just the fluctuations for the R-R four-form in detail in the next subsection. 

As discussed in \cite{Vilenkin:1982ni}, the breaking of a global $U(1)$ symmetry results in global strings around which the Goldstone boson has a monodromy. Global strings in the Minkowski space associated with the broken global symmetry were discussed in the AdS/CFT context for the conifold case in \cite{Klebanov:2007cx}. It was shown that a D3 brane wrapping the two-cycle in the bottom of the resolved cone sources fluctuations that contain the Goldstone boson. Furthermore, this boson has a monodromy around the string-like wrapped D3 brane in the Minkowski space. 

Coming back to the $\C^3/\Z_3$ theory, we will show that a global string, obtained by wrapping a D3 brane on a two-cycle, appears in supergravity \emph{just} for three critical values of the B-field, as expected from the field theory analysis. For other values of the B-field the massless fields are still there but the wrapped D3 brane is not SUSY, and this suggests that indeed these fields cannot be interpreted as Goldstone bosons in these cases. This is expected since for non-critical values of the B-field the only ``broken'' baryonic symmetries are anomalous. 

We now wish to discuss the wrapped D3 brane in greater detail. Notice that the blown-up four-cycle, being a $\mathbb{CP}^2$, contains a topologically non-trivial $\mathbb{CP}^1$. So we may consider a D3 brane wrapping this two-cycle by taking the world-volume of the brane to be $\{t,\, x,\,\mathbb{CP}^1\}$. This brane, having half of its world-volume on the Minkowski space and half in the internal space, is not sensitive to the warp factor. Thus, its energy at $\rho=a$ is a constant, proportional to $a^2$. From the Minkowski space point of view, it looks like a string. 

Notice that if we were considering a D5 brane wrapping the four-cycle, similar arguments show that the tension of this brane blows up in the IR. This shows that in the IR of the field theory the corresponding string is infinitely massive and therefore completely decoupled. It is interesting that such D5 branes source $C_2$ fluctuations. Such a fluctuation that solves the supergravity equations, and corresponds to a massless field in $\R^4$, should have an interpretation in the field theory. One might consider this as the fluctuation that contains the Goldstone boson coming from the broken emerging baryonic symmetry. However, since in this case the global string, namely a D5 brane wrapped on four-cycle, is not part of the field theory spectrum, we do not want to interpret it as such. Similarly, if we consider the R-R four-form fluctuation for non-critical B-field period discussed above, since in such a background the D3 brane wrapping the two-cycle is not SUSY the global string is absent, and the interpretation of the corresponding massless field is not clear to us.

To determine the criteria under which the D3 brane wrapping the two-cycle in the resolved cone is SUSY we follow \cite{Martucci:2005ht}. In their notations, the two supersymmetry parameters $\epsilon_{1,2}$ are Majorana-Weyl spinors of positive ten-dimensional chirality, and may be written as
\bea
\epsilon_a(y)=\xi_+ \otimes \eta_+^{(a)}(y)+\xi_- \otimes \eta_-^{(a)}(y)\ , \quad a=1,2~,
\eea
where $\xi$ and $\eta$ are the spinors in the internal four-dimensional and external six-dimensional space, respectively.
 
The conditions for the D3 brane to preserve the supersymmetry of the background are given in (3.6) and (3.7) of \cite{Martucci:2005ht}, which are reproduced here for convenience
\bea \label{cond1}
\gamma_{\underline{0..q}} \xi_+\,=\,\alpha^{-1}\,\xi_{(-)^{q+1}}~;
\quad
\quad
\hat{\gamma}^{'}_{(p-q)} \eta^{(2)}_{(-)^{p+1}}\,=\,\alpha\,\eta^{(1)}_{(-)^{q+1}}~,
\eea
where
\bea
\hat{\gamma}^{'}_{(r)}=\frac{1}{\sqrt{\det(g+\mathcal{F})}}\sum_{2l+s=r}\frac{\epsilon^{\alpha_1...\alpha_{2l}\beta_1...\beta_s}}{l!s!2^l}\mathcal{F}_{\alpha_1\alpha_2}\cdots\,\mathcal{F}_{\alpha_{2l-1}\alpha_{2l}}\hat{\gamma}_{\beta_1...\beta_s}~.
\eea
The four-dimensional gamma matrices $\gamma$ are real, the six-dimensional ones $\hat{\gamma}$ are antisymmetric and purely imaginary, and underlined indices represent an orthonormal frame. $\mathcal{F} \equiv 2\pi\ell_s^2\,F-B$ here stands for the gauge-invariant two-form that lives on the D3 brane. The D3 brane wrapped over a two-cycle corresponds to $q = 1$ and $p = 3$. In addition, from the discussion after (3.7) in \cite{Martucci:2005ht}, $\alpha = \pm 1$ in our case. 
       
We want to rephrase the conditions in terms of the geometrical objects $\Psi^+$ and $\Psi^-$ introduced in \cite{Martucci:2005ht}. For the branes which are strings in the $(3+1)$d Minkowski space the second equation in \eqref{cond1} reduces to 
\bea \label{con1b}
\lbrace P[Re(i\,\Psi^+)] \wedge \me^{\mathcal{F}} \rbrace_{(2)}=0~; \ \ \
\lbrace P[(\imath_m + g_{mn}\,\diff\,x^n\,\wedge)\Psi^-]\wedge\,\me^{\mathcal{F}} \rbrace_{(2)}=0~.
\eea
Here $P$ is the pull back to the two-cycle and $\lbrace ... \rbrace_{(2)}$ denotes that just the two-forms inside the brackets should be considered. The warped background of interest is a special case with $SU(3)$ structure\footnote{Read section 5 in \cite{Martucci:2005ht} for more details} in which, from (5.2) in \cite{Martucci:2005ht}, $\Psi^{+}$ and $\Psi^{-}$ reads
\bea
\Psi^+\,=\,\frac{a\,\bar{b}}{8}\,\me^{-iJ}~, \qquad \Psi^-\,=\,-\,\frac{i\,a\,b}{8}\,\Omega~, \qquad \frac{a}{b}=\me^{i\,\phi}~.
\eea
The $\me^{i\,\phi}$ phase parametrizes the embedding of a $U(1)$ family of $\mathcal{N} = 1$ algebras inside the bulk $\mathcal{N} = 2$ superalgebra. Following for example \cite{Gomis:2005wc}, it is easy to see that in the presence of a non-trivial warp factor, $\phi$ takes a fixed value such that $\me^{i\,\phi}$ is purely imaginary\footnote{Note that in \cite{Gomis:2005wc} $\me^{i\,\phi}=-1$. However, there the authors are working in Euclidean signature in which $\gamma_{\underline{0123}}^2=1$, while here we have $\gamma_{\underline{0123}}^2=-1$.}. Therefore the equations in \eqref{con1b} reduce to
\bea\label{con1c}
\lbrace P[Re(\me^{-i\,J})] \wedge \me^{\mathcal{F}} \rbrace_{(2)}=0
~; \ \ \
\lbrace P[(\imath_m + g_{mn}\,\diff\,x^n\,\wedge)\Omega]\wedge\,\me^{\mathcal{F}} \rbrace_{(2)}=0~,
\eea
respectively. Here $J$ is the almost complex structure with respect to which the six-dimensional metric
$g_{mn}$ is Hermitian and $\Omega$ is a $(3,0)$-form constructed from the spinor as explained in \cite{Martucci:2005ht}. The two equations in \eqref{con1c} are easily translated to the SUSY conditions that read 
\bea 
\mathcal{F}\,=\,0 \quad
\text{and} \quad
\imath_m \Omega\,=\,0 \ .
\eea
The second condition is equivalent to demanding that the two-cycle is holomorphically embedded. In addition, we see that we have to set $\mathcal{F}=0$ to have a supersymmetric brane wrapped over the two-cycle. Going back to the first equation in (\ref{cond1}), this condition is satisfied for half of the background spinors, so our strings are one-half BPS.

The $\mathcal{F}=0$ condition reduces to  
\bea \label{Mequ}
\int_{\mathbb{CP}^1}\,2\pi\ell_s^2\,F-B=4\pi^2\ell_s^2\,n-3b=0\ , \quad  n\in \Z~,
\eea
where we used \eqref{mult3} and denoted the quantized period of $F$ as $n$. From \eqref{bperiod} and using the fundamental string tension $T_F=1/2\pi\ell^2$, we see that $b \in [0,4\pi^2\ell_s^2)$. Thus, we find that $\mathcal{F}=0$ for $(b=0,n=0)$, $(b=\frac{4\pi^2\ell_s^2}{3},n=1)$ and $(b=\frac{8\pi^2\ell_s^2}{3},n=2)$. We would like to identify these three special $B$-field values with the three values of the $B$-field implied by the gauge theory analysis. 

Before we turn our attention to the R-R four-form fluctuations that are sourced by this wrapped brane, we want to point out that an E4 brane wrapping the blown-up four-cycle is another interesting brane that one might consider. We studied in \cite{Benishti:2011ab} the SUSY conditions for such a brane and concluded that the brane is SUSY for three values of the B-field period. As we explained in \secref{sec:5} this instanton does not generate a non-perturbative superpotential due to the wrong number of zero modes. However, this brane might contribute to higher F-terms. As discussed in \cite{Beasley:2004ys,Beasley:2005iu}, instantons with four zero modes generate quantum deformations to the moduli space that correspond to the $N_f=N_c$ corrections in the corresponding gauge field theories. In our case we indeed observe such corrections in the field theory; however, these corrections are relevant in the middle of the RG flow. More precisely, these corrections are relevant at the scale of the given VEV and not in the far IR, which is what we argued our supergravity solution corresponds to. This seems consistent as the volume of the E4 brane blows-up due to the warp factor. Since the contribution of the E4 brane to the F-term is proportional to $\me^{-V}$, where $V$ is the warped volume of the four-cycle, one expects such contributions to indeed vanish. 

\subsection{The fluctuation containing the Goldstone mode} \label{sec:fluctuation}

We now intend to examine further the global string we have found. Specifically, we want to study the linearized backreaction
in the background due to the presence of this probe D3 brane. To linearized order this probe sources fluctuations in the R-R four-form potential containing the term $a_2(x) \wedge W$, where $a_2$ is a two-form in $\R^{3,1}$ and $W$ is a closed two-form in the resolved cone. The latter is proportional to the volume form of the two-cycle in the bottom of the resolved cone wrapped by the D3 brane. The linearized equations of motion that should be satisfied are
\bea
\diff \delta G_5 = 0 \ , \quad \delta G_5 = \star\,\delta G_5~. 
\eea
The self-duality condition is satisfied by taking
\bea \label{flac}
\delta G_5=(1+\star)\diff(a_2(x)\,\wedge\,W)~.
\eea
Then, the equations of motion reduce to
\bea
\diff\,\star_4\,\diff\,a_2=0 
\quad
\text{and}
\quad
\diff\,(h\,\star_6\,W)=0~,
\eea
where $\star_4$ and $\star_6$ are the Hodge duals with respect to the unwarped Minkowski and resolved orbifold metrics,
respectively. 

\subsection*{Solving for $W$}
From \cite{Krishnan:2008kv} we learn that the warp factor is a function of $\rho$ and $\sigma$ when the $N$ D3 branes sit at the bottom of the resolved cone. We start with the ansatz
\bea
W=\diff(f_1(\rho,\sigma)\,g_5+f_2(\rho,\sigma)e_\beta)~.
\eea
The $\diff(h(\rho,\sigma)\star_6 W)=0$ condition reduces to
\bea \label{harm1}
\nn
&&\frac{4}{3}(3\,f_1-\cot\sigma\,f_2)h\,\cot\sigma+\frac{2}{3}\left(2\cot(2\sigma)f_2-6\,f_1+\frac{\partial\,f_2}{\partial\sigma}\right)h\,\cot\sigma+\\ \nn
&&+\frac{\partial}{\partial\sigma}\left(\frac{\rho\,h}{3}\left(2\cot(2\sigma)f_2-6\,f_1+\frac{\partial\,f_2}{\partial\sigma}\right)\right)+\frac{\partial}{\partial\rho}\left(\frac{(\rho^6-a^6)h\,\frac{\partial\,f_2}{\partial\rho}}{3\rho^3}\right)=0~, \\
\eea
and
\bea \label{harm2}
\nn
&&4\rho(\cot\sigma\,f_2-3\,f_1)h+\frac{3\rho^7\,h(-3\cot\sigma+\tan\sigma)\frac{\partial\,f_1}{\partial\sigma}}{a^6-\rho^6}-\frac{\partial}{\partial\sigma}\left(\frac{3\rho^7\,h\,\frac{\partial\,f_1}{\partial\sigma}}{a^6-\rho^6}\right)+\\
&&+\frac{\partial}{\partial\rho}\left(3\,\rho^3\,h\,\frac{\partial\,f_1}{\partial\rho}\right)+2\rho(-6\,f_1+2\,\cot(2\sigma)f_2+\frac{\partial\,f_2}{\partial\sigma})h=0~.
\eea
We have two equations and two functions, and together with the right boundary conditions these equations are expected to have a solution. We take the boundary conditions to be $f_1(\rho,\sigma) \sim f_1(\sigma)\rho^m$ and $f_2(\rho,\sigma) \sim f_2(\sigma)\rho^n$ with $m,n<0$ when $\rho$ goes to infinity. Since the warp factor behaves as $h(\rho,\sigma) \sim 1/\rho^4$ when $\rho$ goes to infinity, the equations \eqref{harm1} and \eqref{harm2} reduce to
\bea
\nn
&&f_2''(\sigma)+f_2'(\sigma) (3 \cot\sigma-\tan\sigma)-6\rho^{m-n} f_1'(\sigma)+\\
&&+f_2(\sigma) \left((n-2) (n-3) \csc^2\sigma-\sec ^2\sigma\right)=0~,
\eea
and
\bea
\nn
&&f_1''(\sigma)+f_1'(\sigma)\,3\cot\sigma-\tan\sigma+(m-4)(m+2)f_1(\sigma)+\\
&&+\frac{2}{3}\rho^{n-m}
   \left(f_2'(\sigma)+f_2(\sigma) (3\cot\sigma-\tan\sigma)\right)=0~.
\eea
These equations have decaying solutions when $m=-2>n$ and $f_1(\sigma)$ is a constant. Therefore for large $\rho$
\bea
W \sim \frac{3\rho(\diff\sigma\wedge e_{\beta}-e_{\theta}\wedge e_{\phi})+\diff\rho\wedge g_5}{\rho ^3}~.
\eea

On the other hand, when taking $\rho=a$ it is straightforward to see that the equations are solved for constant $f_1(a,\sigma)$ and $f_2(a,\sigma)=0$, and thus the corresponding two-form is proportional to that in \eqref{2f-unwarped}. 

These are indeed the boundary conditions that one expects from the discussion in \cite{Martelli:2008cm}. From the discussion there we learn that $H^2_{L^2}(X,hg_X)\cong H^2(X,\R)$. The following exact sequence
\bea
\nn
&&0\cong H^1(Y_5;\R)\rightarrow H^2(X,Y_5;\R)\rightarrow H^2(X;\R)\rightarrow \\
&&\rightarrow H^2(Y_5;\R) \rightarrow H^3(X,Y_5;\R)\cong 0~,
\eea
together with the fact that $b_4(X)=b_2(X)-b_3(Y_5)=b_2(X)=1$ shows that the two-forms in $H^2(X;\R)$ restrict to trivial two-form classes in $H^2(Y_5;\R)$. Recall that the pull back of the Thom class\footnote{The Thom class is represented by a form in $H^2_{cpt}(X;\R)$ that gives 1 when integrated over the fibre.} to the zero section $\mathbb{CP}^2$ is the Euler class, which in our case is just $3[\diff\vol_{\mathbb{CP}^1}]$, where $\mathbb{CP}^1$ is the two-cycle inside $\mathbb{CP}^2$. Thus, the pull back of the $L^2$ normalizable two-form with respect to the warped metric on $\mathbb{CP}^2$ is also a two-form in this class. This then shows that the $a_2 \wedge W$ part in this fluctuation is sourced electrically by a D3 brane wrapping the $\mathbb{CP}^1$.

Following \cite{Klebanov:2007cx} we want to show that the R-R four-form fluctuations just discussed contain the Goldstone boson. We start by introducing the field $p(x)$ by dualizing the two-form $a_2$ 
\bea
 \star_4\,\diff\,a_2 = \diff\,p~.
\eea
Substituting this back into \eqref{flac}, the fluctuation in the five-form field strength reads
\bea
\delta\,G_5=\diff\,a_2\,\wedge\,W\,+\,\diff\,p\,\wedge\,h\,\star_6\,W~, 
\eea
and thus the fluctuation of the four-form potential is
\bea \label{dc4}
\delta\,C_4=a_2(x)\,\wedge\,W\,+\,p\,h\,\star_6\,W~.
\eea
For the conifold \cite{Klebanov:2007cx} it was argued that $p(x)$ in \eqref{dc4} is the Goldstone boson for the broken baryonic $U(1)$. To show that, the authors demonstrated that the fluctuation couples through the Wess-Zumino term to the E4 condensate that corresponds to the VEV of the baryonic operator. Moreover, it was shown that the asymptotic behaviour of \eqref{dc4} in the UV corresponds to a VEV for the baryonic current in the field theory. 

For the broken emerging baryonic symmetry these checks cannot be repeated due to the fact that the gravity solution does not capture the intermediate non-conformal theory. Nevertheless, we want to interpret the $p(x)$ massless field as the Goldstone boson coming from the broken emerging baryonic symmetry. This is strongly suggested by the presence of a global string for critical B-field values. The massless field $p(x)$ has a monodromy around this global string\footnote{For a more elaborate discussion on this type of global string we refer the reader to \cite{Klebanov:2007cx}.} since the wrapped D3 brane electrically sources $a_2$. This is exactly the sector that is expected to appear whenever a global symmetry is broken.

The global string we have found is crucial for the interpretation of the massless field as a Goldstone boson. For generic B-field VEV this global string is not supersymmetric and thus the massless field cannot be interpreted as a Goldstone boson. This is of course expected from the field theory analysis as for generic B-field no non-anomalous baryonic symmetries emerge during the RG flow. Adding this to the fact that the baryonic symmetries in the UV are anomalous, it is clear that the discussed symmetry-breaking sector should not be observed. We summarize our findings in the next chapter and comment on similar results that were obtained by studying other examples in our original paper.

\chapter{Final comments and summary} \label{sec:B4}
In this part of the thesis we studied the moduli space of field theories that are dual to D3 branes probing toric CY three-folds that contain four-cycles in their resolution. We have shown that there is a remarkable agreement between the moduli space in the field theory and supergravity in the orbifold example. One new result is the fact that the directions in the moduli space of the field theory which are dual to the B-field moduli are indeed periodic. This was demonstrated for the orbifold in this manuscript and for two more examples in the original paper that this thesis is based on \cite{Benishti:2011ab}. For a general resolved CY space $X$, one expects $b_4(X)$ such directions.

The main aim of this part of the thesis was to show that there is extra information in the field theory moduli space that singles out critical values of the compactly-supported background B-field. In the $\C^3/\Z_3$ example, these critical values of the B-field were identified with the blue lines in the FI parameter space in Figure~\ref{GKZ}. In the field theory these critical values correspond to RG flows that result in intermediate non-conformal phases. This seems to falsify the speculation in \cite{Krishnan:2008kv}, made for the $\C^3/\Z_3$ orbifold, arguing that such RG flows are induced by giving different VEVs that also end with $\mathcal{N}=4$ $SU(N)$ SYM theory in the deep IR. 

We claimed that some of the surviving anomalous baryonic symmetries in the UV become non-anomalous during the RG flow, depending on the B-field in the background. To make our claim more rigorous it will be very important to understand what is the exact mechanism that explains this result. Here and in \cite{Benishti:2011ab} we focused on the fully resolved geometries. However, there are non-conformal phases that appear when the geometry is just partly resolved. The identifications of the corresponding VEVs with critical values of the B-field seems to generalize to those cases too. It will be interesting to study this in more detail. 

One can consider giving non-vanishing VEVs to mesonic operators in the confined theories. This was done in \cite{Krishnan:2008kv} for the $\C^3/\Z_3$ orbifold. It seems to us that such VEVs correspond to moving the branes from the bottom of the resolved cone. However, this should be studied more carefully. More interesting scenarios seem to appear when one keeps the branes at the bottom of the cone by choosing vanishing VEVs for mesonic operators and non-vanishing VEVs for baryonic operators in the confined theory. In this case the emerging baryonic symmetry is broken, and the existence of the confined theory leaves a trace in the IR in the form of a Goldstone boson and global string.

The next thing we did was to look to see if we could find critical values of the B-field from the supergravity analysis. Indeed we have shown that just for three critical values of the B-field in the supergravity background one can wrap D3 branes on the blown-up $\mathbb{CP}^1$. Those D3 branes source fluctuations in the R-R four-form that we suggest contain the Goldstone boson from the broken emergent symmetry. These D3 branes are also global strings, and the Goldstone has monodromy around them as expected for a broken global symmetry. For generic B-field VEV we do not see those global strings and thus the zero mode for generic B-field value cannot be interpreted as a Goldstone boson. This is an important difference, and is a strongly supportive argument in favour of interpreting the mode coming from the four-form at critical B-field values as a Goldstone boson.
With these results we completed a field theory and a supergravity analysis that showed that three B-field VEVs are special and generate different RG flows in the $\C^3/\Z_3$ case. 

We repeated this analysis in \cite{Benishti:2011ab} for the Sasaki-Einstein five-manifold $T^{2,2}$, which is a $U(1)$ fibration over $\mathbb{CP}^1 \times \mathbb{CP}^1$. There we made a distinction between two different types of D3 branes that wrap two-cycles and form global strings in the Minkowski space. The first type corresponds to D3 branes that source fluctuations that contain the Goldstone bosons coming from the broken non-anomalous baryonic symmetries in the UV. There are $b_3(Y_5)$ such branes corresponding with the $U(1)^{b_3(Y_5)}$ baryonic symmetries in the field theory \cite{Martelli:2008cm}. The second type of D3 branes, we suggest, source fluctuations that contain the Goldstone bosons coming from broken emerging non-anomalous baryonic symmetries. In general, one expects to see $b_4(X)=b_2(X)-b_3(Y_5)$ such D3 branes. These branes can be wrapped just for critical values of the $b_4(X)$ B-field moduli, as explained. 

There are several interesting directions for future work. As we have mentioned in the main text, one should go beyond the supergravity analysis to study the intermediate non-conformal theory, which is not captured by the supergravity solution. In fact, this is a general issue that repeats itself in other known non-conformal theories with gravity duals. Another direction that was not discussed in this thesis is the computation of condensates of the baryonic operators. Such condensates, which correspond to Euclidean D3 branes in string theory, were studied in general CY backgrounds in \cite{Martelli:2008cm}. The formula that they derive implies that indeed the world-volume action of the Euclidean D3 brane depends on the B-field moduli. This is of course expected from the field theory since the spectrum of the operators that obtain VEVs, for critical and non-critical B-field periods, is different. However, more work should be done to make this map precise. An additional interesting research avenue concerns the generalization of our findings to more complicated geometries. In this thesis and in \cite{Benishti:2011ab} we confined ourselves to the $\C^3/\Z_3$ orbifold and $\mathbb{F}_0=\mathcal{C}(T^{2,2})$ examples in which the moduli space, together with the critical B-field VEVs, seem to match nicely with the supergravity results. As we have shown in \cite{Benishti:2011ab}, when there are more four-cycles in the resolved geometry the structure of the moduli space becomes much richer. This was demonstrated in \cite{Benishti:2011ab} for the $\C^3/\Z_5$ orbifold. In this example the resolved supergravity background is not known and the interpretation of the structure of the critical walls in the moduli space cannot be easily understood. It will be important to develop methods with which the string theory dual can be investigated without dependence on the explicit supergravity solution. As a final remark, besides being of great theoretical interest, Calabi-Yau cones with vanishing four-cycles are interesting for model building. This was recently investigated in \cite{Dolan:2011qu}. We hope that our study will contribute to the exploration of such constructions in the future.
\appendix
\chapter{Un-Higgsing in QCS theories} \label{sec:unhiggsing}
At present there is no known general method for constructing QCS theories for orbifolds $\C^4/\Gamma$ as a projection of the ABJM theory (For recent work on orbifolds of the ABJM theory, see \cite{Berenstein:2009ay}). This leads us to construct an un-Higgsing algorithm where one starts with a QCS theory for a singularity $X$, and then enlarges the quiver in a specific way, corresponding to ``blowing up'' $X$. We begin with a discussion of the known constraints on QCS theories, and then describe a general un-Higgsing algorithm. 
\subsection{Calabi-Yau, toric and tiling conditions}
We begin by reviewing the conditions which should be satisfied by a QCS theory on M2 branes probing a non-compact toric Calabi-Yau four-fold. In order that the VMS is Calabi-Yau, we require that for each node in the quiver the number of arrows entering and leaving the node should be equal. Notice this is the same condition as gauge anomaly cancellation in the $(3+1)$d YM parent. The superpotential satisfies the \emph{toric condition} if each chiral multiplet appears precisely twice in $W$: once with a positive sign and once with a negative sign. This ensures that the solution to the F-term equations is a toric variety. The last condition that we want to impose is the so-called \emph{tiling condition}. All known quiver theories related to toric Calabi-Yaus, in both $(3+1)$ and $(2+1)$ dimensions, obey this condition due to their brane-tiling/dimer model description \cite{Yamazaki:2008bt,Hanany:2008cd,Hanany:2008fj}. This leads to the elegant condition
\begin{equation}\label{euler}
G - E + N_T = 0~,
\end{equation}
where $G$ is the number of nodes, $E$ is the number of fields and $N_T$ is the number of terms in the superpotential. Whenever the rule is broken in the resulting theory the dimension of its VMS is no longer four.
\begin{figure}[ht]
\center
\includegraphics[scale=1.7]{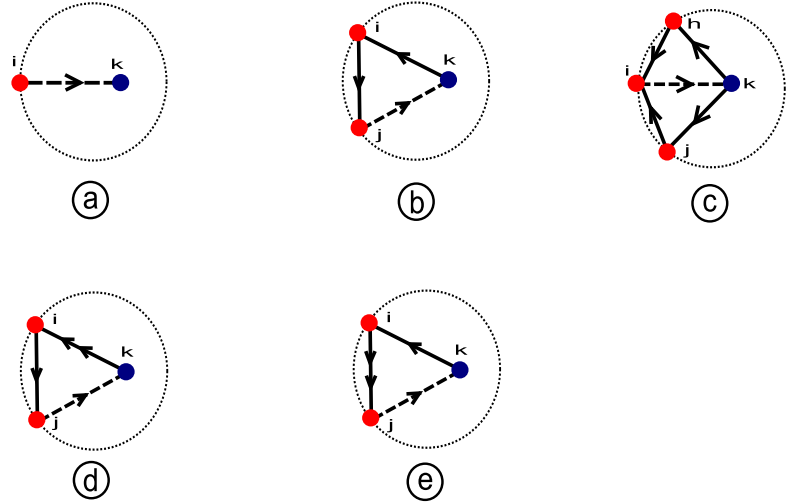}
\caption{The un-Higgsing process. Gauge nodes of the original theory (in red) appear on the boundary of the circle, while the gauge node that has been introduced in the un-Higgsing appears (in blue) inside the circle: (a) adding one field; (b) adding three fields; (c) adding five fields; (d) and (e) adding four fields.}
\label{f:unhiggsingdiagram}
\end{figure}
%
\subsection{The un-Higgsing algorithm}
The un-Higgsing procedure for quiver gauge theories was studied in \cite{Feng:2002fv} in the context of D3 branes probing complex cones over del Pezzo surfaces. Here we wish to systematize this method and use it as a guide for constructing QCS theories living on M2 brane world-volumes. 
As we will explain shortly, the un-Higgsing process for theories with toric Calabi-Yau four-folds as VMS is quite restrictive.  The basic idea is that by adding one gauge node at a time we can obtain theories whose VMSs contain the original toric diagram as a sub-diagram; thus this will be a QCS form of ``blow-up''. 

Let us begin with the simplest case: un-Higgsing by adding one field to the quiver. This step is shown schematically in Figure~\ref{f:unhiggsingdiagram} (a). The gauge nodes which sit on the circumference of the dotted circle are those in the original theory which is being un-Higgsed. The gauge node sitting inside the circle is that being added to the theory. We have indicated the original node in red, indexed by $i$,  and the new node in blue, indexed by $k$. We shall say that $i$ \emph{participates} in the un-Higgsing process, because it will be attached to $k$, while all other nodes in the original theory are non-participatory.

Next, we add to the original quiver a bifundamental field $X_{i\,k}$ charged under $(\square_i,\bar{\square}_k)$; this is an arrow connecting node $i$ to node $k$. The key point in un-Higgsing is that we must be able to Higgs the new theory to the original one by letting $X_{i\,k}$ acquire a non-zero VEV. To continue to satisfy the toric condition, the field $X_{i\,k}$ must be added simultaneously to a positive and a negative term which already appears in the superpotential, and no extra terms should be introduced. In order to exhaust all possiblities for constructing new consistent theories the $i$ index should run over all values between $1$ and $G$, where $G$ is the number of gauge nodes in the original quiver. Moreover, the field $X_{i\,k}$ must be inserted to all possible pairs of negative and positive terms in the superpotential. 

However, notice that after adding $X_{i\,k}$ to the quiver, the Calabi-Yau condition mentioned in the previous subsection is broken: the number of arrows that enter node $i$ or node $k$ is not equal to the number of those that leave. To remedy this we need to relocate the heads and tails of arrows in the original quiver between node $i$ and node $k$. For example, for a three-noded quiver with nodes $i$, $i_1$ and $i_2$ we can do this by changing the tail of $X_{i\,i_1}$ to $k$. Finally, we assign CS levels to nodes $i$ and $k$ such that their sum is equal to the original CS level\footnote{We refer the reader to \cite{Benishti:2009ky,Franco:2008um} for a discussion on the Higgsing of QCS theories.} of node $i$.

Next we turn to more complicated un-Higgsing possibilities. Adding more than one field forces us to add terms to the superpotential, instead of simply adjoining the fields to existing terms, as was the case above; otherwise, it would be impossible to obtain the original quiver by Higgsing. The only possibility is that after introducing such new terms to the superpotential, some of the fields will become massive after the Higgsing and will be integrated out. Therefore, we see immediately that it is not possible to un-Higgs the theory by adding only two fields: insertion of a term that contains two fields is not a valid un-Higgsing step as these fields would be integrated out even before Higgsing because we would be adding a quadratic mass term.

Hence, let us move on to consider introducing three new fields. In accordance with the labelling in Figure~\ref{f:unhiggsingdiagram} (b), the three fields are denoted $X_{i\,j}$, $X_{j\,k}$ and $X_{k\,i}$, where $X_{j\,k}$ is the field which we wish to Higgs in order to reproduce the original theory in the IR. Since the three fields should disappear from the IR theory after Higgsing, there must be a cubic term in the superpotenital which contains all three. This new cubic term should be gauge invariant, and thus the fields which we add must form a closed loop. Notice that after Higgsing $X_{j\,k}$ we are left with a term that contains two fields: $X_{i\,j}$ and $X_{k\,i}$. Those fields should be integrated out in the IR as they give rise to a quadratic mass term. 

To satisfy the toric condition, $X_{i\,j}$, $X_{j\,k}$ and $X_{j\,k}$ should also appear in other terms in the superpotential and have opposite sign with respect to the cubic term. Furthermore, we must satisfy\footnote{{\it A priori}, violating this condition is not a problem. However, in all cases that we have studied the resulting theory will then have a five complex-dimensional VMS.} the tiling condition \eqref{euler}. 
Now, since we have added one node and three fields, we must add two terms to the superpotential. The cubic term mentioned above is one of them. What about the other? There are two options: to add a new term or to split one of the existing terms into two. The first option would just be the cubic term with opposite sign, which would simply cancel in the Abelian theory and hence is ineffective. We must therefore take the second option and split an existing term, inserting $X_{i\,j}$ and $X_{k\,i}$ separately into the two split terms. This guarantees that after integrating out these fields the split terms are united. 
To see this in more detail, suppose the original superpotential contains a term $AB$, where $A$ and $B$ are monomials in bifundamental fields; that is: $W = AB + \ldots$. Then our procedure would change this superpotential to
$W = A\ X_{i\,j} + B\ X_{k\,i} - X_{i\,j}X_{j\,k}X_{k\,i} + \ldots$. When $X_{j\,k}$ acquires a VEV (say $\langle X_{j\,k}\rangle=1$ for convenience), the equation of motion for $X_{i\,j}$ becomes $A = X_{k\,i}$, and the first and third terms cancel while the middle term becomes $AB$, as required.
Finally, $X_{j\,k}$ can be added to an arbitrary term with the opposite sign to the cubic term. 

In order to exhaust all possibilities we split terms, insert fields, assign CS levels and vary $i$ and $j$ in all possible combinations (notice that $i$ could equal $j$ for the case of adjoint fields). Furthermore, we insert the cubic term both with positive and negative signs and allow relocation of heads or tails of arrows involving nodes $i$ and $k$ in ways that satisfy the Calabi-Yau condition, as in the case of adding one field.   

The next possibility is un-Higgsing by introducing four new fields. As shown in Figure~\ref{f:unhiggsingdiagram}  (d) and (e), this can be done in two different ways. Let us discuss (d) first. Notice that the only way this can be achieved is by insertion of two new cubic terms into the superpotential: $X_{j\,k}X_{k\,i}^1 X_{i\,j}-X_{j\,k}X^2_{k\,i} X_{i\,j}$. However, this violates the Calabi-Yau condition on both nodes $i$ and $k$. Since the field to be Higgsed is $X_{j\,k}$, we can transform heads and tails of arrows between nodes $j$ and $k$ only and cannot fix the Calabi-Yau condition on node $i$. Therefore this un-Higgsing step is allowed only when $i$ is equal to $j$. The same analysis can be applied for (e), and the result is the same. With this constraint, since we have introduced four new fields, one gauge node, and two new terms to the superpotential, the tiling condition is violated. In the theories that we have checked this results in five-dimensional VMSs. We hence cannot introduce four fields.

The final un-Higgsing process involves the insertion of five new fields. 
Careful examination implies that this can be done by introducing two cubic terms into the superpotential with opposite signs. If we use the notation of Figure~\ref{f:unhiggsingdiagram} (c), we can write the terms as follows: $X_{i\,k}X_{k\,h}X_{h\,i}-X_{i\,k}X_{k\,j}X_{j\,i}$ ($i$, $j$ and $h$ can be equal). 
Notice that by Higgsing $X_{i\,k}$ we obtain two terms in the superpotential that contain two fields each, and therefore four fields should be integrated out. 
By a similiar analysis to the above, after satisfying the tiling condition by splitting terms in the superpotential it can be seen that $X_{k\,h}$ and $X_{h\,i}$ should appear in different split negative terms. 
Similarly, $X_{k\,j}$ and $X_{j\,i}$ should appear in different split positive terms. 

Finally, note that five fields is the maximum number of fields that can be introduced if one wants to obtain the original theory by Higgsing only one field. This concludes the discussion of our un-Higgsing algorithm.
\chapter{The warp factor for resolutions of $\mathcal{C}(Q^{111})$ and $\mathcal{C}(Q^{222})$} \label{sec:D}
We are interested in studying supergravity backgrounds corresponding to M2 branes localized on the space $X$, which will be the resolution of either $\mathcal{C}(Q^{111})$ or $\mathcal{C}(Q^{222})$. After placing $N$ spacetime-filling M2 branes at a point $y$ in the resolved space $X$ we must solve the Green's equation \eqref{green} for the warp factor $h=h[y]$. Using the explicit form of the Laplacian we can write
\begin{equation}
\frac{1}{\sqrt{\det g}}\, 
\partial_i\Big(\sqrt{\det g}\, g^{ij}\partial_j h\Big)=-\frac{(2\pi \ell_p)^6N}{\sqrt{\det g}} \delta^8(x-y) \ .
\end{equation}

Since we can choose coordinates such that the metrics we are considering are formally identical and given by \eqref{resolvedQ111}, the Laplacian in both cases can be written as
\begin{equation}
\frac{1}{\sqrt{\det g}}\,\partial_i\Big(\sqrt{\det g}\, g^{ij}\partial_j h\Big)=\frac{\partial_r\Big(r^3(2a+r^2)(2b+r^2)\kappa\,\partial_rh\Big)}{r^3(2a+r^2)(2b+r^2)}+\mathbf{A}h \ ,
\end{equation}
where the angular Laplacian $\mathbf{A}$ is
\begin{equation}
\label{AngLap}
\mathbf{A}h=\frac{8}{r^2}\Delta_1h+\frac{8}{2a+r^2}\Delta_2 h+\frac{8}{2b+r^2}\Delta_3h+\frac{16}{r^2\kappa}\partial^2_{\psi}h \ ,
\end{equation}
with
\begin{equation}
\Delta_i=\frac{1}{\sin\theta_i}\partial_{\theta_i}(\sin\theta_i\partial_{\theta_i})+\Big(\frac{1}{\sin\theta_i}\partial_{\phi_i}-\cot\theta_i\partial_{\psi}\Big)^2 \ .
\end{equation}
As we show in (\ref{delta}) in the next subsection, we can expand the delta function in terms of eigenfunctions of the Laplacian such that
\begin{equation}
\frac{1}{\sqrt{\det g}} \delta^8(x-y)=\frac{1}{r^3(r^2+2a)(r^2+2b)}\delta(r-r_0)\sum_I\, Y_I(\xi_0)^*Y_I(\xi) \ ,
\end{equation}
where we denote collectively the angular coordinates as $\xi$, and define $x=(r,\xi)$ and $y=(r_0,\xi_0)$. Then, the equation for the warp factor reads
\begin{equation}
\frac{1}{f}\partial_r\Big(f\,\partial_rh\Big)+\kappa^{-1}\,\mathbf{A}h=- \frac{(2\pi \ell_p)^6N}{f}\delta(r-r_0)\sum_I\, Y_I(\xi_0)^*Y_I(\xi) \ ,
\end{equation}
where we have defined for simplicity
\begin{equation}
\label{f-def}
f=r^3(2a+r^2)(2b+r^2)\kappa \ .
\end{equation}
We can now expand $h$ in eigenfunctions of the angular Laplacian. Since $\xi_0$ is just a fixed point (not a variable), we can write
\begin{equation}
h=\sum_I \psi_I(r)\, Y_I(\xi_0)^*\, Y_I(\xi) \ .
\label{general-warp}
\end{equation}
Then the radial equation we have to solve reduces to
\begin{equation}
\frac{1}{f}\partial_r\Big(f\,\partial_r\psi_I\Big)-\kappa^{-1}\,E_I\,\psi_I=- \frac{(2\pi \ell_p)^6N}{f}\delta(r-r_0) \ ,
\label{radial-to-solve}
\end{equation}
where $E_I$ is the angular eigenvalue of $Y_I$, to which we now turn.

\section{Angular eigenfunctions in $Q^{111}$}

We want to consider (\ref{AngLap}) with fixed $r$ and construct eigenfunctions of such an operator. For this we first concentrate on each of the $\Delta_i$ operators. For each $\theta$, these look like
\begin{equation}
\Delta_{i} =\frac{1}{\sin\theta }\partial_{\theta}(\sin\theta\partial_{\theta})+\Big(\frac{1}{\sin\theta}\partial_{\phi}-\cot\theta\partial_{\psi}\Big)^2 \ .
\end{equation}
Note that these angular Laplacians are the same as those for the conifold. As such, many technical details and results can be borrowed from \cite{Klebanov:2007us}.

We consider the following function $Y=J(\theta)\, \me^{\ii\, m\, \phi}\, \me^{\ii\, {R\psi}/{2}}$. It is obvious that
\begin{equation}
\Delta_{i} Y=\Big\{\frac{1}{\sin\theta}\partial_{\theta}(\sin\theta\partial_{\theta} J)-\Big(\frac{m}{\sin\theta}-\cot\theta\frac{R}{2}\Big)^2 J\Big\}J^{-1} Y \ .
\end{equation}
Therefore it is interesting to consider the following eigenfunctions
\begin{equation}
\frac{1}{\sin\theta}\partial_{\theta}(\sin\theta\partial_{\theta} J)-\Big(\frac{m}{\sin\theta}-\cot\theta\frac{R}{2}\Big)^2 J=-E J \ .
\end{equation}
This equation has two solutions, given in terms of hypergeometric functions
\begin{equation}
\label{sol_A}
J^A_{l,m,R}=\sin^m\theta\, \cot^{\frac{R}{2}}\frac{\theta}{2}\, \,_2F_1\Big(-l+m,1+l+m;1+m-\frac{R}{2};\sin^2\frac{\theta}{2}\Big) ~,
\end{equation}
and
\begin{equation}
\label{sol_B}
J^B_{l,m,R}=\sin^{\frac{R}{2}}\theta\, \cot^{m}\frac{\theta}{2}\, \,_2F_1\Big(-l+\frac{R}{2},1+l+\frac{R}{2};1-m+\frac{R}{2};\sin^2\frac{\theta}{2}\Big) \ ,
\end{equation}
where we have introduced a labelling for the quantum numbers distinguishing the eigenfunctions. If $m\ge \frac{R}{2}$ solution (\ref{sol_A}) is non-singular while if $m\le \frac{R}{2}$ it is (\ref{sol_B}) that is the non-singular solution. Both have eigenvalue under each $\Delta_i$ operator given by $E=l(l+1)-R^2/4$. Given these results, we can consider the functions
\begin{equation}
\label{Y-function}
Y_{I}=\mathcal{C}_I\,J_{l_1,m_1,R}(\theta_1)J_{l_2,m_2,R}(\theta_2)J_{l_3,m_3,R}(\theta_3)\, \me^{\ii\, (m_1\phi_1+m_2\phi_2+m_3\phi_3)}\, \me^{\ii\, {R\psi}/{2}} \ ,
\end{equation}
where the multi-index $I$ stands for $\{(l_1,m_1),(l_2,m_2),(l_3,m_3),R\}$ and where $\mathcal{C}_I$ is just a normalization factor such that the norm of $Y_I$ is one.  It is now clear that $\mathbf{A}Y_I=-E_I Y_I$, with
\begin{equation}
E_I=\frac{8\,l_1(l_1+1)}{r^2}+\frac{8\,l_2(l_2+1)}{r^2+2a}+\frac{8\,l_3(l_3+1)}{r^2+2b}+2R^2\Big(\frac{2}{r^2\kappa}-\frac{1}{r^2}-\frac{1}{r^2+2a}-\frac{1}{r^2+2b}\Big) \ .
\end{equation}
We now note that the $Y_I$ are also eigenfunctions of the singular cone. Indeed, we can consider the Laplacian on the unit $Q^{111}$, namely $\mathbf{\tilde{A}}|_{a=0=b;r=1}$. Then $\mathbf{\tilde{A}}Y_I=-\tilde{E}_I Y_I$, with
\begin{equation}
\tilde{E}_I=8\,l_1(l_1+1)+8\,l_2(l_2+1)+8\,l_3(l_3+1)-2R^2 \ .
\end{equation}
Therefore, the $Y_I$ are also normalized eigenfunctions for the $\mathbf{\tilde{A}}$ operator. Being eigenfunctions of a Hermitian operator, these satisfy
\begin{equation}
\int \dd^7\xi\, \sqrt{\det \tilde{g}}\, Y_I(\xi)^*Y_J(\xi)=\delta_{I-J} \ ,
\end{equation}
and therefore
\begin{equation}
\sum_I\, Y_I(\xi_1)^*Y_I(\xi_2)=\frac{1}{\sqrt{\det \tilde{g}}}\, \delta^7(\xi_1-\xi_2) \ ,
\end{equation}
where we use $\xi$ to generically parametrize the angular coordinates and $\tilde{g}$ stands for the angular part of the metric. One can check very easily that
\begin{equation}
\sqrt{\det g}=r^3(r^2+2a)(r^2+2b)\, \sqrt{\det \tilde{g}} \ .
\end{equation}
Therefore, if we denote $x=(r,\xi)$ and $y=(r_0,\xi_0)$ and use the completeness relation we have
\begin{equation}
\label{delta}
\frac{1}{\sqrt{\det g}}\, \delta^8(x-y)=\frac{1}{r^3(r^2+2a)(r^2+2b)}\delta(r-r_0)\sum_I\, Y_I(\xi_0)^*Y_I(\xi) \ .
\end{equation}
\section{Angular eigenfunctions in $Q^{222}$}

Since $\mathcal{C}(Q^{222})$ is a $\mathbb{Z}_2$ orbifold of $\mathcal{C}(Q^{111})$ along $\psi$ it is clear that the local computation of the previous subsection will not be changed. Thus, we just have to take care of global issues. Recall that the wavefunctions in $\mathcal{C}(Q^{111})$ are
\begin{equation}
Y_{I}=\mathcal{C}_I\,J_{l_1,m_1,R}(\theta_1)J_{l_2,m_2,R}(\theta_2)J_{l_3,m_3,R}(\theta_3)\, \me^{\ii\, (m_1\phi_1+m_2\phi_2+m_3\phi_3)}\, \me^{\ii\, {R\psi}/{2}} \ .
\end{equation}
Since now $\psi\in [0,2\pi]$, it is clear that the well-behaved $Y_I$ will be those for which $R$ is even; that is, $R=2\, \tilde{R}$. Therefore, dropping the tilde, the angular wavefunctions in $\mathcal{C}(Q^{222})$ are
\begin{equation}
Y_{I}=\mathcal{C}_I\,J_{l_1,m_1,R}(\theta_1)J_{l_2,m_2,R}(\theta_2)J_{l_3,m_3,R}(\theta_3)\, \me^{\ii\, (m_1\phi_1+m_2\phi_2+m_3\phi_3)}\, \me^{\ii\, R\psi} \ ,
\end{equation}
such that $\mathbf{A}Y_I=-E_I Y_I$, with
\begin{equation}
E_I=\frac{8\,l_1(l_1+1)}{r^2}+\frac{8\,l_2(l_2+1)}{r^2+2a}+\frac{8\,l_3(l_3+1)}{r^2+2b}+8R^2\Big(\frac{2}{r^2\kappa}-\frac{1}{r^2}-\frac{1}{r^2+2a}-\frac{1}{r^2+2b}\Big) \ .
\end{equation}

\section{The warp factor for $Q^{111}$} \label{sec:Q111-warp}

We now want to use the results derived so far to compute explicitly the warp factor for the resolution of the $\mathcal{C}(Q^{111})$ space. We will consider the stack of branes to be sitting on the exceptional locus, where both the $U(1)$ fibre and the $(\theta_1,\phi_1)$ sphere shrink to zero size. This means that $h=h(r,\theta_2,\theta_3)$, which in turn implies that $R$ and $l_1$ in (\ref{Y-function}) vanish. Then, under these assumptions, the multi-index $I$ takes the values $I=\{(l_2,m_2),(l_3,m_3)\}$. Indeed, we will assume the branes are located at the north pole of each of the two two-spheres. As such, we should consider also $m_2=m_3=0$. Therefore, for such cases the angular eigenfunctions $J^A$ and $J^B$ coincide and reduce, for each sphere, to Legendre polynomials $J_{l,0,0}=P_l(\cos\theta)$,
such that 
\bea
\nn
Y_{l_2,l_3}=\mathcal{C}_{l_2,l_3}\, P_{l_2}(\cos\theta_2)\, P_{l_3}(\cos\theta_3) \quad \text{and} \quad E_I=\frac{8\,l_2(l_2+1)}{r^2+2a}+\frac{8\,l_3(l_3+1)}{r^2+2b} \ .
\eea
Thus, from (\ref{radial-to-solve}) we see that the equation to solve reads
\begin{equation}
\frac{1}{f}\partial_r\Big(f\,\partial_r\psi_I\Big)-\Big(\frac{8\,l_2(l_2+1)}{r^2+2a}+\frac{8\,l_3(l_3+1)}{r^2+2b}\Big)\kappa^{-1}\,\psi_I=-\frac{(2\pi \ell_p)^6N}{f}\, \delta(r) \ .
\label{tosolve}
\end{equation}
We are interested in the simplified case in which, say, only $b\ne 0$. Under such assumption, also the $(\theta_2,\phi_2)$ sphere shrinks to zero, so that we can also consider $l_2=0$. Then the corresponding angular wavefunctions are $Y_{l_3}=\mathcal{C}_{l_3}\, P_{l_3}(\cos\theta_3)$. Also from (\ref{kappa}) and (\ref{f-def}) we see that
\begin{equation}
\kappa=\frac{r^2+\frac{8b}{3}}{r^2+2b}\ ,\qquad f=r^5\, (r^2+\frac{8b}{3}) \ .
\end{equation}
After substituting these back to \eqref{tosolve} the two solutions read
\begin{eqnarray}
\psi_I^{(1)}& \sim &\Big(\frac{8b}{3r^2}\Big)^{\frac{3}{2}(1-\beta)}\, _2F_1\left(-\frac{1}{2}-\frac{3}{2}\beta,\frac{3}{2}-\frac{3}{2}\beta,1-3\beta,-\frac{8b}{3r^2}\right)\ , \\ \nonumber
\psi_I^{(2)}& \sim &\Big(\frac{8b}{3r^2}\Big)^{\frac{3}{2}(1+\beta)}\, _2F_1\left(-\frac{1}{2}+\frac{3}{2}\beta,\frac{3}{2}+\frac{3}{2}\beta,1+3\beta,-\frac{8b}{3r^2}\right) \ ,
\end{eqnarray}
with $\beta=\sqrt{1+\frac{8}{9}l_3(l_3+1)}$. Since $\beta\ge 1$, for large $r$ only the $\psi_I^{(2)}$ solutions decay at infinity, and these are therefore the solutions of interest.

We can now state the result for the warp factor, which turns out to be
\begin{equation}
h=\sum_{l_3}\, \mathcal{C}_{l_3}\, \Big(\frac{8b}{3r^2}\Big)^{\frac{3}{2}(1+\beta)}\, _2F_1\Big(-\frac{1}{2}+\frac{3}{2}\beta,\frac{3}{2}+\frac{3}{2}\beta,1+3\beta,-\frac{8b}{3r^2}\Big)\, P_{l_3}(\cos\theta_3) \ ,
\end{equation}
where we collect all normalization factors in $\mathcal{C}_{l_3}$.

\section{The warp factor for $Q^{222}$} \label{s:warp_factor-Q222}

We will compute the warp factor for $N$ M2 branes in arbitrary location. As was shown above, the equation to solve is (\ref{radial-to-solve}) where now we should use \eqref{kappaQ222} for $\kappa$. We will be interested in the simpler case in which $a=b=0$. Moreover, as we explain in the main text, the interesting contribution is that coming from $R=l_i=0$. Under this simplification, the equation to solve now reads
\begin{equation}
\partial_r\Big(r^7\, (1-\frac{r_{\star}^8}{r^8})\, \partial_r\psi_I\Big)=-(2\pi \ell_p)^6N\, \delta(r-r_0) \ .
\end{equation}
Solving for $r>r_0$ and $r<r_0$ we obtain
\begin{equation}
\psi_>=\frac{1}{r^6}\, _2F_1\Big(\frac{6}{8},\, 1,\, \frac{7}{4},\frac{r_{\star}^8}{r^8}\Big) \ ,
\quad
\psi_<=\frac{1}{r_0^6}\, _2F_1\Big(\frac{3}{4},\, 1,\, \frac{7}{4},\frac{r_{\star}^8}{r_0^8}\Big) \, _2F_1\Big(0,\, \frac{6}{8},\, \frac{3}{4},\frac{r^8}{r_{\star}^8}\Big) \ .
\label{warp-Q222-I}
\end{equation}
Indeed, the leading term for large $r$ corresponds to $l_i=0$, and in this limit 
\begin{equation}
h\sim \frac{ |\mathcal{C}_0|^2}{r^6} \equiv \frac{R^6}{r^6}\ , \, \text{where} \quad R^6= \frac{2^9\, \pi^2\, N\, l_p^6}{3} \ .
\label{warp-Q222-II}
\end{equation}

\addcontentsline{toc}{chapter}{Bibliography}

\end{document}